\documentclass[aps,prd,onecolumn]{revtex4-2}
\usepackage{amsmath,amsthm}
\usepackage{amsfonts}
\usepackage{amssymb}
\usepackage{amsbsy,bm}
\usepackage{times}

\usepackage[us,long,24hr]{datetime}
\usepackage{mathdots}

\theoremstyle{plain}
\newtheorem{theorem}{Theorem}

\theoremstyle{definition}

\theoremstyle{remark}

\newtheorem{example}[theorem]{Example}

\renewcommand{\mathbf}{\bm}
\numberwithin{equation}{section}

\begin{document}

\title{Polyadic braid operators and higher braiding gates}

\author{Steven Duplij}
\email{douplii@uni-muenster.de, sduplij@gmail.com, https://ivv5hpp.uni-muenster.de/u/douplii}
\author{Raimund Vogl}
\email{rvogl@uni-muenster.de}

\affiliation{ Center for Information Technology (WWU IT),
University of M\"unster,
D-48149 M\"unster,
Deutschland}

\date{\textit{start} February 14, 2021,
\textit{completion} July 5, 2021
}

\begin{abstract}

Higher braiding gates, a new kind of quantum gate, are introduced. These are matrix solutions of the polyadic braid equations (which differ from the generalized Yang-Baxter
equations).  Such gates support a special kind of multi-qubit
entanglement which can speed up key distribution and accelerate the execution of algorithms. Ternary braiding gates acting on three qubit states are studied in detail. We
also consider exotic non-invertible gates which can be related to qubit loss,
and define partial identities (which can be orthogonal), partial unitarity,
and partially bounded operators (which can be non-invertible). We define two
classes of matrices, the star and circle types, and find that the magic matrices
(connected with the Cartan decomposition) belong to the star class. The
general algebraic structure of the classes introduced here is described in terms of
semigroups, ternary and $5$-ary groups and modules. The higher braid group and
its representation by higher braid operators are given. Finally, we show
that for each multi-qubit state there exist higher braiding gates which are not
entangling, and the concrete conditions to be non-entangling are given for the
binary and ternary gates discussed.

\end{abstract}

\maketitle
\tableofcontents
\newpage

\section{Introduction}

The modern development of the quantum computing technique implies various
extensions of its foundational concepts \cite{nie/chu,kay/laf/mos,wil/cle}.
One of the main problems in the physical realization of quantum computers is
presence of errors, which implies that it is desirable that quantum
computations be provided with error correction, or that ways be found to make
the states more stable, which leads to the concept of topological quantum
computation (for reviews, see, e.g.,
\cite{fre/kit/lar/wan,nay/sim/ste/das,row/wan}, and references therein). In
the Turaev approach \cite{tur88}, link invariants can be obtained from the
solutions of the constant Yang-Baxter equation (the braid equation). It was
realized that the topological entanglement of knots and links is deeply
connected with quantum entanglement \cite{ara97,kau/lom2002}. Indeed, if the
solutions to the constant Yang-Baxter equation \cite{lam/rad} (Yang-Baxter
operators/maps \cite{buk98,ves2002}) are interpreted as a special class of
quantum gate, namely braiding quantum gates \cite{kau/lom2004,mel/mir/mor},
then the inclusion of non-entangling gates does not change the relevant
topological invariants \cite{ala/jar/jor,kau/meh}. For further properties and
applications of braiding quantum gates, see
\cite{mel/mir/mor2019,bal/wu,kol/mor,kol/mir/mor}.

In this paper we obtain and study the solutions to the higher arity (polyadic)
braid equations introduced in \cite{dup2021a,dup2021b}, as a polyadic
generalization of the constant Yang-Baxter equation (which is considerably
different from the generalized Yang-Baxter equation of
\cite{row/zha/wu/ge,kit/wan,vas/wan/won,pad/sug/tra2020}). We introduce
special classes of matrices (star and circle types), to which most of the
solutions belong, and find that the so-called magic matrices
\cite{kha/gla,kra/cir,bal/wu} belong to the star class. We investigate their
general non-trivial group properties and polyadic generalizations. We then
consider the invertible and non-invertible matrix solutions to the higher
braid equations as the corresponding higher braiding gates acting on
multi-qubit states. It is important that multi-qubit entanglement can speed up
quantum key distribution \cite{epp/kam/mac/bru} and accelerate various
algorithms \cite{var/nis/nak/sal}. As an example, we have made detailed
computations for the ternary braiding gates as solutions to the ternary braid
equations \cite{dup2021a,dup2021b}. A particular solution to the $n$-ary braid
equation is also presented. It is shown, that for each multi-qubit state there
exist higher braiding gates which are not entangling, and the concrete
relations for that are obtained, which can allow us to build non-entangling networks.

\section{Yang-Baxter operators}

Recall here \cite{kau/lom2002,kau/lom2004} the standard construction of the
special kind of gates we will consider, the braiding gates, in terms of
solutions to the \textit{constant Yang-Baxter equation} \cite{lam/rad} (called
also \textit{algebraic} Yang-Baxter equation \cite{dye}), or the (binary)
\textit{braid equation} \cite{dup2021a}.

\subsection{\label{sec-YBmaps}Yang-Baxter maps and braid group}

First we consider a general abstract construction of the (binary) braid
equation. Let $\mathrm{V}$ be a vector space over a field $\mathbb{K}$ and the
mapping $\mathrm{C}_{\mathrm{V}^{2}}:\mathrm{V}\otimes\mathrm{V}%
\rightarrow\mathrm{V}\otimes\mathrm{V}$, where $\otimes=\otimes_{\mathbb{K}}$
is the tensor product over $\mathbb{K}$. A linear operator (\textit{braid
operator}) $\mathrm{C}_{\mathrm{V}^{2}}$ is called a \textit{Yang-Baxter
operator} (denoted by $R$ in \cite{kau/lom2004} and by $B$ in \cite{lam/rad})
or \textit{Yang-Baxter map} \cite{ves2002} (denoted by $F$ in \cite{buk98}),
if it satisfies the \textit{braid equation} \cite{dri89,dri92,kassel}
\begin{equation}
\left(  \mathrm{C}_{\mathrm{V}^{2}}\otimes\operatorname*{id}%
\nolimits_{\mathrm{V}}\right)  \circ\left(  \operatorname*{id}%
\nolimits_{\mathrm{V}}\otimes\mathrm{C}_{\mathrm{V}^{2}}\right)  \circ\left(
\mathrm{C}_{\mathrm{V}^{2}}\otimes\operatorname*{id}\nolimits_{\mathrm{V}%
}\right)  =\left(  \operatorname*{id}\nolimits_{\mathrm{V}}\otimes
\mathrm{C}_{\mathrm{V}^{2}}\right)  \circ\left(  \mathrm{C}_{\mathrm{V}^{2}%
}\otimes\operatorname*{id}\nolimits_{\mathrm{V}}\right)  \circ\left(
\operatorname*{id}\nolimits_{\mathrm{V}}\otimes\mathrm{C}_{\mathrm{V}^{2}%
}\right)  , \label{ri}%
\end{equation}
where $\operatorname*{id}\nolimits_{\mathrm{V}}:\mathrm{V}\rightarrow
\mathrm{V}$, is the identity operator in $\mathrm{V}$. The connection of
$\mathrm{C}_{\mathrm{V}^{2}}$ with the $R$-matrix $\mathrm{R}$ is given by
$\mathrm{C}_{\mathrm{V}^{2}}=\tau\circ\mathrm{R}$, where $\tau$ is the flip
operation \cite{dri89,buk98,lam/rad}.

Let us introduce the operators $\mathrm{A}_{1,2}:\mathrm{V}\otimes
\mathrm{V}\otimes\mathrm{V}\rightarrow\mathrm{V}\otimes\mathrm{V}%
\otimes\mathrm{V}$ by%
\begin{equation}
\mathrm{A}_{1}=\mathrm{C}_{\mathrm{V}^{2}}\otimes\operatorname*{id}%
\nolimits_{\mathrm{V}},\ \ \ \mathrm{A}_{2}=\operatorname*{id}%
\nolimits_{\mathrm{V}}\otimes\mathrm{C}_{\mathrm{V}^{2}}, \label{ar}%
\end{equation}
It follows from (\ref{ri}) that%
\begin{equation}
\mathrm{A}_{1}\circ\mathrm{A}_{2}\circ\mathrm{A}_{1}=\mathrm{A}_{2}%
\circ\mathrm{A}_{1}\circ\mathrm{A}_{2}. \label{aaa}%
\end{equation}

If $\mathrm{C}_{\mathrm{V}^{2}}$ is invertible, then $\mathrm{C}%
_{\mathrm{V}^{2}}^{-1}$ is also the Yang-Baxter map with $\mathrm{A}_{1}^{-1}$
and $\mathrm{A}_{2}^{-1}$. Therefore, the operators $\mathrm{A}_{i}$ represent
the braid group $\mathcal{B}_{3}=\left\{  e,\sigma_{1},\sigma_{2}\mid
\sigma_{1}\sigma_{2}\sigma_{1}=\sigma_{2}\sigma_{1}\sigma_{2}\right\}  $ by
the mapping $\pi_{3}$ as%
\begin{equation}
\mathcal{B}_{3}\overset{\pi_{3}}{\longrightarrow}\operatorname*{End}\left(
\mathrm{V}\otimes\mathrm{V}\otimes\mathrm{V}\right)  ,\ \ \sigma_{1}%
\overset{\pi_{3}}{\mapsto}\mathrm{A}_{1},\ \ \sigma_{2}\overset{\pi_{3}%
}{\mapsto}\mathrm{A}_{2},\ \ e\overset{\pi_{3}}{\mapsto}\operatorname*{id}%
\nolimits_{\mathrm{V}}. \label{b3}%
\end{equation}

The representation $\pi_{m}$ of the braid group with $m$ strands%
\begin{equation}
\mathcal{B}_{m}=\left\{  e,\sigma_{1},\ldots,\sigma_{m-1}\right.  \left\vert
\begin{array}
[c]{c}%
\sigma_{i}\sigma_{i+1}\sigma_{i}=\sigma_{i+1}\sigma_{i}\sigma_{i+1}%
,\ \ i=1,\ldots,m-1,\\
\sigma_{i}\sigma_{j}=\sigma_{j}\sigma_{i},\ \ \ \left\vert i-j\right\vert
\geq2,
\end{array}
\right\}  \label{bm}%
\end{equation}
can be obtained using operators $\mathrm{A}_{i}\left(  m\right)  :$%
\textrm{V}$^{\otimes m}\rightarrow$\textrm{V}$^{\otimes m}$ analogous to
(\ref{ar})%
\begin{equation}
\mathrm{A}_{i}\left(  m\right)  =\overset{i-1}{\overbrace{\operatorname*{id}%
\nolimits_{\mathrm{V}}\otimes\ldots\otimes\operatorname*{id}%
\nolimits_{\mathrm{V}}}}\otimes\mathrm{C}_{\mathrm{V}^{2}}\otimes
\overset{m-i-1}{\overbrace{\operatorname*{id}\nolimits_{\mathrm{V}}%
\otimes\ldots\operatorname*{id}\nolimits_{\mathrm{V}}}},\ \ \ \mathrm{A}%
_{0}\left(  m\right)  =\left(  \operatorname*{id}\nolimits_{\mathrm{V}%
}\right)  ^{\otimes m},\ \ i=1,\ldots,m-1, \label{am}%
\end{equation}
by the mapping $\pi_{m}:\mathcal{B}_{m}\rightarrow\operatorname*{End}%
$\textrm{V}$^{\otimes m}$ in the following way%
\begin{equation}
\pi_{m}\left(  \sigma_{i}\right)  =\mathrm{A}_{i}\left(  m\right)
,\ \ \ \ \ \pi_{m}\left(  e\right)  =\mathrm{A}_{0}\left(  m\right)  .
\end{equation}

In this notation (\ref{ar}) is $\mathrm{A}_{i}=\mathrm{A}_{i}\left(  2\right)
$, $i=1,2$, and therefore (\ref{aaa}) represents $\mathcal{B}_{3}$ by
(\ref{b3}).

\subsection{Constant matrix solutions to the Yang-Baxter equation}

Consider next a concrete version of the vector space $\mathrm{V}$ which is
used in the quantum computation, a $d$-dimensional euclidean vector space
$V_{d}$ over complex numbers $\mathbb{C}$ with a basis $\left\{
e_{i}\right\}  $, $i=1,\ldots,d$. A linear operator $V_{d}\rightarrow V_{d}$
is given by a complex $d\times d$ matrix, the identity operator
$\operatorname*{id}\nolimits_{\mathrm{V}}$ becomes the identity $d\times d$
matrix $I_{d}$, and the Yang-Baxter map $\mathrm{C}_{\mathrm{V}^{2}}$ is a
$d^{2}\times d^{2}$ matrix $C_{d^{2}}$ (denoted by $R$ in \cite{dye})
satisfying the matrix algebraic Yang-Baxter equation%
\begin{equation}
\left(  C_{d^{2}}\otimes I_{d}\right)  \left(  I_{d}\otimes C_{d^{2}}\right)
\left(  C_{d^{2}}\otimes I_{d}\right)  =\left(  I_{d}\otimes C_{d^{2}}\right)
\left(  C_{d^{2}}\otimes I_{d}\right)  \left(  I_{d}\otimes C_{d^{2}}\right)
, \label{rd}%
\end{equation}
being an equality between two matrices of size $d^{3}\times d^{3}$. We use the
unified notations which can be straightforwardly generalized for higher braid
operators. In components%
\begin{equation}
C_{d^{2}}\circ\left(  e_{i_{1}}\otimes e_{i_{2}}\right)  =\sum_{j_{1}^{\prime
},j_{2}^{\prime}=1}^{d}c_{i_{1}i_{2}}^{\ \ \ \ j_{1}^{\prime}j_{2}^{\prime}%
}\cdot e_{j_{1}^{\prime}}\otimes e_{j_{2}^{\prime}}, \label{rr}%
\end{equation}
the Yang-Baxter equation (\ref{rd}) has the shape (where summing is by primed
indices)%
\begin{equation}
\sum_{j_{1}^{\prime},j_{2}^{\prime},j_{3}^{\prime}=1}^{d}c_{i_{1}i_{2}%
}^{\ \ \ \ j_{1}^{\prime}j_{2}^{\prime}}\cdot c_{j_{2}^{\prime}i_{3}%
}^{\ \ \ \ j_{3}^{\prime}k_{3}}\cdot c_{j_{1}^{\prime}j_{3}^{\prime}%
}^{\ \ \ \ k_{1}k_{2}}=\sum_{l_{1}^{\prime},l_{2}^{\prime},l_{3}^{\prime}%
=1}^{d}c_{i_{2}i_{3}}^{\ \ \ \ l_{2}^{\prime}l_{3}^{\prime}}\cdot
c_{i_{1}l_{2}^{\prime}}^{\ \ \ \ k_{1}l_{1}^{\prime}}\cdot c_{l_{1}^{\prime
}l_{3}^{\prime}}^{\ \ \ \ k_{2}k_{3}}\equiv q_{i_{1}i_{2}i_{3}}%
^{\ \ \ \ \ \ k_{1}k_{2}k_{3}}. \label{rrr}%
\end{equation}

The system (\ref{rrr}) is highly overdetermined, because the matrix $C_{d^{2}%
}$ contains $d^{4}$ unknown entries, while there are $d^{6}$ cubic polynomial
equations for them. So for $d=2$ we have $64$ equations for $16$ unknowns,
while for $d=3$ there are $729$ equations for the $81$ unknown entries of
$C_{d^{2}}$. The unitarity of $C_{d^{2}}$ imposes a further $d^{2}$ quadratic
equations, and so for $d=2$ we have in total $68$ equations for $16$ unknowns.
This makes the direct discovery of solutions for the matrix Yang-Baxter
equation (\ref{rrr}) very cumbersome. Nevertheless, using a conjugation
classes method, the unitary solutions and their classification for $d=2$ were
presented in \cite{dye}.

In the standard matrix form (\ref{rr}) can be presented by introducing the
$4$-dimensional vector space $\tilde{V}_{4}=V\otimes V$ with the natural basis
$\tilde{e}_{\tilde{k}}=\left\{  e_{1}\otimes e_{1},e_{1}\otimes e_{2}%
,e_{2}\otimes e_{1},e_{2}\otimes e_{2}\right\}  $, where $\tilde{k}%
=1,\ldots,8$ is a cumulative index. The linear operator $\tilde{C}_{4}%
:\tilde{V}_{4}\rightarrow\tilde{V}_{4}$ corresponding to (\ref{rr}) is given
by $4\times4$ matrix $\tilde{c}_{\tilde{\imath}\tilde{j}}$ as $\tilde{C}%
_{4}\circ\tilde{e}_{\tilde{\imath}}=\sum_{\tilde{j}=1}^{4}\tilde{c}%
_{\tilde{\imath}\tilde{j}}\cdot\tilde{e}_{\tilde{j}}$. The operators
(\ref{ar}) become two $8\times8$ matrices $\tilde{A}_{1,2}$ as%
\begin{equation}
\tilde{A}_{1}=\tilde{c}\otimes_{K}I_{2},\ \ \ \tilde{A}_{2}=I_{2}\otimes
_{K}\tilde{c}, \label{aci}%
\end{equation}
where $\otimes_{K}$ is the Kronecker product of matrices and $I_{2}$ is the
$2\times2$ identity matrix. In this notation (which is universal and also used
for higher braid equations) the operator binary braid equations (\ref{ci})
become a single matrix equation%
\begin{equation}
\tilde{A}_{1}\tilde{A}_{2}\tilde{A}_{1}=\tilde{A}_{2}\tilde{A}_{1}\tilde
{A}_{2}, \label{a123}%
\end{equation}
which we call the \textit{matrix binary braid equation} (and also the constant
Yang-Baxter equation \cite{dye}). In component form (\ref{a123}) is a highly
overdetermined system of $64$ cubic equations for $16$ unknowns, the entries
of $\tilde{c}$.

The matrix equation (\ref{a123}) has the following \textquotedblleft gauge
invariance\textquotedblright, which allows a classification of Yang-Baxter
maps \cite{hie1}. Introduce an invertible operator $Q:V\rightarrow V$ in the
two-dimensional vector space $V\equiv V_{d=2}$. In the basis $\left\{
e_{1},e_{2}\right\}  $ its $2\times2$ matrix $q$ is given by $Q\circ
e_{i}=\sum_{j=1}^{2}q_{ij}\cdot e_{j}$. In the natural $4$-dimensional basis
$\tilde{e}_{\tilde{k}}$ the tensor product of operators $Q\otimes Q$ is
presented by the Kronecker product of matrices $\tilde{q}_{4}=q\otimes_{K}q$.
If the $4\times4$ matrix $\tilde{c}$ is a fixed solution to the Yang-Baxter
equation (\ref{a123}), then the family of solutions $\tilde{c}\left(
q\right)  $ corresponding to the invertible $2\times2$ matrix $q$ is the
conjugation of $\tilde{c}$ by $\tilde{q}_{4}$ such that%
\begin{equation}
\tilde{c}\left(  q\right)  =\tilde{q}_{4}\tilde{c}\tilde{q}_{4}^{-1}=\left(
q\otimes_{K}q\right)  \tilde{c}\left(  q^{-1}\otimes_{K}q^{-1}\right)  ,
\label{cq4}%
\end{equation}
which follows from conjugating (\ref{a123}) by $q\otimes_{K}q$ and using
(\ref{aci}). If we include the obvious invariance of (\ref{a123}) with respect
to an overall factor $t\in\mathbb{C}$, the general family of solutions becomes
(cf. the Yang-Baxter equation \cite{hie1})%
\begin{equation}
\tilde{c}\left(  q,t\right)  =t\tilde{q}_{4}\tilde{c}\tilde{q}_{4}%
^{-1}=t\left(  q\otimes_{K}q\right)  \tilde{c}\left(  q^{-1}\otimes_{K}%
q^{-1}\right)  . \label{cq4t}%
\end{equation}
It follows from (\ref{cq4}) that the matrix $q\in\mathrm{GL}\left(
2,\mathbb{C}\right)  $ is defined up to a complex non-zero factor. In this
case we can put%
\begin{equation}
q=\left(
\begin{array}
[c]{cc}%
a & 1\\
c & d
\end{array}
\right)  , \label{q}%
\end{equation}
and the manifest form of $\tilde{q}_{4}$ is%
\begin{equation}
\tilde{q}_{4}=\left(
\begin{array}
[c]{cccc}%
a^{2} & a & a & 1\\
ac & ad & c & d\\
ac & c & ad & d\\
c^{2} & cd & cd & d^{2}%
\end{array}
\right)  . \label{q4}%
\end{equation}
The matrix $\tilde{q}_{4}^{\mathbf{\star}}\tilde{q}_{4}$ (where $\mathbf{\star
}$ represents Hermitian conjugation) is diagonal (this case is important in a
further classification similar to the binary one \cite{dye}), when the
condition%
\begin{equation}
c=-a/d^{\ast} \label{abc}%
\end{equation}
holds, and so the matrix $q$ takes the special form (depending on $2$ complex
parameters)%
\begin{equation}
q=\left(
\begin{array}
[c]{cc}%
a & 1\\
-a/d^{\ast} & d
\end{array}
\right)  . \label{qc}%
\end{equation}

We call two solutions $\tilde{c}_{1}$ and $\tilde{c}_{2}$ of the constant
Yang-Baxter equation (\ref{a123}) $q$-\textit{conjugated}, if
\begin{equation}
\tilde{c}_{1}\tilde{q}_{4}=\tilde{q}_{4}\tilde{c}_{2}, \label{cqq}%
\end{equation}
and we will not distinguish between them. The $q$-conjugation in the form
(\ref{cqq}) does not require the invertibility of the matrix $q$, and
therefore the solutions of different ranks (or invertible and not invertible)
can be $q$-conjugated (for the invertible case, see
\cite{hie1,ala/bap/jor,pad/sug/tra2021}).

The matrix equation (\ref{a123}) does not imply the invertibility of
solutions, i.e. matrices $\tilde{c}$ being of full rank (in the binary
Yang-Baxter case of rank $4$ and $d=2$). Therefore, below we introduce in a
unified way invertible and non-invertible solutions to the matrix Yang-Baxter
equation (\ref{rrr}) for any rank of the corresponding matrices.

\subsection{Partial identity and unitarity}

To be as close as possible to the invertible case, we introduce
\textquotedblleft non-invertible analogs\textquotedblright\ of identity and
unitarity. Let $M$ be a diagonal $n\times n$ matrix of rank $r\leq n$, and
therefore with $n-r$ zeroes on the diagonal. If the other diagonal elements
are units, such a diagonal $M$ can be reduced by row operations to a block
matrix, being a direct sum of the identity matrix $I_{r\times r}$ and the zero
matrix $Z_{\left(  n-r\right)  \times\left(  n-r\right)  }$. We call such a
diagonal matrix a \textit{block }$r$-\textit{partial identity} $I_{n}^{\left(
block\right)  }\left(  r\right)  =\operatorname*{diag}\left\{  \overset
{r}{\overbrace{1,\ldots1},}\overset{n-r}{\overbrace{0,\ldots,0}}\right\}  $,
and without the block reduction---a \textit{shuffle }$r$-\textit{partial}
\textit{identity} $I_{n}^{\left(  shuffle\right)  }\left(  r\right)  $ (these
are connected by conjugation). We will use the term partial identity and
$I_{n}\left(  r\right)  $ to denote any matrix of this form. Obviously, with
the full rank $r=n$ we have $I_{n}\left(  n\right)  \equiv I_{n}$, where
$I_{n}$ is the identity $n\times n$ matrix. As with the invertible case and
identities, the partial identities (of the corresponding form) are
\textit{trivial solutions} of the Yang-Baxter equation.

If a matrix $M=M\left(  r\right)  $ of size $n\times n$ and rank $r$ satisfies
the following $r$-\textit{partial unitarity} condition%
\begin{align}
M\left(  r\right)  ^{\mathbf{\star}}M\left(  r\right)   &  =I_{n}^{\left(
1\right)  }\left(  r\right)  ,\label{mm}\\
M\left(  r\right)  M\left(  r\right)  ^{\mathbf{\star}}  &  =I_{n}^{\left(
2\right)  }\left(  r\right)  , \label{mm1}%
\end{align}
where $M\left(  r\right)  ^{\mathbf{\star}}$ is the conjugate-transposed
matrix and $I_{n}^{\left(  1\right)  }\left(  r\right)  $, $I_{n}^{\left(
2\right)  }\left(  r\right)  $ are partial identities (of any kind, they can
be different), then $M\left(  r\right)  $ is called a $r$-\textit{partial
unitary matrix}. In the case, when $I_{n}^{\left(  1\right)  }\left(
r\right)  =I_{n}^{\left(  2\right)  }\left(  r\right)  $, the matrix $M\left(
r\right)  $ is called \textit{normal}. If $M\left(  r\right)  ^{\mathbf{\star
}}=M\left(  r\right)  $, then it is called $r$-\textit{partial self-adjoint}.
In the case of full rank $r=n$, the conditions (\ref{mm})--(\ref{mm1}) become
ordinary unitarity, and $M\left(  n\right)  $ becomes an unitary (and normal)
matrix, while a $r$-partial self-adjoint matrix becomes a self-adjoint matrix
or Hermitian matrix.

As an example, we consider a $4\times4$ matrix of rank $3$%
\begin{equation}
M\left(  3\right)  =\left(
\begin{array}
[c]{cccc}%
0 & 0 & 0 & 0\\
0 & e^{i\beta} & 0 & 0\\
0 & 0 & 0 & e^{i\gamma}\\
e^{i\alpha} & 0 & 0 & 0
\end{array}
\right)  ,\ \ \ \ \alpha,\beta,\gamma\in\mathbb{R}, \label{m3}%
\end{equation}
which satisfies the $3$-partial unitarity conditions (\ref{mm})--(\ref{mm1})
with two different $3$-partial identities on the r.h.s.%
\begin{equation}
M\left(  3\right)  ^{\mathbf{\star}}M\left(  3\right)  =\left(
\begin{array}
[c]{cccc}%
1 & 0 & 0 & 0\\
0 & 1 & 0 & 0\\
0 & 0 & 0 & 0\\
0 & 0 & 0 & 1
\end{array}
\right)  =I_{4}^{\left(  1\right)  }\left(  3\right)  \neq I_{4}^{\left(
2\right)  }\left(  3\right)  =\left(
\begin{array}
[c]{cccc}%
0 & 0 & 0 & 0\\
0 & 1 & 0 & 0\\
0 & 0 & 1 & 0\\
0 & 0 & 0 & 1
\end{array}
\right)  =M\left(  3\right)  M\left(  3\right)  ^{\mathbf{\star}}.
\end{equation}

For a non-invertible matrix $M\left(  r\right)  $ one can define a
\textit{pseudoinverse} $M\left(  r\right)  ^{\mathbf{+}}$ (or the
\textit{Moore-Penrose inverse}) \cite{nashed} by%
\begin{equation}
M\left(  r\right)  M\left(  r\right)  ^{\mathbf{+}}M\left(  r\right)
=M\left(  r\right)  ,\ \ M\left(  r\right)  ^{\mathbf{+}}M\left(  r\right)
M\left(  r\right)  ^{\mathbf{+}}=M\left(  r\right)  ^{\mathbf{+}}, \label{mp1}%
\end{equation}
and $M\left(  r\right)  M\left(  r\right)  ^{\mathbf{+}}$, $M\left(  r\right)
^{\mathbf{+}}M\left(  r\right)  $ are Hermitian. In the case of (\ref{m3}) the
partial unitary matrix $M\left(  3\right)  $ coincides with its pseudoinverse%
\begin{equation}
M\left(  3\right)  ^{\mathbf{\star}}=M\left(  3\right)  ^{\mathbf{+}},
\label{mp}%
\end{equation}
which is similar to the standard unitarity $M_{inv}^{\mathbf{\star}}%
=M_{inv}^{-1}$ for an invertible matrix $M_{inv}$. It is important that
(\ref{m3}) is a solution of the matrix Yang-Baxter equation (\ref{a123}), and
so is an example of a non-invertible Yang-Baxter map.

If only the first (second) of the conditions (\ref{mm})--(\ref{mm1}) holds, we
call such $M\left(  r\right)  $ a \textit{left (right) }$r$-\textit{partial
unitary matrix}. An example of such a non-invertible Yang-Baxter map of rank
$2$ is the left $2$-partial unitary matrix%
\begin{equation}
M\left(  2\right)  =\frac{1}{\sqrt{2}}\left(
\begin{array}
[c]{cccc}%
0 & 0 & 0 & e^{i\alpha}\\
0 & e^{i\beta} & 0 & 0\\
0 & e^{i\beta} & 0 & 0\\
0 & 0 & 0 & e^{i\beta}%
\end{array}
\right)  ,\ \ \ \ \alpha,\beta\in\mathbb{R},
\end{equation}
which satisfies (\ref{mm}), but not (\ref{mm1}), and so $M\left(  2\right)  $
is not normal%
\begin{equation}
M\left(  2\right)  ^{\mathbf{\star}}M\left(  2\right)  =\left(
\begin{array}
[c]{cccc}%
0 & 0 & 0 & 0\\
0 & 1 & 0 & 0\\
0 & 0 & 0 & 0\\
0 & 0 & 0 & 1
\end{array}
\right)  \neq\left(
\begin{array}
[c]{cccc}%
1 & 0 & 0 & e^{i\left(  \alpha-\beta\right)  }\\
0 & 1 & 1 & 0\\
0 & 1 & 1 & 0\\
e^{i\left(  \beta-\alpha\right)  } & 0 & 0 & 1
\end{array}
\right)  =M\left(  2\right)  M\left(  2\right)  ^{\mathbf{\star}}.
\end{equation}

Nevertheless, the property (\ref{mp}) still holds and $M\left(  2\right)
^{\mathbf{\star}}=M\left(  2\right)  ^{\mathbf{+}}$.

\subsection{Permutation and parameter-permutation $4$-vertex Yang-Baxter
maps\label{subsec-perm}}

The system (\ref{a123}) with respect to all $16$ variables is too cumbersome
for direct solution. The classification of all solutions can only be
accomplished in special cases, e.g. for matrices over finite fields
\cite{hie1} or for fewer than $16$ vertices. Here we will start from
$4$-vertex permutation and parameter-permutation matrix solutions and
investigate their group structure. It was shown \cite{dye,kau/lom2004} that
the special $8$-vertex solutions to the Yang-Baxter equation are most
important for further applications including braiding gates. We will therefore
study the $8$-vertex solutions in the most general way: over $\mathbb{C}$ and
in various configurations, invertible and not invertible, and also consider
their group structure.

First, we introduce the \textit{permutation Yang-Baxter maps} which are
presented by the permutation matrices (binary matrices with a single $1$ in
each row and column), i.e. $4$-vertex solutions. In total, there are $64$
permutation matrices of size $4\times4$, while only $4$ of them have the full
rank $4$ and simultaneously satisfy the Yang-Baxter equation (\ref{a123}).
These are the following%
\begin{align}
\tilde{c}_{bisymm}^{perm}  &  =\left(
\begin{array}
[c]{cccc}%
1 & 0 & 0 & 0\\
0 & 0 & 1 & 0\\
0 & 1 & 0 & 0\\
0 & 0 & 0 & 1
\end{array}
\right)  ,\ \left(
\begin{array}
[c]{cccc}%
0 & 0 & 0 & 1\\
0 & 1 & 0 & 0\\
0 & 0 & 1 & 0\\
1 & 0 & 0 & 0
\end{array}
\right)  ,\ \
\begin{array}
[c]{c}%
\operatorname*{tr}\tilde{c}=2,\ \ \det\tilde{c}=-1,\\
\text{eigenvalues: }\left\{  1\right\}  ^{\left[  2\right]  },\left\{
-1\right\}  ^{\left[  2\right]  },
\end{array}
\label{cp}\\
\tilde{c}_{90symm}^{perm}  &  =\left(
\begin{array}
[c]{cccc}%
0 & 1 & 0 & 0\\
0 & 0 & 0 & 1\\
1 & 0 & 0 & 0\\
0 & 0 & 1 & 0
\end{array}
\right)  ,\ \left(
\begin{array}
[c]{cccc}%
0 & 0 & 1 & 0\\
1 & 0 & 0 & 0\\
0 & 0 & 0 & 1\\
0 & 1 & 0 & 0
\end{array}
\right)  ,\ \
\begin{array}
[c]{c}%
\operatorname*{tr}\tilde{c}=0,\ \ \det\tilde{c}=-1,\\
\text{eigenvalues: }1,i,-1,-i.
\end{array}
\label{cpc}%
\end{align}

Here and next we list eigenvalues to understand which matrices are conjugated,
and after that, if and only if the conjugation matrix is of the form
(\ref{q4}), then such solutions to the Yang-Baxter equation (\ref{a123})
coincide. The traces are important in the construction of corresponding link
invariants \cite{tur88} and local invariants \cite{bal/san,sud2001}, and
determinants are connected with the concurrence \cite{wal/gro/eis,jaf/oed}.
Note that the first matrix in (\ref{cp}) is the SWAP quantum gate
\cite{nie/chu}.

To understand symmetry properties of (\ref{cp})--(\ref{cpc}), we introduce the
so called \textit{reverse matrix} $J\equiv J_{n}$ of size $n\times n$ by
$\left(  J_{n}\right)  _{ij}=\delta_{i,n+1-i}$. For $n=4$ it is%
\begin{equation}
J_{4}=\left(
\begin{array}
[c]{cccc}%
0 & 0 & 0 & 1\\
0 & 0 & 1 & 0\\
0 & 1 & 0 & 0\\
1 & 0 & 0 & 0
\end{array}
\right)  . \label{j4}%
\end{equation}

For any $n\times n$ matrix $M\equiv M_{n}$ the matrix $JM$ is the matrix $M$
reflected vertically, and the product $MJ$ is $M$ reflected horizontally. In
addition to the standard \textit{symmetric matrix} satisfying $M=M^{T}$ ($T$
is the transposition), one can introduce%
\begin{align}
M\text{ is \textit{persymmetric},\textit{ }if }JM  &  =\left(  JM\right)
^{T},\label{mt1}\\
M\text{ is \textit{90}}^{\circ}\text{\textit{-symmetric},\textit{ }if }M^{T}
&  =JM. \label{mt2}%
\end{align}

Thus, a persymmetric matrix is symmetric with respect to the minor diagonal,
while a 90$^{\circ}$-symmetric matrix is symmetric under 90$^{\circ}%
$-rotations. A \textit{bisymmetric matrix} is symmetric and persymmetric
simultaneously. In this notation, the first family of the permutation
solutions (\ref{cp}) are bisymmetric, but not 90$^{\circ}$-symmetric, while
the second family of the solutions (\ref{cpc}) are, oppositely, 90$^{\circ}%
$-symmetric, but not symmetric and not persymmetric (which explains their notation).

In the next step, we define the corresponding \textit{parameter-permutation
solutions} replacing the units in (\ref{cp}) by parameters. We found the
following four $4$-vertex solutions to the Yang-Baxter equation (\ref{a123})
over $\mathbb{C}$%
\begin{align}
\tilde{c}_{rank=4}^{perm,star}\left(  x,y,z,t\right)   &  =\left(
\begin{array}
[c]{cccc}%
x & 0 & 0 & 0\\
0 & 0 & y & 0\\
0 & z & 0 & 0\\
0 & 0 & 0 & t
\end{array}
\right)  ,\ \left(
\begin{array}
[c]{cccc}%
0 & 0 & 0 & y\\
0 & x & 0 & 0\\
0 & 0 & t & 0\\
z & 0 & 0 & 0
\end{array}
\right)  ,%
\begin{array}
[c]{c}%
\operatorname*{tr}\tilde{c}=x+t,\\
\det\tilde{c}=-xyzt,\ \ x,y,z,t\neq0,\\
\text{eigenvalues: }x,t,\sqrt{yz},-\sqrt{yz},
\end{array}
\label{cp1}\\
\tilde{c}_{rank=4}^{perm,circ}\left(  x,y\right)   &  =\left(
\begin{array}
[c]{cccc}%
0 & 0 & x & 0\\
y & 0 & 0 & 0\\
0 & 0 & 0 & x\\
0 & y & 0 & 0
\end{array}
\right)  ,\ \left(
\begin{array}
[c]{cccc}%
0 & x & 0 & 0\\
0 & 0 & 0 & x\\
y & 0 & 0 & 0\\
0 & 0 & y & 0
\end{array}
\right)  ,%
\begin{array}
[c]{c}%
\operatorname*{tr}\tilde{c}=0,\\
\det\tilde{c}=-x^{2}y^{2},\ \ x,y\neq0,\\
\text{eigenvalues: }\sqrt{xy},-\sqrt{xy},i\sqrt{xy},-i\sqrt{xy}.
\end{array}
\label{cp2}%
\end{align}

The first pair of solutions (\ref{cp1}) correspond to the bi-symmetric
permutation matrices (\ref{cp}), and we call them \textit{star-like
solutions}, while the second two solutions (\ref{cp2}) correspond to the
90$^{\circ}$-symmetric matrices (\ref{cp}) which are called
\textit{circle-like solutions}.

The first (second) star-like solution in (\ref{cp1}) with $y=z$ ($x=t$)
becomes symmetric (persymmetric), while on the other hand with $x=t$ ($y=z$)
it becomes persymmetric (symmetric). They become bisymmetric
parameter-permutation solutions if all the parameters are equal $x=y=z=t$. The
circle-like solutions (\ref{cp2}) are 90$^{\circ}$-symmetric when $x=y$.

Using $q$-conjugation (\ref{cq4t}) one can next get families of solutions
depending from the entries of $q$, the additional complex parameters in
(\ref{q}).

\subsection{Group structure of $4$-vertex and $8$-vertex matrices}

Let us analyze the group structure of $4$-vertex matrices (\ref{cp1}%
)--(\ref{cp2}) with respect to matrix multiplication, i.e. which kinds of
subgroups in $\mathrm{GL}\left(  4,\mathbb{C}\right)  $ they can form. For
this we introduce four $4$-vertex $4\times4$ matrices over $\mathbb{C}$: two
star-like matrices%
\begin{equation}
N_{star1}=\left(
\begin{array}
[c]{cccc}%
x & 0 & 0 & 0\\
0 & 0 & y & 0\\
0 & z & 0 & 0\\
0 & 0 & 0 & t
\end{array}
\right)  ,\ \ \ \ N_{star2}=\left(
\begin{array}
[c]{cccc}%
0 & 0 & 0 & y\\
0 & x & 0 & 0\\
0 & 0 & t & 0\\
z & 0 & 0 & 0
\end{array}
\right)  ,\ \ \ \ \
\begin{array}
[c]{c}%
\operatorname*{tr}N=x+t,\\
\det N=-xyzt,\ \ x,y,z,t\neq0,\\
\text{eigenvalues: }x,t,\sqrt{yz},-\sqrt{yz},
\end{array}
\label{m4}%
\end{equation}
and two circle-like matrices%
\begin{equation}
N_{circ1}=\left(
\begin{array}
[c]{cccc}%
0 & 0 & x & 0\\
y & 0 & 0 & 0\\
0 & 0 & 0 & z\\
0 & t & 0 & 0
\end{array}
\right)  ,\ \ \ \ N_{circ2}=\left(
\begin{array}
[c]{cccc}%
0 & x & 0 & 0\\
0 & 0 & 0 & y\\
z & 0 & 0 & 0\\
0 & 0 & t & 0
\end{array}
\right)  ,\ \ \
\begin{array}
[c]{c}%
\operatorname*{tr}N=0,\\
\det N=-xyzt,\ \ x,y,z,t\neq0,\\
\text{eigenvalues: }\sqrt[4]{xyzt},-\sqrt[4]{xyzt},i\sqrt[4]{xyzt}%
,-\sqrt[4]{xyzt},
\end{array}
. \label{mc4}%
\end{equation}

Denoting the corresponding sets by $\mathsf{N}_{star1,2}=\left\{
N_{star1,2}\right\}  $ and $\mathsf{N}_{circ1,2}=\left\{  N_{circ1,2}\right\}
$, these do not intersect and are closed with respect to the following
multiplications%
\begin{align}
\mathsf{N}_{star1}\mathsf{N}_{star1}\mathsf{N}_{star1}  &  =\mathsf{N}%
_{star1},\label{ns1}\\
\mathsf{N}_{star2}\mathsf{N}_{star2}\mathsf{N}_{star2}  &  =\mathsf{N}%
_{star2},\label{ns2}\\
\mathsf{N}_{circ1}\mathsf{N}_{circ1}\mathsf{N}_{circ1}\mathsf{N}%
_{circ1}\mathsf{N}_{circ1}  &  =\mathsf{N}_{circ1},\label{nc1}\\
\mathsf{N}_{circ2}\mathsf{N}_{circ2}\mathsf{N}_{circ2}\mathsf{N}%
_{circ2}\mathsf{N}_{circ2}  &  =\mathsf{N}_{circ2}. \label{nc2}%
\end{align}

Note that there are no closed binary multiplications among the sets of
$4$-vertex matrices (\ref{m4})--(\ref{mc4}).

To give a proper group interpretation of (\ref{ns1})--(\ref{nc2}), we
introduce a $k$-\textit{ary (polyadic) general linear semigroup}
$\mathrm{GLS}^{\left[  k\right]  }\left(  n,\mathbb{C}\right)  =\left\{
\mathsf{M}_{full}\mid\mu^{\left[  k\right]  }\right\}  $, where $\mathsf{M}%
_{full}=\left\{  M_{n\times n}\right\}  $ is the set of $n\times n$ matrices
over $\mathbb{C}$ and $\mu^{\left[  k\right]  }$ is an ordinary product of $k$
matrices. The full semigroup $\mathrm{GLS}^{\left[  k\right]  }\left(
n,\mathbb{C}\right)  $ is \textit{derived} in the sense that its product can
be obtained by repeating the binary products which are (binary) closed at each
step. However, $n\times n$ matrices of special shape can form $k$-ary
subsemigroups of $\mathrm{GLS}^{\left[  k\right]  }\left(  n,\mathbb{C}%
\right)  $ which can be closed with respect to the product of at minimum $k$
matrices, but not of $2$ matrices, and we call such semigroups $k$%
-\textit{non-derived}. Moreover, we have for the sets $\mathsf{N}_{star1,2}$
and $\mathsf{N}_{circ1,2}$%
\begin{equation}
\mathsf{M}_{full}=\mathsf{N}_{star1}\cup\mathsf{N}_{star2}\cup\mathsf{N}%
_{circ1}\cup\mathsf{N}_{circ2},\ \ \ \mathsf{N}_{star1}\cap\mathsf{N}%
_{star2}\cap\mathsf{N}_{circ1}\cap\mathsf{N}_{circ2}=\varnothing. \label{mn4}%
\end{equation}

A simple example of a $3$-nonderived subsemigroup of the full semigroup
$\mathrm{GLS}^{\left[  k\right]  }\left(  n,\mathbb{C}\right)  $ is the set of
antidiagonal matrices $\mathsf{M}_{\text{adiag}}=\left\{  M_{\text{adiag}%
}\right\}  $ (having nonzero elements on the minor diagonal only): the product
$\mu^{\left[  3\right]  }$ of $3$ matrices from $\mathsf{M}_{\text{adiag}}$ is
closed, and therefore $\mathsf{M}_{\text{adiag}}$ is a subsemigroup
$\mathcal{S}_{\text{adiag}}^{\left[  3\right]  }=\left\{  \mathsf{M}%
_{\text{adiag}}\mid\mu^{\left[  3\right]  }\right\}  $ of the full ternary
general linear semigroup $\mathrm{GLS}^{\left[  3\right]  }\left(
n,\mathbb{C}\right)  $ with the multiplication $\mu^{\left[  3\right]  }$ as
the ordinary triple matrix product.

In the theory of polyadic groups \cite{dor3} an analog of the binary inverse
$M^{-1}$ is given by the \textit{querelement}, which is denoted by $\bar{M}$
and in the matrix $k$-ary case is defined by%
\begin{equation}
\overset{k-1}{\overbrace{M\ldots M}}\bar{M}=M, \label{mqm}%
\end{equation}
where $\bar{M}$ can be on any place. If each element of the $k$-ary semigroup
$\mathrm{GLS}^{\left[  k\right]  }\left(  n,\mathbb{C}\right)  $ (or its
subsemigroup) has its querelement $\bar{M}$, then this semigroup is a
$k$-\textit{ary general linear group} $\mathrm{GL}^{\left[  k\right]  }\left(
n,\mathbb{C}\right)  $.

In the set of $n\times n$ matrices the binary (ordinary) product is defined
(even it is not closed), and for invertible matrices we formally determine the
standard inverse $M^{-1}$, but for arity $k\geq4$ it does not coincide with
the querelement $\bar{M}$, because, as follows from (\ref{mqm}) and
cancellativity in $\mathbb{C}$ that%
\begin{equation}
\bar{M}=M^{2-k}. \label{mmk}%
\end{equation}

The $k$-\textit{ary (polyadic) identity} $I_{n}^{\left[  k\right]  }$ in
$\mathrm{GLS}^{\left[  k\right]  }\left(  n,\mathbb{C}\right)  $ is defined by%
\begin{equation}
\overset{k-1}{\overbrace{I_{n}^{\left[  k\right]  }\ldots I_{n}^{\left[
k\right]  }}}M=M, \label{km}%
\end{equation}
which holds when $M$ in the l.h.s. is on any place. If $M$ is only on one or
another side (but not in the middle places) in (\ref{km}), $I_{n}^{\left[
k\right]  }$ is called \textit{left (right) polyadic identity}. For instance,
in the subsemigroup (in $\mathrm{GLS}^{\left[  k\right]  }\left(
n,\mathbb{C}\right)  $) of antidiagonal matrices $\mathcal{S}_{\text{adiag}%
}^{\left[  3\right]  }$ the ternary identity $I_{n}^{\left[  3\right]  }$ can
be chosen as the $n\times n$ reverse matrix (\ref{j4}) having units on the
minor diagonal, while the ordinary $n\times n$ unit matrix $I_{n}$ is not in
$\mathcal{S}_{\text{adiag}}^{\left[  3\right]  }$. It follows from (\ref{km}),
that for matrices over $\mathbb{C}$ the (left, right) polyadic identity
$I_{n}^{\left[  k\right]  }$ is%
\begin{equation}
\left(  I_{n}^{\left[  k\right]  }\right)  ^{k-1}=I_{n}, \label{ik}%
\end{equation}
which means that for the ordinary matrix product $I_{n}^{\left[  k\right]  }$
is a $\left(  k-1\right)  $-root of $I_{n}$ (or $I_{n}^{\left[  k\right]  }$
is a reflection of $\left(  k-1\right)  $ degree), while both sides cannot
belong to a subsemigroup $\mathcal{S}^{\left[  k\right]  }$ of $\mathrm{GLS}%
^{\left[  k\right]  }\left(  n,\mathbb{C}\right)  $ under consideration (as in
$\mathcal{S}_{\text{adiag}}^{\left[  3\right]  }$). As the solutions of
(\ref{ik}) are not unique, there can be many $k$-ary identities in a $k$-ary
matrix semigroup. We denote the set of $k$-ary identities by $\mathsf{I}%
_{n}^{\left[  k\right]  }=\left\{  I_{n}^{\left[  k\right]  }\right\}  $. In
the case of $\mathcal{S}_{\text{adiag}}^{\left[  3\right]  }$ the ternary
identity $I_{n}^{\left[  3\right]  }$ can be chosen as any of the $n\times n$
reverse matrices (\ref{j4}) with unit complex numbers $e^{i\alpha_{j}}$,
$j=1,\ldots,n$ on the minor diagonal, where $\alpha_{j}$ satisfy additional
conditions depending on the semigroup. In the concrete case of $\mathcal{S}%
_{\text{adiag}}^{\left[  3\right]  }$ the conditions, giving (\ref{ik}), are
$\left(  k-1\right)  \alpha_{j}=1+2\pi r_{j}$, $r_{j}\in\mathbb{Z}$,
$j=1,\ldots,n$.

In the framework of the above definitions, we can interpret the closed
products (\ref{ns1})--(\ref{ns2}) as the multiplications $\mu^{\left[
3\right]  }$ of the \textit{ternary semigroups} $\mathcal{S}_{star1,2}%
^{\left[  3\right]  }\left(  4,\mathbb{C}\right)  =\left\{  \mathsf{N}%
_{star1,2}\mid\mu^{\left[  3\right]  }\right\}  $. The corresponding
querelements are given by%
\begin{equation}
\bar{N}_{star1}=N_{star1}^{-1}=\left(
\begin{array}
[c]{cccc}%
\frac{1}{x} & 0 & 0 & 0\\
0 & 0 & \frac{1}{z} & 0\\
0 & \frac{1}{y} & 0 & 0\\
0 & 0 & 0 & \frac{1}{t}%
\end{array}
\right)  ,\ \ \ \bar{N}_{star2}=N_{star2}^{-1}=\left(
\begin{array}
[c]{cccc}%
0 & 0 & 0 & \frac{1}{z}\\
0 & \frac{1}{x} & 0 & 0\\
0 & 0 & \frac{1}{t} & 0\\
\frac{1}{y} & 0 & 0 & 0
\end{array}
\right)  ,\ \ \ \ \ x,y,z,t\neq0. \label{nn}%
\end{equation}

The ternary semigroups having querelements for each element (i.e. the
additional operation $\overline{\left(  \ \right)  }$ defined by (\ref{nn}))
are the \textit{ternary groups} $\mathcal{G}_{star1,2}^{\left[  3\right]
}\left(  4,\mathbb{C}\right)  =\left\{  \mathsf{N}_{star1,2}\mid\mu^{\left[
3\right]  },\overline{\left(  \ \right)  }\right\}  $ which are two
(non-intersecting because $\mathsf{N}_{star1}\cap\mathsf{N}_{star2}%
=\varnothing$) subgroups of the ternary general linear group $\mathrm{GL}%
^{\left[  3\right]  }\left(  4,\mathbb{C}\right)  $. The ternary identities in
$\mathcal{G}_{star1,2}^{\left[  3\right]  }\left(  4,\mathbb{C}\right)  $ are
the following different continuous sets $\mathsf{I}_{star1,2}^{\left[
3\right]  }=\left\{  I_{star1,2}^{\left[  3\right]  }\right\}  $, where%
\begin{align}
I_{star1}^{\left[  3\right]  }  &  =\left(
\begin{array}
[c]{cccc}%
e^{i\alpha_{1}} & 0 & 0 & 0\\
0 & 0 & e^{i\alpha_{2}} & 0\\
0 & e^{i\alpha_{3}} & 0 & 0\\
0 & 0 & 0 & e^{i\alpha_{4}}%
\end{array}
\right)  ,\ \ \ e^{2i\alpha_{1}}=e^{2i\alpha_{4}}=e^{i\left(  \alpha
_{2}+\alpha_{3}\right)  }=1,\ \ \alpha_{j}\in\mathbb{R},\label{is1}\\
I_{star2}^{\left[  3\right]  }  &  =\left(
\begin{array}
[c]{cccc}%
0 & 0 & 0 & e^{i\alpha_{1}}\\
0 & e^{i\alpha_{2}} & 0 & 0\\
0 & 0 & e^{i\alpha_{3}} & 0\\
e^{i\alpha_{4}} & 0 & 0 & 0
\end{array}
\right)  ,\ \ \ e^{2i\alpha_{2}}=e^{2i\alpha_{3}}=e^{i\left(  \alpha
_{1}+\alpha_{4}\right)  }=1,\ \ \alpha_{j}\in\mathbb{R}. \label{is2}%
\end{align}

In the particular case $\alpha_{j}=0$, $j=1,2,3,4$, the ternary identities
(\ref{is1})--(\ref{is2}) coincide with the bisymmetric permutation matrices
(\ref{cp}).

Next we treat the closed set products (\ref{nc1})--(\ref{nc2}) as the
multiplications $\mu^{\left[  5\right]  }$ of the $5$-\textit{ary semigroups}
$\mathcal{S}_{circ1,2}^{\left[  5\right]  }\left(  4,\mathbb{C}\right)
=\left\{  \mathsf{N}_{circ1,2}\mid\mu^{\left[  5\right]  }\right\}  $. The
querelements are%
\begin{equation}
\bar{N}_{circ1}=N_{circ1}^{-3}=\left(
\begin{array}
[c]{cccc}%
0 & 0 & \frac{1}{yzt} & 0\\
\frac{1}{xzt} & 0 & 0 & 0\\
0 & 0 & 0 & \frac{1}{xyt}\\
0 & \frac{1}{xyz} & 0 & 0
\end{array}
\right)  ,\ \ \ \bar{N}_{circ2}=N_{circ2}^{-3}=\left(
\begin{array}
[c]{cccc}%
0 & \frac{1}{yzt} & 0 & 0\\
0 & 0 & 0 & \frac{1}{xzt}\\
\frac{1}{xyt} & 0 & 0 & 0\\
0 & 0 & \frac{1}{xyz} & 0
\end{array}
\right)  ,\ \ \ \ \ x,y,z,t\neq0.
\end{equation}
and the corresponding $5$-\textit{ary groups} $\mathcal{G}_{circ1,2}^{\left[
5\right]  }\left(  4,\mathbb{C}\right)  =\left\{  \mathsf{N}_{circ1,2}\mid
\mu^{\left[  5\right]  },\overline{\left(  \ \right)  }\right\}  $ which are
two (non-intersecting because $\mathsf{N}_{circ1}\cap\mathsf{N}_{circ2}%
=\varnothing$) subgroups of the $5$-ary general linear group $\mathrm{GL}%
^{\left[  5\right]  }\left(  n,\mathbb{C}\right)  $. We have the following
continuous sets of $5$-ary identities $\mathsf{I}_{circ1,2}^{\left[  3\right]
}=\left\{  I_{circ1,2}^{\left[  3\right]  }\right\}  $ in $\mathcal{G}%
_{circ1,2}^{\left[  5\right]  }\left(  4,\mathbb{C}\right)  $ satisfying
\begin{align}
I_{circ1}^{\left[  5\right]  }  &  =\left(
\begin{array}
[c]{cccc}%
0 & 0 & e^{i\alpha_{1}} & 0\\
e^{i\alpha_{2}} & 0 & 0 & 0\\
0 & 0 & 0 & e^{i\alpha_{3}}\\
0 & e^{i\alpha_{4}} & 0 & 0
\end{array}
\right)  ,\ \ \ e^{i\left(  \alpha_{1}+\alpha_{2}+\alpha_{3}+\alpha
_{4}\right)  }=1,\ \ \alpha_{j}\in\mathbb{R},\label{ic1}\\
I_{circ2}^{\left[  5\right]  }  &  =\left(
\begin{array}
[c]{cccc}%
0 & e^{i\alpha_{1}} & 0 & 0\\
0 & 0 & 0 & e^{i\alpha_{2}}\\
e^{i\alpha_{3}} & 0 & 0 & 0\\
0 & 0 & e^{i\alpha_{4}} & 0
\end{array}
\right)  ,\ \ \ e^{i\left(  \alpha_{1}+\alpha_{2}+\alpha_{3}+\alpha
_{4}\right)  }=1,\ \ \alpha_{j}\in\mathbb{R}. \label{ic2}%
\end{align}
In the case $\alpha_{j}=0$, $j=1,2,3,4$, the $5$-ary identities (\ref{ic1}%
)--(\ref{ic2}) coincide with the $90^{\circ}$-symmetric permutation matrices
(\ref{cpc}).

Thus, it follows from (\ref{nn})--(\ref{ic2}) that the $4$-vertex star-like
(\ref{m4}) and circle-like (\ref{mc4}) matrices form subgroups of the $k$-ary
general linear group $\mathrm{GL}^{\left[  k\right]  }\left(  4,\mathbb{C}%
\right)  $ with significantly different properties: they have different
querelements and (sets of) polyadic identities, and even the arities of the
subgroups $\mathcal{G}_{star1,2}^{\left[  3\right]  }\left(  4,\mathbb{C}%
\right)  $ and $\mathcal{G}_{circ1,2}^{\left[  5\right]  }\left(
4,\mathbb{C}\right)  $ do not coincide (\ref{ns1})--(\ref{nc2}). If we take
into account that $4$-vertex star-like (\ref{m4}) and circle-like (\ref{mc4})
matrices are (binary) additive and distributive, then they form (with respect
to the binary matrix addition $\left(  +\right)  $ and the multiplications
$\mu^{\left[  3\right]  }$ and $\mu^{\left[  5\right]  }$) the $\left(
2,3\right)  $-ring $\mathcal{R}_{star1,2}^{\left[  3\right]  }\left(
4,\mathbb{C}\right)  =\left\{  \mathsf{N}_{star1,2}\mid+,\mu^{\left[
3\right]  }\right\}  $ and $\left(  2,5\right)  $-ring $\mathcal{R}%
_{circ1,2}^{\left[  5\right]  }\left(  4,\mathbb{C}\right)  =\left\{
\mathsf{N}_{star1,2}\mid+,\mu^{\left[  5\right]  }\right\}  $.

Next we consider the \textquotedblleft interaction\textquotedblright\ of the
$4$-vertex star-like (\ref{m4}) and circle-like (\ref{mc4}) matrix sets, i.e.
their exotic module structure. For this, let us recall the ternary (polyadic)
module \cite{dup26} and $s$-place action \cite{dup2018a} definitions, which
are suitable for our case. An abelian group $\mathcal{M}$ is a ternary left
(middle, right) $\mathcal{R}$-module (or a module over $\mathcal{R}$), if
there exists a ternary operation $\mathcal{R}\times\mathcal{R}\times
\mathcal{M}\rightarrow\mathcal{M}$ ($\mathcal{R}\times\mathcal{M}%
\times\mathcal{R}\rightarrow\mathcal{M}$, $\mathcal{M}\times\mathcal{R}%
\times\mathcal{R}\rightarrow\mathcal{M}$) which satisfies some compatibility
conditions (associativity and distributivity) which hold in the matrix case
under consideration (and where the module operation is the triple ordinary
matrix product) \cite{dup26}. A $5$-ary left (right) module $\mathcal{M}$ over
$\mathcal{R}$ is a $5$-ary operation $\mathcal{R}\times\mathcal{R}%
\times\mathcal{R}\times\mathcal{R}\times\mathcal{M}\rightarrow\mathcal{M}$
($\mathcal{M}\times\mathcal{R}\times\mathcal{R}\times\mathcal{R}%
\times\mathcal{R}\rightarrow\mathcal{M}$) with analogous conditions (and where
the module operation is the pentuple matrix product) \cite{dup2018a}.

First, we have the triple relations \textquotedblleft inside\textquotedblright%
\ star and circle matrices%

\begin{align}
\mathsf{N}_{star1}\left(  \mathsf{N}_{star2}\right)  \mathsf{N}_{star1}  &
=\left(  \mathsf{N}_{star2}\right)  ,\ \ \ \mathsf{N}_{circ1}\mathsf{N}%
_{circ2}\mathsf{N}_{circ1}=\mathsf{N}_{circ1},\label{sss1}\\
\mathsf{N}_{star1}\mathsf{N}_{star1}\left(  \mathsf{N}_{star2}\right)   &
=\left(  \mathsf{N}_{star2}\right)  ,\ \ \ \mathsf{N}_{circ1}\mathsf{N}%
_{circ1}\mathsf{N}_{circ2}=\mathsf{N}_{circ1},\label{sss2}\\
\left(  \mathsf{N}_{star2}\right)  \mathsf{N}_{star1}\mathsf{N}_{star1}  &
=\left(  \mathsf{N}_{star2}\right)  ,\ \ \ \mathsf{N}_{circ2}\mathsf{N}%
_{circ1}\mathsf{N}_{circ1}=\mathsf{N}_{circ1},\label{sss3}\\
\mathsf{N}_{star2}\mathsf{N}_{star2}\left(  \mathsf{N}_{star1}\right)   &
=\left(  \mathsf{N}_{star1}\right)  ,\ \ \ \mathsf{N}_{circ2}\mathsf{N}%
_{circ2}\mathsf{N}_{circ1}=\mathsf{N}_{circ2},\label{sss4}\\
\mathsf{N}_{star2}\left(  \mathsf{N}_{star1}\right)  \mathsf{N}_{star2}  &
=\left(  \mathsf{N}_{star1}\right)  ,\ \ \ \mathsf{N}_{circ2}\mathsf{N}%
_{circ1}\mathsf{N}_{circ2}=\mathsf{N}_{circ2},\label{sss5}\\
\left(  \mathsf{N}_{star1}\right)  \mathsf{N}_{star2}\mathsf{N}_{star2}  &
=\left(  \mathsf{N}_{star1}\right)  ,\ \ \ \mathsf{N}_{circ1}\mathsf{N}%
_{circ2}\mathsf{N}_{circ2}=\mathsf{N}_{circ2}. \label{sss6}%
\end{align}
We observe the following module structures on the left column above (elements
of the corresponding module are in brackets, and we informally denote modules
by their sets): 1) from (\ref{sss1})--(\ref{sss3}), the set $\mathsf{N}%
_{star2}$ is a middle, right and left module over $\mathsf{N}_{star1}$; 2)
from (\ref{sss4})--(\ref{sss6}), the set $\mathsf{N}_{star1}$ is a middle,
right and left module over $\mathsf{N}_{star2}$;
\begin{align}
\mathsf{N}_{star1}\mathsf{N}_{circ1}\mathsf{N}_{star1}  &  =\mathsf{N}%
_{circ2},\ \ \ \mathsf{N}_{star1}\mathsf{N}_{circ2}\mathsf{N}_{star1}%
=\mathsf{N}_{circ1},\label{scs1}\\
\mathsf{N}_{star2}\mathsf{N}_{circ1}\mathsf{N}_{star2}  &  =\mathsf{N}%
_{circ2},\ \ \ \mathsf{N}_{star2}\mathsf{N}_{circ2}\mathsf{N}_{star2}%
=\mathsf{N}_{circ1},\label{scs2}\\
\mathsf{N}_{star1}\mathsf{N}_{star1}\left(  \mathsf{N}_{circ1}\right)   &
=\left(  \mathsf{N}_{circ1}\right)  ,\ \ \ \left(  \mathsf{N}_{circ1}\right)
\mathsf{N}_{star1}\mathsf{N}_{star1}=\left(  \mathsf{N}_{circ1}\right)
,\label{scs3}\\
\mathsf{N}_{star1}\mathsf{N}_{star1}\left(  \mathsf{N}_{circ2}\right)   &
=\left(  \mathsf{N}_{circ2}\right)  ,\ \ \ \left(  \mathsf{N}_{circ2}\right)
\mathsf{N}_{star1}\mathsf{N}_{star1}=\left(  \mathsf{N}_{circ2}\right)
,\label{scs4}\\
\mathsf{N}_{star2}\mathsf{N}_{star2}\left(  \mathsf{N}_{circ1}\right)   &
=\left(  \mathsf{N}_{circ1}\right)  ,\ \ \ \left(  \mathsf{N}_{circ1}\right)
\mathsf{N}_{star2}\mathsf{N}_{star2}=\left(  \mathsf{N}_{circ1}\right)
,\label{scs5}\\
\mathsf{N}_{star2}\mathsf{N}_{star2}\left(  \mathsf{N}_{circ2}\right)   &
=\left(  \mathsf{N}_{circ2}\right)  ,\ \ \ \left(  \mathsf{N}_{circ2}\right)
\mathsf{N}_{star2}\mathsf{N}_{star2}=\left(  \mathsf{N}_{circ2}\right)  ,
\label{scs6}%
\end{align}
3) from (\ref{scs3})--(\ref{scs6}), the sets $\mathsf{N}_{circ1,2}$ are a
right and left module over $\mathsf{N}_{star1,2}$;%
\begin{align}
\mathsf{N}_{circ1}\left(  \mathsf{N}_{star1}\right)  \mathsf{N}_{circ1}  &
=\left(  \mathsf{N}_{star1}\right)  ,\ \ \ \mathsf{N}_{circ1}\left(
\mathsf{N}_{star2}\right)  \mathsf{N}_{circ1}=\left(  \mathsf{N}%
_{star2}\right)  ,\label{csc1}\\
\mathsf{N}_{circ2}\left(  \mathsf{N}_{star1}\right)  \mathsf{N}_{circ2}  &
=\left(  \mathsf{N}_{star1}\right)  ,\ \ \ \mathsf{N}_{circ2}\left(
\mathsf{N}_{star2}\right)  \mathsf{N}_{circ2}=\left(  \mathsf{N}%
_{star2}\right)  ,\label{csc2}\\
\mathsf{N}_{circ1}\mathsf{N}_{circ1}\mathsf{N}_{star1}  &  =\mathsf{N}%
_{star2},\ \ \ \mathsf{N}_{star1}\mathsf{N}_{circ1}\mathsf{N}_{circ1}%
=\mathsf{N}_{star2},\label{csc3}\\
\mathsf{N}_{circ1}\mathsf{N}_{circ1}\mathsf{N}_{star2}  &  =\mathsf{N}%
_{star1},\ \ \ \mathsf{N}_{star2}\mathsf{N}_{circ1}\mathsf{N}_{circ1}%
=\mathsf{N}_{star1},\label{csc4}\\
\mathsf{N}_{circ2}\mathsf{N}_{circ2}\mathsf{N}_{star1}  &  =\mathsf{N}%
_{star2},\ \ \ \mathsf{N}_{star1}\mathsf{N}_{circ2}\mathsf{N}_{circ2}%
=\mathsf{N}_{star2},\label{csc5}\\
\mathsf{N}_{circ2}\mathsf{N}_{circ2}\mathsf{N}_{star2}  &  =\mathsf{N}%
_{star1},\ \ \ \mathsf{N}_{star2}\mathsf{N}_{circ2}\mathsf{N}_{circ2}%
=\mathsf{N}_{star1}, \label{csc6}%
\end{align}
4) from (\ref{csc1})--(\ref{csc2}), the sets $\mathsf{N}_{star1,2}$ are a
middle ternary module over $\mathsf{N}_{circ1,2}$;%
\begin{align}
\mathsf{N}_{circ1}\mathsf{N}_{circ1}\mathsf{N}_{circ1}\mathsf{N}%
_{circ1}\left(  \mathsf{N}_{star1}\right)   &  =\left(  \mathsf{N}%
_{star1}\right)  ,\ \mathsf{N}_{circ1}\mathsf{N}_{circ1}\mathsf{N}%
_{circ1}\mathsf{N}_{circ1}\left(  \mathsf{N}_{star2}\right)  =\left(
\mathsf{N}_{star2}\right)  ,\label{ccc1}\\
\left(  \mathsf{N}_{star1}\right)  \mathsf{N}_{circ1}\mathsf{N}_{circ1}%
\mathsf{N}_{circ1}\mathsf{N}_{circ1}  &  =\left(  \mathsf{N}_{star1}\right)
,\ \left(  \mathsf{N}_{star2}\right)  \mathsf{N}_{circ1}\mathsf{N}%
_{circ1}\mathsf{N}_{circ1}\mathsf{N}_{circ1}=\left(  \mathsf{N}_{star2}%
\right)  ,\label{ccc2}\\
\mathsf{N}_{circ2}\mathsf{N}_{circ2}\mathsf{N}_{circ2}\mathsf{N}%
_{circ2}\left(  \mathsf{N}_{star1}\right)   &  =\left(  \mathsf{N}%
_{star1}\right)  ,\ \mathsf{N}_{circ2}\mathsf{N}_{circ2}\mathsf{N}%
_{circ2}\mathsf{N}_{circ2}\left(  \mathsf{N}_{star2}\right)  =\left(
\mathsf{N}_{star2}\right)  ,\label{ccc3}\\
\left(  \mathsf{N}_{star1}\right)  \mathsf{N}_{circ2}\mathsf{N}_{circ2}%
\mathsf{N}_{circ2}\mathsf{N}_{circ2}  &  =\left(  \mathsf{N}_{star1}\right)
,\ \left(  \mathsf{N}_{star2}\right)  \mathsf{N}_{circ2}\mathsf{N}%
_{circ2}\mathsf{N}_{circ2}\mathsf{N}_{circ2}=\left(  \mathsf{N}_{star2}%
\right)  ,\label{ccc4}\\
\mathsf{N}_{circ1}\mathsf{N}_{circ1}\mathsf{N}_{circ1}\mathsf{N}%
_{circ1}\left(  \mathsf{N}_{circ2}\right)   &  =\left(  \mathsf{N}%
_{circ2}\right)  ,\ \left(  \mathsf{N}_{circ2}\right)  \mathsf{N}%
_{circ1}\mathsf{N}_{circ1}\mathsf{N}_{circ1}\mathsf{N}_{circ1}=\left(
\mathsf{N}_{circ2}\right)  ,\label{ccc5}\\
\mathsf{N}_{circ2}\mathsf{N}_{circ2}\mathsf{N}_{circ2}\mathsf{N}%
_{circ2}\left(  \mathsf{N}_{circ1}\right)   &  =\left(  \mathsf{N}%
_{circ1}\right)  ,\ \left(  \mathsf{N}_{circ1}\right)  \mathsf{N}%
_{circ2}\mathsf{N}_{circ2}\mathsf{N}_{circ2}\mathsf{N}_{circ2}=\left(
\mathsf{N}_{circ1}\right)  . \label{ccc6}%
\end{align}
5) from (\ref{ccc1})--(\ref{ccc6}), the sets $\mathsf{N}_{circ1,2}$ are right
and left $5$-ary modules over $\mathsf{N}_{circ2,1}$ and $\mathsf{N}%
_{star1,2}$.

Note that the sum of $4$-vertex star solutions of the Yang-Baxter equations
(\ref{cp1}) (with different parameters) gives the shape of $8$-vertex
matrices, and the same with the $4$-vertex circle solutions (\ref{cp2}). Let
us introduce two kind of $8$-vertex $4\times4$ matrices over $\mathbb{C}$: an
$8$-vertex \textit{star matrix} $M_{star}$ and an $8$-vertex \textit{circle
matrix }$M_{circ}$ as%
\begin{align}
M_{star}  &  =\left(
\begin{array}
[c]{cccc}%
x & 0 & 0 & y\\
0 & z & s & 0\\
0 & t & u & 0\\
v & 0 & 0 & w
\end{array}
\right)  ,\ \ \ \det M_{star}=\left(  xw-yv\right)  \left(  st-uz\right)
,\ \ \ \operatorname*{tr}M_{star}=x+z+u+w,\label{star}\\
M_{circ}  &  =\left(
\begin{array}
[c]{cccc}%
0 & x & y & 0\\
z & 0 & 0 & s\\
t & 0 & 0 & u\\
0 & v & w & 0
\end{array}
\right)  ,\ \ \ \det M_{circ}=\left(  xw-yv\right)  \left(  st-uz\right)
,\ \ \ \operatorname*{tr}M_{circ}=0. \label{circ}%
\end{align}

If $M_{star}$ and $M_{circ}$ are invertible (the determinants in
(\ref{star})--(\ref{circ}) are non-vanishing), then%
\begin{equation}
M_{star}^{-1}=\left(
\begin{array}
[c]{cccc}%
\frac{w}{xw-yv} & 0 & 0 & -\frac{v}{xw-yv}\\
0 & -\frac{u}{st-uz} & \frac{t}{st-uz} & 0\\
0 & \frac{s}{st-uz} & -\frac{z}{st-uz} & 0\\
-\frac{y}{xw-yv} & 0 & 0 & \frac{x}{xw-yv}%
\end{array}
\right)  ,\ \ \ \ M_{circ}^{-1}=\left(
\begin{array}
[c]{cccc}%
0 & \frac{w}{xw-yv} & -\frac{v}{xw-yv} & 0\\
-\frac{u}{st-uz} & 0 & 0 & \frac{t}{st-uz}\\
\frac{s}{st-uz} & 0 & 0 & -\frac{z}{st-uz}\\
0 & -\frac{y}{xw-yv} & \frac{x}{xw-yv} & 0
\end{array}
\right)  , \label{m1}%
\end{equation}
and therefore the parameter conditions for invertibility are the same in both
$M_{star}$ and $M_{circ}$%
\begin{equation}
xw-yv\neq0,\ \ \ \ \ \ \ \ \ st-uz\neq0. \label{i}%
\end{equation}

The corresponding sets $\mathsf{M}_{star}=\left\{  M_{star}\right\}  $ and
$\mathsf{M}_{circ}=\left\{  M_{circ}\right\}  $ are closed under the following
multiplications%
\begin{align}
\mathsf{M}_{star}\mathsf{M}_{star}  &  =\mathsf{M}_{star},\label{mmm}\\
\mathsf{M}_{star}\mathsf{M}_{circ}  &  =\mathsf{M}_{circ},\ \ \ \ \ \mathsf{M}%
_{circ}\mathsf{M}_{star}=\mathsf{M}_{circ},\label{mmm1}\\
\mathsf{M}_{circ}\mathsf{M}_{circ}  &  =\mathsf{M}_{star}, \label{mmm2}%
\end{align}
and in terms of sets we can write $\mathsf{M}_{star}=\mathsf{N}_{star1}%
\cup\mathsf{N}_{star2}$ and $\mathsf{M}_{circ}=\mathsf{N}_{circ1}%
\cup\mathsf{N}_{circ2}$, while $\mathsf{N}_{star1}\cap\mathsf{N}%
_{star2}=\varnothing$ and $\mathsf{N}_{circ1}\cap\mathsf{N}_{circ2}%
=\varnothing$ (see (\ref{mn4})). Note that, if $\mathsf{M}_{star}$ and
$\mathsf{M}_{circ}$ are treated as elements of an algebra, then (\ref{mmm}%
)--(\ref{mmm2}) are reminiscent of the Cartan decomposition (see, e.g.,
\cite{Helg2}), but we will consider them from a more general viewpoint, which
will treat such structures as semigroups, ternary groups and modules.

Thus the set $\mathsf{M}_{8vertex}=\mathsf{M}_{star}\cup\mathsf{M}_{circ}$ is
closed, and because of the associativity of matrix multiplication,
$\mathsf{M}_{8vertex}$ forms a non-commutative semigroup which we call a
$8$-\textit{vertex matrix semigroup} $\mathcal{S}_{8vertex}\left(
4,\mathbb{C}\right)  $, which contains the zero matrix $Z\in\mathcal{S}%
_{8vertex}\left(  4,\mathbb{C}\right)  $ and is a subsemigroup of the (binary)
general linear semigroup $\mathrm{GLS}\left(  4,\mathbb{C}\right)  $. It
follows from (\ref{mmm}), that $\mathsf{M}_{star}$ is its subsemigroup
$\mathcal{S}_{8vertex}^{star}\left(  4,\mathbb{C}\right)  $. Moreover, the
invertible elements of $\mathcal{S}_{8vertex}\left(  4,\mathbb{C}\right)  $
form a $8$-\textit{vertex matrix group} $\mathcal{G}_{8vertex}\left(
4,\mathbb{C}\right)  $, because its identity is a unit $4\times4$ matrix
$I_{4}\in\mathsf{M}_{star}$, and so $\mathsf{M}_{star}$ is a subgroup
$\mathcal{G}_{8vertex}^{star}\left(  4,\mathbb{C}\right)  $ of $\mathcal{G}%
_{8vertex}\left(  4,\mathbb{C}\right)  $ and a subgroup of the (binary)
general linear group $\mathrm{GL}\left(  4,\mathbb{C}\right)  $. The structure
of $\mathcal{S}_{8vertex}\left(  4,\mathbb{C}\right)  $ (\ref{mmm}) is similar
to that of block-diagonal and block-antidiagonal matrices (of the necessary
sizes). So the $8$-vertex (binary) matrix semigroup $S_{8vertex}\left(
4,\mathbb{C}\right)  $ in which the parameters satisfy (\ref{i}) is a
$8$-vertex (binary) matrix group $\mathcal{G}_{8vertex}\left(  4,\mathbb{C}%
\right)  $, having a subgroup $\mathcal{G}_{8vertex}^{star}\left(
4,\mathbb{C}\right)  =\left\langle \mathsf{M}_{star}\mid\cdot,I_{4}%
\right\rangle $, where $\left(  \cdot\right)  $ is an ordinary matrix product,
and $I_{4}$ is its identity.

The group structure of the circle matrices $\mathsf{M}_{circ}$ (\ref{circ})
follows from%
\begin{equation}
\mathsf{M}_{circ}\mathsf{M}_{circ}\mathsf{M}_{circ}=\mathsf{M}_{circ},
\label{mc}%
\end{equation}
which means that $\mathsf{M}_{circ}$ is closed with respect to the product of
three matrices (the product of two matrices from $\mathsf{M}_{circ}$ is
outside the set (\ref{mmm2})). We define a ternary multiplication
$\nu^{\left[  3\right]  }$ as the ordinary triple product of matrices, then
$\mathcal{S}_{8vertex}^{circ\left[  3\right]  }\left(  4,\mathbb{C}\right)
=\left\langle \mathsf{M}_{circ}\mid\nu^{\left[  3\right]  }\right\rangle $ is
a ternary ($3$-nonderived) semigroup with the zero $Z\in\mathsf{M}_{circ}$
which is a subsemigroup of the ternary (derived) general linear semigroup
$\mathrm{GLS}^{\left[  3\right]  }\left(  4,\mathbb{C}\right)  $. Instead of
the inverse, for each invertible element $M_{circ}\in\mathsf{M}_{circ}%
\setminus Z$ we introduce the unique querelement $\bar{M}_{circ}$ \cite{dor3}
by (\ref{mqm}), and since the ternary product is the triple ordinary product,
we have $\bar{M}_{circ}=M_{circ}^{-1}$ from (\ref{mmk}). Thus, if the
conditions of invertibility (\ref{i}) hold valid, then the ternary semigroup
$\mathcal{S}_{8vertex}^{circ\left(  3\right)  }\left(  4,\mathbb{C}\right)  $
becomes the ternary group $\mathcal{G}_{8vertex}^{circ\left(  3\right)
}\left(  4,\mathbb{C}\right)  =\left\langle \mathsf{M}_{circ}\mid\nu^{\left[
3\right]  },\overline{\left(  {}\right)  }\right\rangle $ which does not
contain the ordinary (binary) identity, since $I_{4}\notin\mathsf{M}_{circ}$.
Nevertheless, the ternary group of circle matrices $\mathcal{G}_{8vertex}%
^{circ\left[  3\right]  }\left(  4,\mathbb{C}\right)  $ has the following set
$\mathsf{I}_{circ}^{\left[  3\right]  }=\left\{  I_{circ}^{\left[  3\right]
}\right\}  $ of left-right $6$-vertex and $8$-vertex ternary identities (see
(\ref{km})--(\ref{ik}))%
\begin{equation}
I_{circ}^{\left[  3\right]  }=\left(
\begin{array}
[c]{cccc}%
0 & \frac{1}{a} & b & 0\\
a & 0 & 0 & -\frac{ab}{c}\\
0 & 0 & 0 & \frac{1}{c}\\
0 & 0 & c & 0
\end{array}
\right)  ,\ \ \left(
\begin{array}
[c]{cccc}%
0 & -\frac{ab}{c} & \frac{1}{c} & 0\\
0 & 0 & 0 & \frac{1}{b}\\
c & 0 & 0 & a\\
0 & b & 0 & 0
\end{array}
\right)  ,\ \ \left(
\begin{array}
[c]{cccc}%
0 & -\frac{ab}{c} & \frac{1-ad}{c} & 0\\
-\frac{cd}{b} & 0 & 0 & \frac{1-ad}{b}\\
c & 0 & 0 & a\\
0 & b & d & 0
\end{array}
\right)  , \label{i3c}%
\end{equation}
which (without additional conditions) depend upon the free parameters
$a,b,c,d\in\mathbb{C}$, $b,c\neq0$, and $\left(  I_{circ}^{\left[  3\right]
}\right)  ^{2}=I_{4}$, $I_{circ}^{\left[  3\right]  }\in\mathsf{M}_{circ}$. In
the binary sense, the matrices from (\ref{i3c}) are mutually similar, but as
ternary identities they are different.

If we consider the second operation for matrices (as elements of a general
matrix ring), the binary matrix addition $\left(  +\right)  $, the structure
of $\mathsf{M}_{8vertex}=\mathsf{M}_{star}\cup\mathsf{M}_{circ}$ becomes more
exotic: the set $\mathsf{M}_{star}$ is a $\left(  2,2\right)  $-ring
$\mathcal{R}_{8vertex}^{star\left[  2,2\right]  }=\left\langle \mathsf{M}%
_{star}\mid+,\cdot\right\rangle $ with the binary addition $\left(  +\right)
$ and binary multiplication $\left(  \cdot\right)  $ from the semigroup
$\mathcal{S}_{8vertex}^{star}$, while $\mathsf{M}_{circ}$ is a $\left(
2,3\right)  $-ring $\mathcal{R}_{8vertex}^{circ\left[  2,3\right]
}=\left\langle \mathsf{M}_{circ}\mid+,\nu^{\left[  3\right]  }\right\rangle $
with the binary matrix addition $\left(  +\right)  $, the ternary matrix
multiplication $\nu^{\left[  3\right]  }$ and the zero $Z$.

Moreover, because of the distributivity and associativity of binary matrix
multiplication, the relations (\ref{mmm1}) mean that the set $\mathsf{M}%
_{circ}$ (being an abelian group under binary addition) can be treated as a
left and right binary module $\mathcal{M}_{8vertex}^{circ}$ over the ring
$\mathcal{R}_{8vertex}^{star\left(  2,2\right)  }$ with an operation $\left(
\ast\right)  $: the module action $M_{circ}\ast M_{star}=M_{circ}$,
$M_{star}\ast M_{circ}=M_{circ}$ (coinciding with the ordinary matrix product
(\ref{mmm1})). The left and right modules are compatible, since the
associativity of ordinary matrix multiplication gives the compatibility
condition $\left(  M_{circ}M_{star}\right)  M_{circ}^{\prime}=M_{circ}\left(
M_{star}M_{circ}^{\prime}\right)  $, $M_{star}\in\mathcal{R}_{8vertex}%
^{star\left(  2,2\right)  }$, $M_{circ},M_{circ}^{\prime}\in\mathcal{R}%
_{8vertex}^{circ\left(  2,3\right)  }$, and therefore $M_{circ}$ (as an
abelian group under the binary addition $\left(  +\right)  $ and the module
action $\left(  \ast\right)  $) is a $\mathcal{R}_{8vertex}^{star\left(
2,2\right)  }$-bimodule $\mathcal{M}_{8vertex}^{circ}$. The last relation
(\ref{mmm2}) shows another interpretation of $\mathsf{M}_{circ}$ as a formal
\textquotedblleft square root\textquotedblright\ of $\mathsf{M}_{star}$ (as sets).

\subsection{Star $8$-vertex and circle $8$-vertex Yang-Baxter maps}

Let us consider the star $8$-vertex solutions $\tilde{c}$ to the Yang-Baxter
equation (\ref{a123}), having the shape (\ref{star}), in the most general
setting, over $\mathbb{C}$ and for different ranks (i.e. including
noninvertible ones). In components they are determined by%
\begin{equation}%
\begin{array}
[c]{l}%
vy(u-z)=0,\ \ \ y\left(  t^{2}-wz-x^{2}+xz\right)
=0,\ \ \ y(s(x-z)+t(u-x))=0,\ \ \ y\left(  u(w-x)+x^{2}-s^{2}\right)  =0,\\
svy-tuz=0,\ \ \ tvy-suz=0,\ \ \ vwy+xz(x-z)-stz=0,\ \ \ y\left(
w^{2}-wz+xz-s^{2}\right)  =0,\\
uz(z-u)=0,\ \ \ suz-tvy=0,\ \ \ y(s(w-u)+t(z-w))=0,\ \ \ v\left(
s^{2}-wz-x^{2}+xz\right)  =0,\\
tuz-svy=0,\ \ \ stz-vxy+wz(z-w)=0,\ \ \ stu+u^{2}x-ux^{2}%
-vwy=0,\ \ \ v(s(z-w)+t(w-u))=0,\\
uz(u-z)=0,\ \ \ y\left(  t^{2}+u(w-x)-w^{2}\right)
=0,\ \ \ v(s(u-x)+t(x-z))=0,\ \ \ v\left(  s^{2}+u(w-x)-w^{2}\right)  =0,\\
vy(z-u)=0,\ \ \ v\left(  u(w-x)+x^{2}-t^{2}\right)  =0,\ \ \ uw^{2}%
+vxy-stu-u^{2}w=0,\ \ \ v\left(  w^{2}-t^{2}-wz+xz\right)  =0.
\end{array}
\label{eq8s}%
\end{equation}
Solutions from, e.g. \cite{dye,hie1}, etc., should satisfy this overdetermined
system of $24$ cubic equations for $8$ variables.

We search for the $8$-vertex constant solutions to the Yang-Baxter equation
over $\mathbb{C}$ without additional conditions, unitarity, etc. (which will
be considered in the next sections). We also will need the matrix functions
$\operatorname*{tr}$ and $\det$ which are related to link invariants, as well
as eigenvalues which help to find similar matrices and $q$-conjugated
solutions to braid equations. Take into account that the Yang-Baxter maps are
determined up to a general complex factor $t\in\mathbb{C}$ (\ref{cq4t}). For
eigenvalues (which are determined up to the same factor $t$) we use the
notation: $\left\{  \text{eigenvalue}\right\}  ^{\left[  \text{algebraic
multiplicity}\right]  }$.

We found the following $8$-vertex solutions, classified by rank and number of parameters.

$\bullet$ Rank $=4$ (invertible star Yang-Baxter maps) are

1) quadratic in two parameters
\begin{equation}
\tilde{c}_{rank=4}^{par=2}\left(  x,y\right)  =\left(
\begin{array}
[c]{cccc}%
xy & 0 & 0 & y^{2}\\
0 & xy & \pm xy & 0\\
0 & \mp xy & xy & 0\\
-x^{2} & 0 & 0 & xy
\end{array}
\right)  ,\ \ \ \
\begin{array}
[c]{c}%
\operatorname*{tr}\tilde{c}=4xy,\\
\det\tilde{c}=4x^{4}y^{4},\ \ \ \ x\neq0,\ \ y\neq0,\\
\text{eigenvalues: }\left\{  \left(  1+i\right)  xy\right\}  ^{\left[
2\right]  },\left\{  \left(  1-i\right)  xy\right\}  ^{\left[  2\right]  },
\end{array}
\label{c24}%
\end{equation}

2) quadratic in three parameters%
\begin{equation}
\tilde{c}_{rank=4,1}^{par=3}\left(  x,y,z\right)  =\left(
\begin{array}
[c]{cccc}%
xy & 0 & 0 & y^{2}\\
0 & zy & \pm xy & 0\\
0 & \pm xy & zy & 0\\
z^{2} & 0 & 0 & xy
\end{array}
\right)  ,\
\begin{array}
[c]{c}%
\operatorname*{tr}\tilde{c}=2y\left(  x+z\right)  ,\\
\det\tilde{c}=y^{4}\left(  z^{2}-x^{2}\right)  ^{2},\ \ z\neq\pm
x,\ \ y\neq0,\\
\text{eigenvalues:\ }y\left(  x-z\right)  ,-y\left(  x-z\right)  ,\left\{
y\left(  x+z\right)  \right\}  ^{\left[  2\right]  },
\end{array}
\label{c34}%
\end{equation}

3) irrational in three parameters%
\begin{align}
\tilde{c}_{rank=4}^{par=3}\left(  x,y,z\right)   &  =\left(
\begin{array}
[c]{cccc}%
xy & 0 & 0 & y^{2}\\
0 & \frac{x+z}{2}y & \pm y\sqrt{\frac{x^{2}+z^{2}}{2}} & 0\\
0 & \pm y\sqrt{\frac{x^{2}+z^{2}}{2}} & \frac{x+z}{2}y & 0\\
\frac{\left(  x+z\right)  ^{2}}{4} & 0 & 0 & yz
\end{array}
\right)  ,%
\begin{array}
[c]{c}%
\operatorname*{tr}\tilde{c}=2y\left(  x+z\right)  ,\\
\det\tilde{c}=\frac{1}{16}y^{4}\left(  x-z\right)  ^{4},\ \ \ y\neq0,z\neq x,
\end{array}
\label{c34y}\\
\text{eigenvalues}  &  \text{:}\text{\ }\left\{  \frac{1}{2}y\left(
x+z-\sqrt{2}\sqrt{x^{2}+z^{2}}\right)  \right\}  ^{\left[  2\right]
},\left\{  \frac{1}{2}y\left(  x+z+\sqrt{2}\sqrt{x^{2}+z^{2}}\right)
\right\}  ^{\left[  2\right]  }.\nonumber
\end{align}

Note that only the first and the last cases are genuine $8$-vertex Yang-Baxter
maps, because the three-parameter matrices (\ref{c34}) are $q$-conjugated with
the $4$-vertex parameter-permutation solutions (\ref{cp1}). Indeed,%
\begin{align}
\left(
\begin{array}
[c]{cccc}%
xy & 0 & 0 & y^{2}\\
0 & zy & xy & 0\\
0 & xy & zy & 0\\
z^{2} & 0 & 0 & xy
\end{array}
\right)   &  =\left(  q\otimes_{K}q\right)  \left(
\begin{array}
[c]{cccc}%
y\left(  x+z\right)  & 0 & 0 & 0\\
0 & 0 & y\left(  x-z\right)  & 0\\
0 & y\left(  x-z\right)  & 0 & 0\\
0 & 0 & 0 & y\left(  x+z\right)
\end{array}
\right)  \left(  q^{-1}\otimes_{K}q^{-1}\right)  ,\label{xq}\\
q  &  =\left(
\begin{array}
[c]{cc}%
\pm\sqrt{\frac{y}{z}} & b\\
1 & \mp b\sqrt{\frac{z}{y}}%
\end{array}
\right)  , \label{qb}%
\end{align}
where $b\in\mathbb{C}$ is a free parameter. If $b=\frac{y}{z}$ two matrices
$q$ in (\ref{qb}) are similar, and we have the unique $q$-conjugation
(\ref{xq}). Another solution in (\ref{c34}) is $q$-conjugated to the second
$4$-vertex parameter-permutation solutions (\ref{cp1}) such that%
\begin{align}
\left(
\begin{array}
[c]{cccc}%
xy & 0 & 0 & y^{2}\\
0 & zy & -xy & 0\\
0 & -xy & zy & 0\\
z^{2} & 0 & 0 & xy
\end{array}
\right)   &  =\left(  q\otimes_{K}q\right)  \left(
\begin{array}
[c]{cccc}%
0 & 0 & 0 & y\left(  x-z\right) \\
0 & y\left(  x+z\right)  & 0 & 0\\
0 & 0 & y\left(  x+z\right)  & 0\\
y\left(  x-z\right)  & 0 & 0 & 0
\end{array}
\right)  \left(  q^{-1}\otimes_{K}q^{-1}\right)  ,\label{xq1}\\
q  &  =\left(
\begin{array}
[c]{cc}%
i\sqrt{\frac{y}{z}} & \pm i\sqrt{\frac{y}{z}}\\
\pm1 & 1
\end{array}
\right)  ,\left(
\begin{array}
[c]{cc}%
-i\sqrt{\frac{y}{z}} & \pm i\sqrt{\frac{y}{z}}\\
\pm1 & 1
\end{array}
\right)  , \label{qb1}%
\end{align}
where $q$'s are pairwise similar in (\ref{qb1}), and therefore we have $2$
different $q$-conjugations.

$\bullet$ Rank $=2$ (noninvertible star Yang-Baxter maps) are quadratic in
parameters%
\begin{equation}
\tilde{c}_{rank=2}^{par=2}\left(  x,y\right)  =\left(
\begin{array}
[c]{cccc}%
xy & 0 & 0 & y^{2}\\
0 & xy & \pm xy & 0\\
0 & \pm xy & xy & 0\\
x^{2} & 0 & 0 & xy
\end{array}
\right)  ,\ \ \
\begin{array}
[c]{c}%
\operatorname*{tr}\tilde{c}=4xy,\\
\text{eigenvalues:\ }\left\{  2xy\right\}  ^{\left[  2\right]  },\left\{
0\right\}  ^{\left[  2\right]  }.
\end{array}
\label{c22}%
\end{equation}

There are no star $8$-vertex solutions of rank $3$. The above two solutions
for $\tilde{c}_{rank=4}^{par=2}$ with different signs are $q$-conjugated
(\ref{cqq}) with the matrix $q$ being one of the following%
\begin{equation}
q=\left(
\begin{array}
[c]{cc}%
0 & 1\\
\pm i\frac{x}{y} & 0
\end{array}
\right)  .
\end{equation}

Further families of solutions can be obtained from (\ref{c24})--(\ref{c22}) by
applying the general $q$-conjugation (\ref{cq4t}).

Particular cases of the star solutions are called also $X$-type operators
\cite{pad/sug/tra2021} or magic matrices \cite{bal/wu} connected with the
Cartan decomposition of $SU\left(  4\right)  $
\cite{kha/gla,kra/cir,bul2004,bul/bre}.

The circle $8$-vertex solutions $\tilde{c}$ to the Yang-Baxter equation
(\ref{a123}) of the shape (\ref{circ}) are determined by the following system
of $32$ cubic equations for $8$ unknowns over $\mathbb{C}$%
\begin{equation}%
\begin{array}
[c]{l}%
x(ty+z(u-y)-vx)=0,tx^{2}+y^{2}(v-z)-wx^{2}%
=0,y(-st+tx+wy-xz)=0,su(x-y)-sxy+uxy=0,\\
z(t(y-x)-sz+wx)=0,v\left(  sy+x^{2}\right)  -z\left(  s^{2}+ux\right)
=0,swy-s^{2}v+xy(v-z)=0,swx-s^{2}w+yz(u-y)=0,\\
st^{2}-t^{2}x+z^{2}%
(y-u)=0,su(v-z)+x(xz-tu)=0,su(w-v)+xy(z-t)=0,s(tu-uw+yz)-ty^{2}=0,\\
s\left(  tv+z^{2}\right)  -x\left(  v^{2}+wz\right)
=0,svw-vwx+z(xz-wy)=0,sw(w-t)+yz(z-v)=0,s(sz+u(v-w)-vy)=0,\\
t(tu-vy+z(y-x))=0,tx(x-s)+u^{2}v-uvy=0,xy(t-w)+u^{2}w-uvx=0,t\left(
sy+u^{2}\right)  -w\left(  ux+y^{2}\right)  =0,\\
tz(s-x)-sv^{2}+tuv=0,tz(x-y)-svw+uvw=0,u\left(  w^{2}-tz\right)
-swz+tyz=0,s^{2}(t-w)+u^{2}(v-z)=0,\\
tx(w-t)+uv(z-v)=0,tvx-t^{2}y-uvw+vwy=0,tvy-t^{2}%
u+w(wy-uz)=0,u(s(v-w)-tu+wx)=0,\\
twz-tv(w+z)+vwz=0,v(s(w-t)-uw+vx)=0,sw^{2}-uv^{2}+v^{2}y-w^{2}%
x=0,w(sv+u(z-v)-wy)=0.
\end{array}
\label{eq32c}%
\end{equation}

We found the $8$-vertex solutions, classified by rank and number of parameters.

$\bullet$ Rank $=4$ (invertible circle Yang-Baxter map) are quadratic in
parameters%
\begin{equation}
\tilde{c}_{rank=4}^{par=3}\left(  x,y,z\right)  =\left(
\begin{array}
[c]{cccc}%
0 & xy & yz & 0\\
z^{2} & 0 & 0 & xy\\
xz & 0 & 0 & yz\\
0 & z^{2} & xz & 0
\end{array}
\right)  ,%
\begin{array}
[c]{c}%
\operatorname*{tr}\tilde{c}=0,\\
\det\tilde{c}=y^{2}z^{2}\left(  z^{2}-x^{2}\right)  ,y\neq0,\ \ z\neq
0,\ z\neq\pm x,\\
\text{eigenvalues:\ }\sqrt{-yz}\left(  x-z\right)  ,-\sqrt{-yz}\left(
x-z\right)  ,\sqrt{yz}\allowbreak\left(  x+z\right)  ,-\sqrt{yz}\left(
x+z\right)  .
\end{array}
\label{c4c}%
\end{equation}

$\bullet$ Rank $=2$ (noninvertible circle Yang-Baxter map) are linear in
parameters%
\begin{equation}
\tilde{c}_{rank=2}^{par=2}\left(  x,y\right)  =\left(
\begin{array}
[c]{cccc}%
0 & -y & -y & 0\\
-x & 0 & 0 & y\\
-x & 0 & 0 & y\\
0 & x & x & 0
\end{array}
\right)  ,\ \ \text{eigenvalues: }2\sqrt{xy},-2\sqrt{xy},\left\{  0\right\}
^{\left[  2\right]  }. \label{c2c}%
\end{equation}

There are no circle $8$-vertex solutions of rank $3$. The corresponding
families of solutions can be derived from the above by using the
$q$-conjugation (\ref{cq4t}).

A particular case of the $8$-vertex circle solution (\ref{c4c}) was considered
in \cite{asa/kor}.

\subsection{Triangle invertible $9$- and $10$-vertex solutions}

There are some higher vertex solutions to the Yang-Baxter equations which are
not in the above star/circle classification. They are determined by the
following system of $15$ cubic equations for $9$ unknowns over $\mathbb{C}$%
\begin{align}
&  y(-py-x(u+w-y)+v(y+z))+s(v-x)(v+x)=0,\ x(-ty+vz+x(y-z))=0,\nonumber\\
&  sx(t-v)+ty(w-z)+vz(y-u)=0,\ z(pz-t(y+z)+x(u+w-z))+s\left(  x^{2}%
-t^{2}\right)  =0,\nonumber\\
&  ps(-u+w-y+z)+s(-t(u+z)+x(u-w+y-z)+v(w+y))+uwy-uwz-uyz+wyz=0,\nonumber\\
&  t(y(p-t)+u(x-v))=0,\ t(pz-t(u+z)+ux)=0,\nonumber\\
&  p^{2}s+pu(-u+y+z)-t(st+u(u+w))+u^{2}x=0,\ t(z(p-v)-tw+wx)=0,\nonumber\\
&  ps(t-v)+tw(y-u)+uv(w-z)=0,v(-py+v(w+y)-wx)=0,\nonumber\\
&  v(z(v-p)+tw-wx)=0,v(y(t-p)+u(v-x))=0,\nonumber\\
&  p^{2}(-s)+pw(w-y-z)+sv^{2}+w(v(u+w)-wx)=0,\ p(p(w-u)-tw+uv)=0,
\end{align}
We found the following $9$-vertex Yang-Baxter maps%
\begin{align}
\tilde{c}_{rank=4}^{9-vert,1}  &  =\left(
\begin{array}
[c]{cccc}%
x & y & z & s\\
0 & 0 & x & y\\
0 & x & 0 & z\\
0 & 0 & 0 & x
\end{array}
\right)  ,\ \left(
\begin{array}
[c]{cccc}%
x & y & y & z\\
0 & 0 & -x & -y\\
0 & -x & 0 & -y\\
0 & 0 & 0 & x
\end{array}
\right)  ,\ \left(
\begin{array}
[c]{cccc}%
x & y & -y & z\\
0 & 0 & x & -\frac{zx}{y}\\
0 & x & 0 & \frac{zx}{y}\\
0 & 0 & 0 & x
\end{array}
\right)  ,\label{c9}\\
\operatorname*{tr}\tilde{c}  &  =2x,\ \ \ \ \det\tilde{c}=-x^{4}%
,\ \ \ \ x\neq0,\ \ \ \text{eigenvalues: }\left\{  x\right\}  ^{\left[
3\right]  },-x. \label{e9}%
\end{align}
The third matrix in (\ref{c9}) is conjugated with the $4$-vertex
parameter-permutation solutions (\ref{cp1}) of the form (which has the same
the same eigenvalues (\ref{e9}))%
\begin{equation}
\tilde{c}_{rank=4}^{4-vert}\left(  x\right)  =\left(
\begin{array}
[c]{cccc}%
x & 0 & 0 & 0\\
0 & 0 & x & 0\\
0 & x & 0 & 0\\
0 & 0 & 0 & x
\end{array}
\right)  \label{c4x}%
\end{equation}
by the conjugated matrix%
\begin{equation}
U^{9to4}=\left(
\begin{array}
[c]{cccc}%
1 & -\frac{y}{2x} & \frac{y}{2x} & 0\\
0 & 1 & 0 & -\frac{z}{y}\\
0 & 0 & 1 & 0\\
0 & 0 & 0 & 1
\end{array}
\right)  . \label{u9}%
\end{equation}
The matrix (\ref{u9}) cannot be presented as the Kronecker product
$q\otimes_{K}q$ (\ref{q4}), and so the third matrix in (\ref{c9}) and
(\ref{c4x}) are different solutions of the Yang-Baxter equation (\ref{a123}).
Despite the first two matrices in (\ref{c9}) have the same eigenvalues
(\ref{e9}), they are not similar, because they have different from (\ref{c4x})
middle Jordan blocks.

Then we have another $3$-parameter solutions with fractions%
\begin{equation}
\tilde{c}_{rank=4}^{9-vert,2}\left(  x,y,z\right)  =\left(
\begin{array}
[c]{cccc}%
x & y & y & z\\
0 & 0 & -x & y-\frac{2xz}{y}\\
0 & -x & 0 & y-\frac{2xz}{y}\\
0 & 0 & 0 & x\left(  \frac{4xz}{y^{2}}-3\right)
\end{array}
\right)  ,%
\begin{array}
[c]{c}%
\operatorname*{tr}\tilde{c}=2x\frac{2xz-y^{2}}{y^{2}},\\
\det\tilde{c}=x^{4}\left(  3-\frac{4xz}{y^{2}}\right)  ,\ \ \ \ x\neq
0,\ \ y\neq0,\ \ z\neq\frac{3y^{2}}{4x},\\
\text{eigenvalues: }\left\{  x\right\}  ^{\left[  2\right]  },-x,x\left(
\frac{4xz}{y^{2}}-3\right)  ,
\end{array}
\end{equation}
and%
\begin{equation}
\tilde{c}_{rank=4}^{9-vert,3}\left(  x,y,z\right)  =\left(
\begin{array}
[c]{cccc}%
x & y & -y & z\\
0 & 0 & -x & \frac{2zx}{y}+y\\
0 & 3x & 0 & \frac{2zx}{y}-y\\
0 & 0 & 0 & \frac{4zx^{2}}{y^{2}}+x
\end{array}
\right)  ,%
\begin{array}
[c]{c}%
\operatorname*{tr}\tilde{c}=2x\left(  1+2\frac{xz}{y^{2}}\right)  \\
\det\tilde{c}=3x^{4}\left(  \frac{4zx}{y^{2}}+1\right)  ,\ \ \ \ x\neq
0,\ \ y\neq0,\ \ z\neq\frac{y^{2}}{4x}\\
\text{eigenvalues: }x,i\sqrt{3}x,-i\sqrt{3}x,x\left(  1+\frac{4zx}{y^{2}%
}\right)
\end{array}
\end{equation}
The $4$-parameter $9$-vertex solution is%
\begin{align}
\tilde{c}_{rank=4}^{9-vert,par=4}\left(  x,y,z,s\right)   &  =\left(
\begin{array}
[c]{cccc}%
x & y & z & s\\
0 & 0 & -x & y-\frac{2sx}{z}\\
0 & x-\frac{2xy}{z} & 0 & z-\frac{2sx}{z}\\
0 & 0 & 0 & \frac{x(4sx-z(2y+z))}{z^{2}}%
\end{array}
\right)  ,%
\begin{array}
[c]{c}%
\operatorname*{tr}\tilde{c}=2x\frac{2sx-yz}{z^{2}}\\
\det\tilde{c}=\frac{x^{4}(2y-z)(z(2y+z)-4sx)}{z^{3}},x\neq0,y\neq\frac{z}%
{2},z\neq0,
\end{array}
\\
\text{eigenvalues} &  \text{: }x,x\sqrt{\frac{2y}{z}-1},-x\sqrt{\frac{2y}%
{z}-1},\frac{x(4sx-z(2y+z))}{z^{2}}.
\end{align}

We also found $5$-parameter, $9$-vertex solution of the form%
\begin{align}
\tilde{c}_{rank=4}^{9-vert,par=5}\left(  x,y,z,s,t\right)   &  =\left(
\begin{array}
[c]{cccc}%
x & y & z & s\\
0 & 0 & t & \frac{s(t-x)}{z}+y\\
0 & \frac{y(t-x)}{z}+x & 0 & \frac{s(t-x)}{z}+z\\
0 & 0 & 0 & \frac{s(t-x)^{2}+tz(y+z)-xyz}{z^{2}}%
\end{array}
\right)  ,\\
\operatorname*{tr}\tilde{c} &  =\frac{st^{2}+sx^{2}+tz^{2}+xz^{2}%
-2stx+tyz-xyz}{z^{2}},\\
\det\tilde{c} &  =\frac{xt\left(  x\left(  y-z\right)  -ty\right)  \left(
s\left(  t-x\right)  ^{2}+tz\left(  y+z\right)  -xyz\right)  }{z^{3}}%
,\ x\neq0,\ z\neq0,\ t\neq0,\\
\text{eigenvalues} &  \text{: }x,\sqrt{\frac{t}{z}\left(  ty-xy+xz\right)
},-\sqrt{\frac{t}{z}\left(  ty-xy+xz\right)  },\frac{st^{2}-2stx+tz^{2}%
+ytz+sx^{2}-yxz}{z^{2}}.
\end{align}

Finally, we found the following $3$-parameter $10$-vertex solution%
\begin{equation}
\tilde{c}_{rank=4}^{10-vert}\left(  x,y,z\right)  =\left(
\begin{array}
[c]{cccc}%
x & y & y & \frac{y^{2}}{x}\\
0 & 0 & -x & -y\\
0 & -x & 0 & -y\\
z & 0 & 0 & x
\end{array}
\right)  ,%
\begin{array}
[c]{c}%
\operatorname*{tr}\tilde{c}=2x,\\
\det\tilde{c}=-x\left(  x^{3}+zy^{2}\right)  ,\ \ x\neq0,\\
\text{eigenvalues: }\left\{  x\right\}  ^{\left[  2\right]  },\sqrt
{x^{2}+\frac{zy^{2}}{x}},-\sqrt{x^{2}+\frac{zy^{2}}{x}}.
\end{array}
\label{c10}%
\end{equation}

This solution is conjugated with the $4$-vertex parameter-permutation
solutions (\ref{cp1}) of the form (which has the same the same eigenvalues as
(\ref{c10}))%
\begin{equation}
\tilde{c}_{rank=4}^{4-vert}\left(  x,y,z\right)  =\left(
\begin{array}
[c]{cccc}%
x & 0 & 0 & 0\\
0 & 0 & x+\frac{y^{2}z}{x^{2}} & 0\\
0 & x & 0 & 0\\
0 & 0 & 0 & x
\end{array}
\right)  \label{c4}%
\end{equation}
by the conjugated matrix%
\begin{equation}
U^{10to4}=\left(
\begin{array}
[c]{cccc}%
0 & \frac{x}{z} & -\frac{x}{z} & 0\\
-1 & -\frac{x^{2}}{yz} & \frac{x^{2}}{yz} & -\frac{y}{x}\\
1 & -\frac{x^{2}}{yz} & \frac{x^{2}}{yz} & 0\\
0 & 0 & 1 & 1
\end{array}
\right)  . \label{u10}%
\end{equation}
Because the matrix (\ref{u10}) cannot be presented as the Kronecker product
$q\otimes_{K}q$ (\ref{q4}), therefore (\ref{c10}) and (\ref{c4}) are different
solutions of the Yang-Baxter equation (\ref{a123}).

Further families of the higher vertex solutions to the constant Yang-Baxter
equation (\ref{a123}) can be obtained from the above ones by using the
$q$-conjugation (\ref{cq4t}).

\section{Polyadic braid operators and higher braid equations}

The polyadic version of the braid equation (\ref{ri}) was introduced in
\cite{dup2021a}. Here we define higher analog of the Yang-Baxter operator and
develop its connection with higher braid groups and quantum computations.

Let us consider a vector space $\mathrm{V}$ over a field $\mathbb{K}$. A
\textit{polyadic (}$n$\textit{-ary) braid operator} $\mathrm{C}_{\mathrm{V}%
^{n}}$ is defined as the mapping \cite{dup2021a}%
\begin{equation}
\mathrm{C}_{\mathrm{V}^{n}}:\overset{n}{\overbrace{\mathrm{V}\otimes
\ldots\otimes\mathrm{V}}}\rightarrow\overset{n}{\overbrace{\mathrm{V}%
\otimes\ldots\otimes\mathrm{V}}}. \label{cv}%
\end{equation}

The polyadic analog of the braid equation (\ref{ri}) was introduced in
\cite{dup2021a} using the associative quiver technique \cite{dup2018a}.

Let us introduce $n$ operators%
\begin{align}
\mathrm{A}_{p}  &  :\overset{2n-1}{\overbrace{\mathrm{V}\otimes\ldots
\otimes\mathrm{V}}}\rightarrow\overset{2n-1}{\overbrace{\mathrm{V}%
\otimes\ldots\otimes\mathrm{V}}},\label{a1}\\
\mathrm{A}_{p}  &  =\operatorname*{id}\nolimits_{\mathrm{V}}^{\otimes\left(
p-1\right)  }\otimes\mathrm{C}_{\mathrm{V}^{n}}\otimes\operatorname*{id}%
\nolimits_{\mathrm{V}}^{\otimes\left(  n-p\right)  },\ \ \ p=1,\ldots,n,
\label{a2}%
\end{align}
i.e. $p$ is a place of $\mathrm{C}_{\mathrm{V}^{n}}$ instead of one
$\operatorname*{id}_{\mathrm{V}}$ in $\operatorname*{id}\nolimits_{\mathrm{V}%
}^{\otimes n}$. A system of $\left(  n-1\right)  $ \textit{polyadic (}%
$n$\textit{-ary) braid equations} is defined by%
\begin{align}
&  \mathrm{A}_{1}\circ\mathrm{A}_{2}\circ\mathrm{A}_{3}\circ\mathrm{A}%
_{4}\circ\ldots\circ\mathrm{A}_{n-2}\circ\mathrm{A}_{n-1}\circ\mathrm{A}%
_{n}\circ\mathrm{A}_{1}\label{aa1}\\
&  =\mathrm{A}_{2}\circ\mathrm{A}_{3}\circ\mathrm{A}_{4}\circ\mathrm{A}%
_{5}\circ\ldots\circ\mathrm{A}_{n-1}\circ\mathrm{A}_{n}\circ\mathrm{A}%
_{1}\circ\mathrm{A}_{2}\\
&  \vdots\nonumber\\
&  =\mathrm{A}_{n}\circ\mathrm{A}_{1}\circ\mathrm{A}_{2}\circ\mathrm{A}%
_{3}\circ\ldots\circ\mathrm{A}_{n-3}\circ\mathrm{A}_{n-2}\circ\mathrm{A}%
_{n-1}\circ\mathrm{A}_{n}. \label{aa2}%
\end{align}

\begin{example}
In the lowest non-binary case $n=3$, we have the ternary braid operator
$\mathrm{C}_{\mathrm{V}^{3}}:\mathrm{V}\otimes\mathrm{V}\otimes\mathrm{V}%
\rightarrow\mathrm{V}\otimes\mathrm{V}\otimes\mathrm{V}$ and two ternary braid
equations on $\mathrm{V}^{\otimes5}$%
\begin{align}
&  \left(  \mathrm{C}_{\mathrm{V}^{3}}\otimes\operatorname*{id}%
\nolimits_{\mathrm{V}}\otimes\operatorname*{id}\nolimits_{\mathrm{V}}\right)
\circ\left(  \operatorname*{id}\nolimits_{\mathrm{V}}\otimes\mathrm{C}%
_{\mathrm{V}^{3}}\otimes\operatorname*{id}\nolimits_{\mathrm{V}}\right)
\circ\left(  \operatorname*{id}\nolimits_{\mathrm{V}}\otimes\operatorname*{id}%
\nolimits_{\mathrm{V}}\otimes\mathrm{C}_{\mathrm{V}^{3}}\right)  \circ\left(
\mathrm{C}_{\mathrm{V}^{3}}\otimes\operatorname*{id}\nolimits_{\mathrm{V}%
}\otimes\operatorname*{id}\nolimits_{\mathrm{V}}\right) \nonumber\\
&  =\left(  \operatorname*{id}\nolimits_{\mathrm{V}}\otimes\mathrm{C}%
_{\mathrm{V}^{3}}\otimes\operatorname*{id}\nolimits_{\mathrm{V}}\right)
\circ\left(  \operatorname*{id}\nolimits_{\mathrm{V}}\otimes\operatorname*{id}%
\nolimits_{\mathrm{V}}\otimes\mathrm{C}_{\mathrm{V}^{3}}\right)  \circ\left(
\mathrm{C}_{\mathrm{V}^{3}}\otimes\operatorname*{id}\nolimits_{\mathrm{V}%
}\otimes\operatorname*{id}\nolimits_{\mathrm{V}}\right)  \circ\left(
\operatorname*{id}\nolimits_{\mathrm{V}}\otimes\mathrm{C}_{\mathrm{V}^{3}%
}\otimes\operatorname*{id}\nolimits_{\mathrm{V}}\right) \nonumber\\
&  =\left(  \operatorname*{id}\nolimits_{\mathrm{V}}\otimes\operatorname*{id}%
\nolimits_{\mathrm{V}}\otimes\mathrm{C}_{\mathrm{V}^{3}}\right)  \circ\left(
\mathrm{C}_{\mathrm{V}^{3}}\otimes\operatorname*{id}\nolimits_{\mathrm{V}%
}\otimes\operatorname*{id}\nolimits_{\mathrm{V}}\right)  \circ\left(
\operatorname*{id}\nolimits_{\mathrm{V}}\otimes\mathrm{C}_{\mathrm{V}^{3}%
}\otimes\operatorname*{id}\nolimits_{\mathrm{V}}\right)  \circ\left(
\operatorname*{id}\nolimits_{\mathrm{V}}\otimes\operatorname*{id}%
\nolimits_{\mathrm{V}}\otimes\mathrm{C}_{\mathrm{V}^{3}}\right)  . \label{ci}%
\end{align}

\end{example}

Note that the higher braid equations presented above differ from the
generalized Yang-Baxter equations of \cite{row/zha/wu/ge,kit/wan,che2012}.

The higher braid operators (\ref{cv}) satisfying the higher braid equations
(\ref{aa1})--(\ref{aa2}) can represent the higher braid group \cite{dup2021b}
using (\ref{am}) and (\ref{a2}). By analogy with (\ref{am}) we introduce $m$
operators by%
\begin{align}
\mathrm{B}_{i}\left(  m\right)   &  :\overset{m+n-2}{\overbrace{\mathrm{V}%
\otimes\ldots\otimes\mathrm{V}}}\rightarrow\overset{m+n-2}{\overbrace
{\mathrm{V}\otimes\ldots\otimes\mathrm{V}}},\ \ \ \ \mathrm{B}_{0}\left(
m\right)  =\left(  \operatorname*{id}\nolimits_{\mathrm{V}}\right)
^{\otimes\left(  m+n-2\right)  },\label{b1}\\
\mathrm{B}_{i}\left(  m\right)   &  =\operatorname*{id}\nolimits_{\mathrm{V}%
}^{\otimes\left(  i-1\right)  }\otimes\mathrm{C}_{\mathrm{V}^{n}}%
\otimes\operatorname*{id}\nolimits_{\mathrm{V}}^{\otimes\left(  m-i-1\right)
},\ \ \ i=1,\ldots,m-1. \label{b2}%
\end{align}

The representation $\pi_{m}^{\left[  n\right]  }$ of the higher braid group
$\mathcal{B}_{m}^{\left[  n+1\right]  }$ (of $\left(  n+1\right)  $-degree in
the notation of \cite{dup2021b}) (having $m-1$ generators $\mathbf{\sigma}%
_{i}$ and identity $\mathbf{e}$) is given by%
\begin{align}
\pi_{m}^{\left[  n\right]  }  &  :\mathcal{B}_{m}^{\left[  n+1\right]
}\longrightarrow\operatorname*{End}\mathrm{V}^{\otimes\left(  m+n-2\right)
}\mathrm{,}\label{pb1}\\
\pi_{m}^{\left[  n\right]  }\left(  \mathbf{\sigma}_{i}\right)   &
=\mathrm{B}_{i}\left(  m\right)  ,\ \ \ i=1,\ldots,m-1. \label{pb2}%
\end{align}

In this way, the generators $\mathbf{\sigma}_{i}$ of the higher braid group
$\mathcal{B}_{m}^{\left[  n+1\right]  }$ satisfy the relations

$\bullet$ $n$ \textit{higher braid relations}%
\begin{align}
&  \overset{n+1}{\overbrace{\mathbf{\sigma}_{i}\mathbf{\sigma}_{i+1}%
\ldots\mathbf{\sigma}_{i+n-2}\mathbf{\sigma}_{i+n-1}\mathbf{\sigma}_{i}}%
}\label{ss1}\\
&  =\mathbf{\sigma}_{i+1}\mathbf{\sigma}_{i+2}\ldots\mathbf{\sigma}%
_{i+n-1}\mathbf{\sigma}_{i}\mathbf{\sigma}_{i+1}\label{ss2}\\
&  \vdots\nonumber\\
&  =\mathbf{\sigma}_{i+n-1}\mathbf{\sigma}_{i}\mathbf{\sigma}_{i+1}%
\mathbf{\sigma}_{i+2}\ldots\mathbf{\sigma}_{i}\mathbf{\sigma}_{i+1}%
\mathbf{\sigma}_{i+n-1},\label{ss3}\\
\ \ i  &  =1,\ldots,m-n, \label{ibr}%
\end{align}

$\bullet$ $n$-\textit{ary far commutativity}%
\begin{align}
&  \overset{n}{\overbrace{\mathbf{\sigma}_{i_{1}}\mathbf{\sigma}_{i_{2}}%
\ldots\mathbf{\sigma}_{i_{n-2}}\mathbf{\sigma}_{i_{n-1}}\mathbf{\sigma}%
_{i_{n}}}}\label{kc1}\\
&  \vdots\nonumber\\
&  =\mathbf{\sigma}_{\tau\left(  i_{1}\right)  }\mathbf{\sigma}_{\tau\left(
i_{2}\right)  }\ldots\mathbf{\sigma}_{\tau\left(  i_{n-2}\right)
}\mathbf{\sigma}_{\tau\left(  i_{n-1}\right)  }\mathbf{\sigma}_{\tau\left(
i_{n}\right)  },\label{kc2}\\
\text{if all }\left\vert i_{p}-i_{s}\right\vert  &  \geq
n,\ \ \ \ p,s=1,\ldots,n, \label{ip}%
\end{align}
where $\tau$ is an element of the permutation symmetry group $\tau\in S_{n}$.
The relations (\ref{ss1})--(\ref{kc2}) coincide with those from
\cite{dup2021b}, obtained by another method, that is via the polyadic-binary correspondence.

In the case $m=4$ and $n=3$ the higher braid group $\mathcal{B}_{4}^{\left[
4\right]  }$ is represented by (\ref{ci}) and generated by $3$ generators
$\mathbf{\sigma}_{1}$, $\mathbf{\sigma}_{2}$, $\mathbf{\sigma}_{3}$, which
satisfy 2 braid relations only (without far commutativity)%
\begin{equation}
\mathbf{\sigma}_{1}\mathbf{\sigma}_{2}\mathbf{\sigma}_{3}\mathbf{\sigma}%
_{1}=\mathbf{\sigma}_{2}\mathbf{\sigma}_{3}\mathbf{\sigma}_{1}\mathbf{\sigma
}_{2}=\mathbf{\sigma}_{3}\mathbf{\sigma}_{1}\mathbf{\sigma}_{2}\mathbf{\sigma
}_{3}. \label{b44}%
\end{equation}

According to (\ref{kc1})--(\ref{kc2}), the far commutativity relations appear
when the number of elements of the higher braid groups satisfy%
\begin{equation}
m\geq m_{\min}=n\left(  n-1\right)  +2, \label{mn}%
\end{equation}
such that all conditions (\ref{ip}) should hold. Thus, to have the far
commutativity relations in the ordinary (binary) braid group (\ref{bm}) we
need $3$ generators and $\mathcal{B}_{4}$, while for $n=3$ we need at least
$7$ generators $\mathbf{\sigma}_{i}$ and $\mathcal{B}_{8}^{\left[  4\right]
}$ (see \textit{Example} \textbf{7.12} in \cite{dup2021b}).

In the concrete realization of $\mathrm{V}$ as a $d$-dimensional euclidean
vector space $V_{d}$ over the complex numbers $\mathbb{C}$ and basis $\left\{
e_{i}\right\}  $, $i=1,\ldots,d$, the polyadic ($n$-ary) braid operator
$\mathrm{C}_{\mathrm{V}^{n}}$ becomes a matrix $C_{d^{n}}$ of size
$d^{n}\times d^{n}$ which satisfies $n-1$ higher braid equations
(\ref{aa1})--(\ref{aa2}) in matrix form. In the components, the matrix braid
operator is%
\begin{equation}
C_{d^{n}}\circ\left(  e_{i_{1}}\otimes e_{i_{2}}\otimes\ldots\otimes e_{i_{n}%
}\right)  =\sum_{j_{1}^{\prime},j_{2}^{\prime}\ldots j_{n}^{\prime}=1}%
^{d}c_{i_{1}i_{2}\ldots i_{n}}^{\ \ \ \ \ \ \ j_{1}^{\prime}j_{2}^{\prime
}\ldots j_{n}^{\prime}}\cdot e_{j_{1}^{\prime}}\otimes e_{j_{2}^{\prime}%
}\otimes\ldots\otimes e_{j_{n}^{\prime}}. \label{cd}%
\end{equation}

THUS we have $d^{2n}$ entries (unknowns) in $C_{d^{n}}$ satisfying $\left(
n-1\right)  d^{4n-2}$ equations (\ref{aa1})--(\ref{aa2}) in components of
polynomial power $n+1$. In the minimal non-binary case $n=3$, we have
$2d^{10}$ equations of power $4$ for $d^{6}$ unknowns, e.g. even for $d=2$ we
have $2048$ for $64$ components, and for $d=3$ there are $118\,098$ equations
for $729$ components. Thus solving the matrix higher braid equations directly
is cumbersome, and only particular cases are possible to investigate, for
instance by using permutation matrices (\ref{cp}), or the star and circle
matrices (\ref{star})--(\ref{circ}).

\section{Solutions to the ternary braid equations}

Here we consider some special solutions to the minimal ternary version ($n=3$)
of the polyadic braid equation (\ref{aa1})--(\ref{aa2}), the ternary braid
equation (\ref{ci}).

\subsection{Constant matrix solutions}

Let us consider the following two-dimensional vector space $V\equiv V_{d=2}$
(which is important for quantum computations) and the component matrix
realization (\ref{cd}) of the ternary braiding operator $C_{8}:V\otimes
V\otimes V\rightarrow V\otimes V\otimes V$ as%
\begin{equation}
C_{8}\circ\left(  e_{i_{1}}\otimes e_{i_{2}}\otimes e_{i_{3}}\right)
=\sum_{j_{1}^{\prime},j_{2}^{\prime},j_{3}^{\prime}=1}^{2}c_{i_{1}i_{2}i_{3}%
}^{\ \ \ \ \ \ \ j_{1}^{\prime}j_{2}^{\prime}j_{3}^{\prime}}\cdot
e_{j_{1}^{\prime}}\otimes e_{j_{2}^{\prime}}\otimes e_{j_{3}^{\prime}%
},\ \ \ i_{1,2,3},j_{1,2,3}^{\prime}=1,2. \label{c8}%
\end{equation}
We now turn (\ref{c8}) to the standard matrix form (just to fix notations) by
introducing the $8$-dimensional vector space $\tilde{V}_{8}=V\otimes V\otimes
V$ with the natural basis $\tilde{e}_{\tilde{k}}=\left\{  e_{1}\otimes
e_{1}\otimes e_{1},e_{1}\otimes e_{1}\otimes e_{2},\ldots,e_{2}\otimes
e_{2}\otimes e_{2}\right\}  $, where $\tilde{k}=1,\ldots,8$ is a cumulative
index. The linear operator $\tilde{C}_{8}:\tilde{V}_{8}\rightarrow\tilde
{V}_{8}$ corresponding to (\ref{c8}) is given by the $8\times8$ matrix
$\tilde{c}_{\tilde{\imath}\tilde{j}}$ as $\tilde{C}_{8}\circ\tilde{e}%
_{\tilde{\imath}}=\sum_{\tilde{j}=1}^{8}\tilde{c}_{\tilde{\imath}\tilde{j}%
}\cdot\tilde{e}_{\tilde{j}}$. The operators (\ref{a1})--(\ref{a2}) become
three $32\times32$ matrices $\tilde{A}_{1,2,3}$ as%
\begin{equation}
\tilde{A}_{1}=\tilde{c}\otimes_{K}I_{2}\otimes_{K}I_{2},\ \ \ \tilde{A}%
_{2}=I_{2}\otimes_{K}\tilde{c}\otimes_{K}I_{2},\ \ \ \tilde{A}_{3}%
=I_{2}\otimes_{K}I_{2}\otimes_{K}\tilde{c}, \label{ac}%
\end{equation}
where $\otimes_{K}$ is the Kronecker product of matrices and $I_{2}$ is the
$2\times2$ identity matrix. In this notation the operator ternary braid
equations (\ref{ci}) become the matrix equations (cf. (\ref{aa1}%
)--(\ref{aa2})) with $n=3$)%
\begin{equation}
\tilde{A}_{1}\tilde{A}_{2}\tilde{A}_{3}\tilde{A}_{1}=\tilde{A}_{2}\tilde
{A}_{3}\tilde{A}_{1}\tilde{A}_{2}=\tilde{A}_{3}\tilde{A}_{1}\tilde{A}%
_{2}\tilde{A}_{3}, \label{pt}%
\end{equation}
which we call the \textit{total matrix ternary braid equations}. Some weaker
versions of ternary braiding are described by the \textit{partial braid
equations}%
\begin{align}
\text{partial }12\text{-braid equation \ \ \ \ }\tilde{A}_{1}\tilde{A}%
_{2}\tilde{A}_{3}\tilde{A}_{1}  &  =\tilde{A}_{2}\tilde{A}_{3}\tilde{A}%
_{1}\tilde{A}_{2},\label{p1}\\
\text{partial }13\text{-braid equation \ \ \ \ }\tilde{A}_{1}\tilde{A}%
_{2}\tilde{A}_{3}\tilde{A}_{1}  &  =\tilde{A}_{3}\tilde{A}_{1}\tilde{A}%
_{2}\tilde{A}_{3},\label{p2}\\
\text{partial }23\text{-braid equation \ \ \ \ }\tilde{A}_{2}\tilde{A}%
_{3}\tilde{A}_{1}\tilde{A}_{2}  &  =\tilde{A}_{3}\tilde{A}_{1}\tilde{A}%
_{2}\tilde{A}_{3}, \label{p3}%
\end{align}
where, obviously, two of them are independent. It follows from (\ref{aa1}%
)--(\ref{aa2}) that the weaker versions of braiding are possible for $n\geq3$,
only, so for higher than binary braiding (the Yang-Baxter equation (\ref{rd})).

Thus, comparing (\ref{pt}) and (\ref{b44}) we conclude that (for each
invertible matrix $\tilde{c}$ in (\ref{ac}) satisfying (\ref{pt})) the
isomorphism $\tilde{\pi}_{4}^{\left[  4\right]  }:\mathbf{\sigma}_{i}%
\mapsto\tilde{A}_{i}$, $i=1,2,3$ gives a representation of the braid group
$\mathcal{B}_{4}^{\left[  4\right]  }$ by $32\times32$ matrices over
$\mathbb{C}$.

Now we can generate families of solutions corresponding to (\ref{ac}%
)--(\ref{pt}) in the following way. Consider an invertible operator
$Q:V\rightarrow V$ in the two-dimensional vector space $V\equiv V_{d=2}$. In
the basis $\left\{  e_{1},e_{2}\right\}  $ its $2\times2$ matrix $q$ is given
by $Q\circ e_{i}=\sum_{j=1}^{2}q_{ij}\cdot e_{j}$. In the natural
$8$-dimensional basis $\tilde{e}_{\tilde{k}}$ the tensor product of operators
$Q\otimes Q\otimes Q$ is presented by the Kronecker product of matrices
$\tilde{q}_{8}=q\otimes_{K}q\otimes_{K}q$. Let the $8\times8$ matrix
$\tilde{c}$ be a fixed solution to the ternary braid matrix equations
(\ref{pt}). Then the family of solutions $\tilde{c}\left(  q\right)  $
corresponding to the invertible $2\times2$ matrix $q$ is the conjugation of
$\tilde{c}$ by $\tilde{q}_{8}$ so that%
\begin{equation}
\tilde{c}\left(  q\right)  =\tilde{q}_{8}\tilde{c}\tilde{q}_{8}^{-1}=\left(
q\otimes_{K}q\otimes_{K}q\right)  \tilde{c}\left(  q^{-1}\otimes_{K}%
q^{-1}\otimes_{K}q^{-1}\right)  .\label{cq}%
\end{equation}

This also follows directly from the conjugation of the braid equations
(\ref{pt})--(\ref{p3}) by $q\otimes_{K}q\otimes_{K}q\otimes_{K}q\otimes_{K}q$
and (\ref{ac}). If we include the obvious invariance of the braid equations
with the respect of an overall factor $t\in\mathbb{C}$, the general family of
solutions becomes (cf. the Yang-Baxter equation \cite{hie1})%
\begin{equation}
\tilde{c}\left(  q,t\right)  =t\tilde{q}_{8}\tilde{c}\tilde{q}_{8}%
^{-1}=t\left(  q\otimes_{K}q\otimes_{K}q\right)  \tilde{c}\left(
q^{-1}\otimes_{K}q^{-1}\otimes_{K}q^{-1}\right)  . \label{cqt}%
\end{equation}
Let%
\begin{equation}
q=\left(
\begin{array}
[c]{cc}%
a & b\\
c & d
\end{array}
\right)  \in\mathrm{GL}\left(  2,\mathbb{C}\right)  , \label{qg}%
\end{equation}
and then the manifest form of $\tilde{q}_{8}$ is%
\begin{equation}
\tilde{q}_{8}=\left(
\begin{array}
[c]{cccccccc}%
a^{3} & a^{2}b & a^{2}b & ab^{2} & a^{2}b & ab^{2} & ab^{2} & b^{3}\\
a^{2}c & a^{2}d & abc & abd & abc & abd & b^{2}c & b^{2}d\\
a^{2}c & abc & a^{2}d & abd & abc & b^{2}c & abd & b^{2}d\\
ac^{2} & acd & acd & ad^{2} & bc^{2} & bcd & bcd & bd^{2}\\
a^{2}c & abc & abc & b^{2}c & a^{2}d & abd & abd & b^{2}d\\
ac^{2} & acd & bc^{2} & bcd & acd & ad^{2} & bcd & bd^{2}\\
ac^{2} & bc^{2} & acd & bcd & acd & bcd & ad^{2} & bd^{2}\\
c^{3} & c^{2}d & c^{2}d & cd^{2} & c^{2}d & cd^{2} & cd^{2} & d^{3}%
\end{array}
\right)  . \label{q8}%
\end{equation}
It is important that not every conjugation matrix has this very special form
(\ref{q8}), and that therefore, in general, conjugated matrices are different
solutions of the ternary braid equations (\ref{pt}). The matrix $\tilde{q}%
_{8}^{\mathbf{\star}}\tilde{q}_{8}$ ($\mathbf{\star}$ being the Hermitian
conjugation) is diagonal (this case is important for further classification
similar to the binary one \cite{dye}), when the conditions%
\begin{equation}
ab^{\ast}+cd^{\ast}=0 \label{abcd}%
\end{equation}
hold, and so the matrix $q$ has the special form (depending of 3 complex
parameters, for $d\neq0$)%
\begin{equation}
q=\left(
\begin{array}
[c]{cc}%
a & b\\
-a\frac{b^{\ast}}{d^{\ast}} & d
\end{array}
\right)  . \label{q1}%
\end{equation}
We can present the families (\ref{cq}) for different ranks, because the
conjugation by an invertible matrix does not change rank. To avoid demanding
(\ref{abcd}), due to the cumbersome calculations involved, we restrict
ourselves to a triangle matrix for $q$ (\ref{qg}).

In general, there are $8\times8=64$ unknowns (elements of the matrix
$\tilde{c}$), and each partial braid equation (\ref{p1})--(\ref{p3}) gives
$32\times32=1024$ conditions (of power $4$) for the elements of $\tilde{c}$,
while the total braid equations (\ref{pt}) give twice as many conditions
$1024\times2=2048$ (cf. the binary case: $64$ cubic equations for $16$
unknowns (\ref{rd})). This means that even in the ternary case the higher
braid system of equations is hugely overdetermined, and finding even the
simplest solutions is a non-trivial task.

\subsection{Permutation and parameter-permutation $8$-vertex solutions}

First we consider the case when $\tilde{c}$ is a binary (or logical) matrix
consisting of $\left\{  0,1\right\}  $ only, and, moreover, it is a
permutation matrix (see Subsection \ref{subsec-perm}). In the latter case
$\tilde{c}$ can be considered as a matrix over the field $\mathbb{F}_{2}$
(Galois field $GF\left(  2\right)  $). In total, there are $8!=40\,320$
permutation matrices of the size $8\times8$. All of them are invertible of
full rank $8$, because they are obtained from the identity matrix by
permutation of rows and columns.

We have found the following four invertible $8$-vertex permutation matrix
solutions to the ternary braid equations (\ref{pt})%
\begin{align}
\tilde{c}_{rank=8}^{bisymm1}  &  =\left(
\begin{array}
[c]{cccccccc}%
1 & 0 & 0 & 0 & 0 & 0 & 0 & 0\\
0 & 0 & 0 & 0 & 0 & 0 & 1 & 0\\
0 & 0 & 1 & 0 & 0 & 0 & 0 & 0\\
0 & 0 & 0 & 0 & 1 & 0 & 0 & 0\\
0 & 0 & 0 & 1 & 0 & 0 & 0 & 0\\
0 & 0 & 0 & 0 & 0 & 1 & 0 & 0\\
0 & 1 & 0 & 0 & 0 & 0 & 0 & 0\\
0 & 0 & 0 & 0 & 0 & 0 & 0 & 1
\end{array}
\right)  ,\ \ \ \tilde{c}_{rank=8}^{bisymm2}=\left(
\begin{array}
[c]{cccccccc}%
0 & 0 & 0 & 0 & 0 & 0 & 0 & 1\\
0 & 1 & 0 & 0 & 0 & 0 & 0 & 0\\
0 & 0 & 0 & 0 & 0 & 1 & 0 & 0\\
0 & 0 & 0 & 1 & 0 & 0 & 0 & 0\\
0 & 0 & 0 & 0 & 1 & 0 & 0 & 0\\
0 & 0 & 1 & 0 & 0 & 0 & 0 & 0\\
0 & 0 & 0 & 0 & 0 & 0 & 1 & 0\\
1 & 0 & 0 & 0 & 0 & 0 & 0 & 0
\end{array}
\right)  ,%
\begin{array}
[c]{c}%
\operatorname*{tr}\tilde{c}=4,\\
\det\tilde{c}=1,\\
\text{eigenvalues:}\left\{  1\right\}  ^{\left[  4\right]  },\left\{
-1\right\}  ^{\left[  4\right]  },
\end{array}
\label{cp8b}\\
\tilde{c}_{rank=8}^{symm1}  &  =\left(
\begin{array}
[c]{cccccccc}%
1 & 0 & 0 & 0 & 0 & 0 & 0 & 0\\
0 & 0 & 0 & 0 & 1 & 0 & 0 & 0\\
0 & 0 & 0 & 0 & 0 & 0 & 0 & 1\\
0 & 0 & 0 & 1 & 0 & 0 & 0 & 0\\
0 & 1 & 0 & 0 & 0 & 0 & 0 & 0\\
0 & 0 & 0 & 0 & 0 & 1 & 0 & 0\\
0 & 0 & 0 & 0 & 0 & 0 & 1 & 0\\
0 & 0 & 1 & 0 & 0 & 0 & 0 & 0
\end{array}
\right)  ,\ \ \ \tilde{c}_{rank=8}^{symm2}=\left(
\begin{array}
[c]{cccccccc}%
0 & 0 & 0 & 0 & 0 & 1 & 0 & 0\\
0 & 1 & 0 & 0 & 0 & 0 & 0 & 0\\
0 & 0 & 1 & 0 & 0 & 0 & 0 & 0\\
0 & 0 & 0 & 0 & 0 & 0 & 1 & 0\\
0 & 0 & 0 & 0 & 1 & 0 & 0 & 0\\
1 & 0 & 0 & 0 & 0 & 0 & 0 & 0\\
0 & 0 & 0 & 1 & 0 & 0 & 0 & 0\\
0 & 0 & 0 & 0 & 0 & 0 & 0 & 1
\end{array}
\right)  ,%
\begin{array}
[c]{c}%
\operatorname*{tr}\tilde{c}=4,\\
\det\tilde{c}=1,\\
\text{eigenvalues:}\left\{  1\right\}  ^{\left[  4\right]  },\left\{
-1\right\}  ^{\left[  4\right]  }.
\end{array}
\label{cp8s}%
\end{align}

The first two solutions (\ref{cp8b}) are given by bisymmetric permutation
matrices (see (\ref{mt1})), and we call them $8$-vertex $\mathsf{bisymm1}$ and
$\mathsf{bisymm2}$ respectively. The second two solutions (\ref{cp8s}) are
symmetric matrices only (we call them $8$-vertex $\mathsf{symm1}$ and
$\mathsf{symm2}$), but one matrix is a reflection of the other with respect to
the minor diagonal (making them mutually persymmetric). No $90^{\circ}%
$-symmetric (see (\ref{mt2})) solution for the ternary braid equations
(\ref{pt}) was found. The bisymmetric and symmetric matrices have the same
eigenvalues, and are therefore pairwise conjugate, but not $q$-conjugate,
because the conjugation matrices do not have the form (\ref{q8}). Thus they
are $4$ different permutation solutions to the ternary braid equations
(\ref{pt}). Note that the $\mathsf{bisymm1}$ solution (\ref{cp8b}) coincides
with the three-qubit swap operator introduced in \cite{bal/wu}.

All the permutation solutions are reflections (or involutions) $\tilde{c}%
^{2}=I_{8}$ having $\det\tilde{c}=+1$, eigenvalues $\left\{  1,-1\right\}  $,
and are semi-magic squares (the sums in rows and columns are $1$, but not the
sums in both diagonals). The $8$-vertex permutation matrix solutions do not
form a binary or ternary group, because they are not closed with respect to multiplication.

By analogy with (\ref{cp1})--(\ref{cp2}), we obtain the $8$-vertex
parameter-permutation solutions from (\ref{cp8b})--(\ref{cp8s}) by replacing
units with parameters and then solving the ternary braid equations (\ref{pt}).
Each type of the permutation solutions $\mathsf{bisymm1,2}$ and
$\mathsf{symm1,2}$ from (\ref{cp8b})--(\ref{cp8s}) will give a corresponding
series of parameter-permutation solutions over $\mathbb{C}$. The ternary braid
maps are determined up to a general complex factor (see (\ref{cq4t}) for the
Yang-Baxter maps and (\ref{cqt})), and therefore we can present all the
parameter-permutation solutions in polynomial form.

$\bullet$ The $\mathsf{bisymm1}$ series consists of $2$ two-parameter matrices
with and $2$ two-parameter matrices
\begin{equation}
\tilde{c}_{rank=8}^{bisymm1,1}\left(  x,y\right)  =\left(
\begin{array}
[c]{cccccccc}%
xy & 0 & 0 & 0 & 0 & 0 & 0 & 0\\
0 & 0 & 0 & 0 & 0 & 0 & \pm y^{2} & 0\\
0 & 0 & xy & 0 & 0 & 0 & 0 & 0\\
0 & 0 & 0 & 0 & \pm x^{2} & 0 & 0 & 0\\
0 & 0 & 0 & \pm y^{2} & 0 & 0 & 0 & 0\\
0 & 0 & 0 & 0 & 0 & xy & 0 & 0\\
0 & \pm x^{2} & 0 & 0 & 0 & 0 & 0 & 0\\
0 & 0 & 0 & 0 & 0 & 0 & 0 & xy
\end{array}
\right)  ,%
\begin{array}
[c]{c}%
\operatorname*{tr}\tilde{c}=4xy,\\
\det\tilde{c}=x^{8}y^{8},\ \ x,y\neq0,\\
\text{eigenvalues: }\left\{  xy\right\}  ^{\left[  6\right]  },\left\{
-xy\right\}  ^{\left[  2\right]  },
\end{array}
\label{b11}%
\end{equation}%
\begin{equation}
\tilde{c}_{rank=8}^{bisymm1,2}\left(  x,y\right)  =\left(
\begin{array}
[c]{cccccccc}%
xy & 0 & 0 & 0 & 0 & 0 & 0 & 0\\
0 & 0 & 0 & 0 & 0 & 0 & \pm y^{2} & 0\\
0 & 0 & xy & 0 & 0 & 0 & 0 & 0\\
0 & 0 & 0 & 0 & \pm x^{2} & 0 & 0 & 0\\
0 & 0 & 0 & \mp y^{2} & 0 & 0 & 0 & 0\\
0 & 0 & 0 & 0 & 0 & xy & 0 & 0\\
0 & \mp x^{2} & 0 & 0 & 0 & 0 & 0 & 0\\
0 & 0 & 0 & 0 & 0 & 0 & 0 & xy
\end{array}
\right)  ,%
\begin{array}
[c]{c}%
\operatorname*{tr}\tilde{c}=4xy,\\
\det\tilde{c}=x^{8}y^{8},\ \ x,y\neq0,\\
\text{eigenvalues: }\left\{  xy\right\}  ^{\left[  4\right]  },\left\{
ixy\right\}  ^{\left[  2\right]  },\left\{  -ixy\right\}  ^{\left[  2\right]
}.
\end{array}
\label{b12}%
\end{equation}

$\bullet$ The $\mathsf{bisymm2}$ series consists of $4$ two-parameter matrices%
\begin{align}
\tilde{c}_{rank=8}^{bisymm2,1}\left(  x,y\right)   &  =\left(
\begin{array}
[c]{cccccccc}%
0 & 0 & 0 & 0 & 0 & 0 & 0 & x^{6}\\
0 & \pm x^{3}y^{3} & 0 & 0 & 0 & 0 & 0 & 0\\
0 & 0 & 0 & 0 & 0 & x^{4}y^{2} & 0 & 0\\
0 & 0 & 0 & \pm x^{3}y^{3} & 0 & 0 & 0 & 0\\
0 & 0 & 0 & 0 & \pm x^{3}y^{3} & 0 & 0 & 0\\
0 & 0 & x^{2}y^{4} & 0 & 0 & 0 & 0 & 0\\
0 & 0 & 0 & 0 & 0 & 0 & \pm x^{3}y^{3} & 0\\
y^{6} & 0 & 0 & 0 & 0 & 0 & 0 & 0
\end{array}
\right)  ,%
\begin{array}
[c]{c}%
\operatorname*{tr}\tilde{c}=\pm4x^{3}y^{3},\\
\det\tilde{c}_{rank=8}^{bisymm2}\left(  x,y\right)  =x^{24}y^{24}%
,\text{\ }x,y\neq0,
\end{array}
\label{b21}\\
\text{eigenvalues}  &  \text{: }\left\{  x^{3}y^{3}\right\}  ^{\left[
2\right]  },\left\{  -x^{3}y^{3}\right\}  ^{\left[  2\right]  },\left\{  \pm
x^{3}y^{3}\right\}  ^{\left[  4\right]  },\nonumber
\end{align}%
\begin{align}
\tilde{c}_{rank=8}^{bisymm2,2}\left(  x,y\right)   &  =\left(
\begin{array}
[c]{cccccccc}%
0 & 0 & 0 & 0 & 0 & 0 & 0 & x^{6}\\
0 & \pm x^{3}y^{3} & 0 & 0 & 0 & 0 & 0 & 0\\
0 & 0 & 0 & 0 & 0 & x^{4}y^{2} & 0 & 0\\
0 & 0 & 0 & \pm x^{3}y^{3} & 0 & 0 & 0 & 0\\
0 & 0 & 0 & 0 & \pm x^{3}y^{3} & 0 & 0 & 0\\
0 & 0 & -x^{2}y^{4} & 0 & 0 & 0 & 0 & 0\\
0 & 0 & 0 & 0 & 0 & 0 & \pm x^{3}y^{3} & 0\\
-y^{6} & 0 & 0 & 0 & 0 & 0 & 0 & 0
\end{array}
\right)  ,%
\begin{array}
[c]{c}%
\operatorname*{tr}\tilde{c}=\pm4x^{3}y^{3}\\
\det\tilde{c}_{rank=8}^{bisymm2}\left(  x,y\right)  =x^{24}y^{24},x,y\neq0,
\end{array}
\label{b22}\\
\text{eigenvalues}  &  \text{: }\left\{  ix^{3}y^{3}\right\}  ^{\left[
2\right]  },\left\{  -ix^{3}y^{3}\right\}  ^{\left[  2\right]  },\left\{  \pm
x^{3}y^{3}\right\}  ^{\left[  4\right]  }.\nonumber
\end{align}

$\bullet$ The $\mathsf{symm1}$ series consists of $4$ two-parameter matrices%
\begin{equation}
\tilde{c}_{rank=8}^{symm1,1}\left(  x,y\right)  =\left(
\begin{array}
[c]{cccccccc}%
xy & 0 & 0 & 0 & 0 & 0 & 0 & 0\\
0 & 0 & 0 & 0 & \pm xy & 0 & 0 & 0\\
0 & 0 & 0 & 0 & 0 & 0 & 0 & y^{2}\\
0 & 0 & 0 & xy & 0 & 0 & 0 & 0\\
0 & \pm xy & 0 & 0 & 0 & 0 & 0 & 0\\
0 & 0 & 0 & 0 & 0 & xy & 0 & 0\\
0 & 0 & 0 & 0 & 0 & 0 & xy & 0\\
0 & 0 & x^{2} & 0 & 0 & 0 & 0 & 0
\end{array}
\right)  ,%
\begin{array}
[c]{c}%
\operatorname*{tr}\tilde{c}=4xy,\\
\det\tilde{c}=x^{8}y^{8},\ \ x,y\neq0,\\
\text{eigenvalues: }\left\{  xy\right\}  ^{\left[  6\right]  },\left\{
-xy\right\}  ^{\left[  2\right]  },
\end{array}
\label{s11}%
\end{equation}%
\begin{equation}
\tilde{c}_{rank=8}^{symm1,2}\left(  x,y\right)  =\left(
\begin{array}
[c]{cccccccc}%
xy & 0 & 0 & 0 & 0 & 0 & 0 & 0\\
0 & 0 & 0 & 0 & \pm xy & 0 & 0 & 0\\
0 & 0 & 0 & 0 & 0 & 0 & 0 & y^{2}\\
0 & 0 & 0 & xy & 0 & 0 & 0 & 0\\
0 & \mp xy & 0 & 0 & 0 & 0 & 0 & 0\\
0 & 0 & 0 & 0 & 0 & xy & 0 & 0\\
0 & 0 & 0 & 0 & 0 & 0 & xy & 0\\
0 & 0 & -x^{2} & 0 & 0 & 0 & 0 & 0
\end{array}
\right)  ,%
\begin{array}
[c]{c}%
\operatorname*{tr}\tilde{c}=4xy,\\
\det\tilde{c}=x^{8}y^{8},\ \ x,y\neq0,\\
\text{eigenvalues: }\left\{  xy\right\}  ^{\left[  6\right]  },\left\{
-xy\right\}  ^{\left[  2\right]  },
\end{array}
\label{s12}%
\end{equation}

$\bullet$ The $\mathsf{symm2}$ series consists of $4$ two-parameter matrices%
\begin{equation}
\tilde{c}_{rank=8}^{symm2,1}\left(  x,y\right)  =\left(
\begin{array}
[c]{cccccccc}%
0 & 0 & 0 & 0 & 0 & y^{2} & 0 & 0\\
0 & xy & 0 & 0 & 0 & 0 & 0 & 0\\
0 & 0 & xy & 0 & 0 & 0 & 0 & 0\\
0 & 0 & 0 & 0 & 0 & 0 & \pm xy & 0\\
0 & 0 & 0 & 0 & xy & 0 & 0 & 0\\
x^{2} & 0 & 0 & 0 & 0 & 0 & 0 & 0\\
0 & 0 & 0 & \pm xy & 0 & 0 & 0 & 0\\
0 & 0 & 0 & 0 & 0 & 0 & 0 & xy
\end{array}
\right)  ,%
\begin{array}
[c]{c}%
\operatorname*{tr}\tilde{c}=4xy,\\
\det\tilde{c}=x^{8}y^{8},\ \ x,y\neq0,\\
\text{eigenvalues: }\left\{  xy\right\}  ^{\left[  4\right]  },\left\{
ixy\right\}  ^{\left[  2\right]  },\left\{  -ixy\right\}  ^{\left[  2\right]
}.
\end{array}
\label{s21}%
\end{equation}%
\begin{equation}
\tilde{c}_{rank=8}^{symm2,2}\left(  x,y\right)  =\left(
\begin{array}
[c]{cccccccc}%
0 & 0 & 0 & 0 & 0 & y^{2} & 0 & 0\\
0 & xy & 0 & 0 & 0 & 0 & 0 & 0\\
0 & 0 & xy & 0 & 0 & 0 & 0 & 0\\
0 & 0 & 0 & 0 & 0 & 0 & \pm xy & 0\\
0 & 0 & 0 & 0 & xy & 0 & 0 & 0\\
-x^{2} & 0 & 0 & 0 & 0 & 0 & 0 & 0\\
0 & 0 & 0 & \mp xy & 0 & 0 & 0 & 0\\
0 & 0 & 0 & 0 & 0 & 0 & 0 & xy
\end{array}
\right)  ,%
\begin{array}
[c]{c}%
\operatorname*{tr}\tilde{c}=4xy,\\
\det\tilde{c}=x^{8}y^{8},\ \ x,y\neq0,\\
\text{eigenvalues: }\left\{  xy\right\}  ^{\left[  4\right]  },\left\{
ixy\right\}  ^{\left[  2\right]  },\left\{  -ixy\right\}  ^{\left[  2\right]
}.
\end{array}
\label{s22}%
\end{equation}

The above matrices with the same eigenvalues are similar, but their
conjugation matrices do not have the form of the triple Kronecker product
(\ref{q8}), and therefore all of them together are $16$ different
two-parameter invertible solutions to the ternary braid equations (\ref{pt}).
Further families of solutions can be obtained using ternary $q$-conjugation
(\ref{cqt}).

\subsection{Group structure of the star and circle $8$-vertex matrices}

Here we investigate the group structure of $8\times8$ matrices by analogy with
the star-like (\ref{m4}) and circle-like (\ref{mc4}) $4\times4$ matrices which
are connected with our $8$-vertex constant solutions (\ref{b11})--(\ref{s22})
to the ternary braid equations (\ref{pt}).

Let us introduce the \textit{star-like }$8\times8$ \textit{matrices} (cf.
(\ref{m4})) which correspond to the $\mathsf{bisymm}$ series (\ref{b11}%
)--(\ref{b22})%
\begin{align}
N_{star1}^{\prime}  &  =\left(
\begin{array}
[c]{cccccccc}%
x & 0 & 0 & 0 & 0 & 0 & 0 & 0\\
0 & 0 & 0 & 0 & 0 & 0 & y & 0\\
0 & 0 & z & 0 & 0 & 0 & 0 & 0\\
0 & 0 & 0 & 0 & s & 0 & 0 & 0\\
0 & 0 & 0 & t & 0 & 0 & 0 & 0\\
0 & 0 & 0 & 0 & 0 & u & 0 & 0\\
0 & v & 0 & 0 & 0 & 0 & 0 & 0\\
0 & 0 & 0 & 0 & 0 & 0 & 0 & w
\end{array}
\right)  ,\ \ \ \ N_{star2}^{\prime}=\left(
\begin{array}
[c]{cccccccc}%
0 & 0 & 0 & 0 & 0 & 0 & 0 & y\\
0 & x & 0 & 0 & 0 & 0 & 0 & 0\\
0 & 0 & 0 & 0 & 0 & s & 0 & 0\\
0 & 0 & 0 & z & 0 & 0 & 0 & 0\\
0 & 0 & 0 & 0 & u & 0 & 0 & 0\\
0 & 0 & t & 0 & 0 & 0 & 0 & 0\\
0 & 0 & 0 & 0 & 0 & 0 & w & 0\\
v & 0 & 0 & 0 & 0 & 0 & 0 & 0
\end{array}
\right)  ,\label{ns}\\
&
\begin{array}
[c]{c}%
\operatorname*{tr}N^{\prime}=x+z+u+w,\ \ \ \det N^{\prime}%
=stuvwxyz,\ \ s,t,u,v,w,x,y,z\neq0,\\
\text{eigenvalues: }x,z,u,w,-\sqrt{yv},\sqrt{yv},-\sqrt{st},\sqrt{st},
\end{array}
\nonumber
\end{align}
and the \textit{circle-like }$8\times8$ \textit{matrices} (cf. (\ref{mc4}))
which correspond to the $\mathsf{symm}$ series (\ref{s11})--(\ref{s22})%
\begin{align}
N_{circ1}^{\prime}  &  =\left(
\begin{array}
[c]{cccccccc}%
x & 0 & 0 & 0 & 0 & 0 & 0 & 0\\
0 & 0 & 0 & 0 & y & 0 & 0 & 0\\
0 & 0 & 0 & 0 & 0 & 0 & 0 & z\\
0 & 0 & 0 & s & 0 & 0 & 0 & 0\\
0 & t & 0 & 0 & 0 & 0 & 0 & 0\\
0 & 0 & 0 & 0 & 0 & u & 0 & 0\\
0 & 0 & 0 & 0 & 0 & 0 & v & 0\\
0 & 0 & w & 0 & 0 & 0 & 0 & 0
\end{array}
\right)  ,\ \ \ \ N_{circ2}^{\prime}=\left(
\begin{array}
[c]{cccccccc}%
0 & 0 & 0 & 0 & 0 & y & 0 & 0\\
0 & x & 0 & 0 & 0 & 0 & 0 & 0\\
0 & 0 & s & 0 & 0 & 0 & 0 & 0\\
0 & 0 & 0 & 0 & 0 & 0 & z & 0\\
0 & 0 & 0 & 0 & u & 0 & 0 & 0\\
t & 0 & 0 & 0 & 0 & 0 & 0 & 0\\
0 & 0 & 0 & w & 0 & 0 & 0 & 0\\
0 & 0 & 0 & 0 & 0 & 0 & 0 & v
\end{array}
\right)  ,\label{nc}\\
&
\begin{array}
[c]{c}%
\operatorname*{tr}N^{\prime}=x+s+u+v,\ \ \ \det N^{\prime}%
=stuvwxyz,\ \ s,t,u,v,w,x,y,z\neq0,\\
\text{eigenvalues: }x,s,u,v,-\sqrt{ty},\sqrt{ty},-\sqrt{wz},\sqrt{wz}.
\end{array}
\end{align}

Denote the corresponding sets by $\mathsf{N}_{star1,2}^{\prime}=\left\{
N_{star1,2}^{\prime}\right\}  $ and $\mathsf{N}_{circ1,2}^{\prime}=\left\{
N_{circ1,2}^{\prime}\right\}  $, then we have for them (which differs from
$4\times4$ matrix sets (\ref{mn4}))%
\begin{equation}
\mathsf{M}_{full}^{\prime}=\mathsf{N}_{star1}^{\prime}\cup\mathsf{N}%
_{star2}^{\prime}\cup\mathsf{N}_{circ1}^{\prime}\cup\mathsf{N}_{circ2}%
^{\prime},\ \ \ \mathsf{N}_{star1}^{\prime}\cap\mathsf{N}_{star2}^{\prime}%
\cap\mathsf{N}_{circ1}^{\prime}\cap\mathsf{N}_{circ2}^{\prime}=\mathsf{D},
\label{mf8}%
\end{equation}
where $\mathsf{D}$ is the set of diagonal $8\times8$ matrices. Again, as for
$4\times4$ star-like and circle-like matrices, there are no closed binary
multiplications among the sets of $8$-vertex matrices (\ref{ns})--(\ref{nc}).
Nevertheless, we have the following triple set products%
\begin{align}
\mathsf{N}_{star1}^{\prime}\mathsf{N}_{star1}^{\prime}\mathsf{N}%
_{star1}^{\prime}  &  =\mathsf{N}_{star1}^{\prime},\label{nns1}\\
\mathsf{N}_{star2}^{\prime}\mathsf{N}_{star2}^{\prime}\mathsf{N}%
_{star2}^{\prime}  &  =\mathsf{N}_{star2}^{\prime},\label{nns2}\\
\mathsf{N}_{circ1}^{\prime}\mathsf{N}_{circ1}^{\prime}\mathsf{N}%
_{circ1}^{\prime}  &  =\mathsf{N}_{circ1}^{\prime},\label{nnc1}\\
\mathsf{N}_{circ2}^{\prime}\mathsf{N}_{circ2}^{\prime}\mathsf{N}%
_{circ2}^{\prime}  &  =\mathsf{N}_{circ2}^{\prime}, \label{nnc2}%
\end{align}
which should be compared with the analogous $4\times4$ matrices (\ref{nc1}%
)--(\ref{nc2}): note that now we do not have pentuple products.

Using the definitions (\ref{mqm})--(\ref{ik}), we interpret the closed
products (\ref{nns1})--(\ref{nns2}) and (\ref{nnc1})--(\ref{nnc2}) as the
multiplications $\mu^{\left[  3\right]  }$ (being the ordinary triple matrix
product) of the \textit{ternary semigroups} $\mathcal{S}_{star1,2}^{\left[
3\right]  }\left(  8,\mathbb{C}\right)  =\left\{  \mathsf{N}_{star1,2}%
^{\prime}\mid\mu^{\left[  3\right]  }\right\}  $ and $\mathcal{S}%
_{circ1,2}^{\left[  3\right]  }\left(  8,\mathbb{C}\right)  =\left\{
\mathsf{N}_{circ1,2}^{\prime}\mid\mu^{\left[  3\right]  }\right\}  $,
respectively. The corresponding querelements (\ref{mqm}) are given by%
\begin{equation}
\bar{N}_{star1}^{\prime}=N_{star1}^{\prime-1}=\left(
\begin{array}
[c]{cccccccc}%
\frac{1}{x} & 0 & 0 & 0 & 0 & 0 & 0 & 0\\
0 & 0 & 0 & 0 & 0 & 0 & \frac{1}{v} & 0\\
0 & 0 & \frac{1}{z} & 0 & 0 & 0 & 0 & 0\\
0 & 0 & 0 & 0 & \frac{1}{t} & 0 & 0 & 0\\
0 & 0 & 0 & \frac{1}{s} & 0 & 0 & 0 & 0\\
0 & 0 & 0 & 0 & 0 & \frac{1}{u} & 0 & 0\\
0 & \frac{1}{y} & 0 & 0 & 0 & 0 & 0 & 0\\
0 & 0 & 0 & 0 & 0 & 0 & 0 & \frac{1}{w}%
\end{array}
\right)  ,\bar{N}_{star2}^{\prime}=N_{star2}^{\prime-1}=\left(
\begin{array}
[c]{cccccccc}%
0 & 0 & 0 & 0 & 0 & 0 & 0 & \frac{1}{v}\\
0 & \frac{1}{x} & 0 & 0 & 0 & 0 & 0 & 0\\
0 & 0 & 0 & 0 & 0 & \frac{1}{t} & 0 & 0\\
0 & 0 & 0 & \frac{1}{z} & 0 & 0 & 0 & 0\\
0 & 0 & 0 & 0 & \frac{1}{u} & 0 & 0 & 0\\
0 & 0 & \frac{1}{s} & 0 & 0 & 0 & 0 & 0\\
0 & 0 & 0 & 0 & 0 & 0 & \frac{1}{w} & 0\\
\frac{1}{y} & 0 & 0 & 0 & 0 & 0 & 0 & 0
\end{array}
\right)  ,s,t,u,v,w,x,y,z\neq0, \label{nq1}%
\end{equation}
and%
\begin{equation}
\bar{N}_{circ1}^{\prime}=N_{circ1}^{\prime-1}=\left(
\begin{array}
[c]{cccccccc}%
\frac{1}{x} & 0 & 0 & 0 & 0 & 0 & 0 & 0\\
0 & 0 & 0 & 0 & \frac{1}{t} & 0 & 0 & 0\\
0 & 0 & 0 & 0 & 0 & 0 & 0 & \frac{1}{w}\\
0 & 0 & 0 & \frac{1}{s} & 0 & 0 & 0 & 0\\
0 & \frac{1}{y} & 0 & 0 & 0 & 0 & 0 & 0\\
0 & 0 & 0 & 0 & 0 & \frac{1}{u} & 0 & 0\\
0 & 0 & 0 & 0 & 0 & 0 & \frac{1}{v} & 0\\
0 & 0 & \frac{1}{z} & 0 & 0 & 0 & 0 & 0
\end{array}
\right)  ,\bar{N}_{circ2}^{\prime}=N_{circ2}^{\prime-1}=\left(
\begin{array}
[c]{cccccccc}%
0 & 0 & 0 & 0 & 0 & \frac{1}{t} & 0 & 0\\
0 & \frac{1}{x} & 0 & 0 & 0 & 0 & 0 & 0\\
0 & 0 & \frac{1}{s} & 0 & 0 & 0 & 0 & 0\\
0 & 0 & 0 & 0 & 0 & 0 & \frac{1}{w} & 0\\
0 & 0 & 0 & 0 & \frac{1}{u} & 0 & 0 & 0\\
\frac{1}{y} & 0 & 0 & 0 & 0 & 0 & 0 & 0\\
0 & 0 & 0 & \frac{1}{z} & 0 & 0 & 0 & 0\\
0 & 0 & 0 & 0 & 0 & 0 & 0 & \frac{1}{v}%
\end{array}
\right)  ,s,t,u,v,w,x,y,z\neq0. \label{nq2}%
\end{equation}

The ternary semigroups $\mathcal{S}_{star1,2}^{\left[  3\right]  }\left(
8,\mathbb{C}\right)  =\left\{  \mathsf{N}_{star1,2}^{\prime}\mid\mu^{\left[
3\right]  }\right\}  $ and $\mathcal{S}_{circ1,2}^{\left[  3\right]  }\left(
8,\mathbb{C}\right)  =\left\{  \mathsf{N}_{circ1,2}^{\prime}\mid\mu^{\left[
3\right]  }\right\}  $ in which every element has its querelement given by
(\ref{nq1})--(\ref{nq2}) become the \textit{ternary groups} $\mathcal{G}%
_{star1,2}^{\left[  3\right]  }\left(  8,\mathbb{C}\right)  =\left\{
\mathsf{N}_{star1,2}^{\prime}\mid\mu^{\left[  3\right]  },\overline{\left(
\ \right)  }\right\}  $ and $\mathcal{G}_{circ1,2}^{\left[  3\right]  }\left(
8,\mathbb{C}\right)  =\left\{  \mathsf{N}_{circ1,2}^{\prime}\mid\mu{}^{\left[
3\right]  },\overline{\left(  \ \right)  }\right\}  $, which are four
different ($3$-nonderived) ternary subgroups of the derived ternary general
linear group $\mathrm{GL}^{\left[  3\right]  }\left(  8,\mathbb{C}\right)  $.
The ternary identities in $\mathcal{G}_{star1,2}^{\left[  3\right]  }\left(
8,\mathbb{C}\right)  $ and $\mathcal{G}_{circ1,2}^{\left[  3\right]  }\left(
8,\mathbb{C}\right)  $ are the following different continuous sets
$\mathsf{I}_{star1,2}^{\prime\left[  3\right]  }=\left\{  I_{star1,2}%
^{\prime\left[  3\right]  }\right\}  $ and $\mathsf{I}_{circ1,2}%
^{\prime\left[  3\right]  }=\left\{  I_{circ1,2}^{\prime\left[  3\right]
}\right\}  $, where%
\begin{align}
I_{star1}^{\prime\left[  3\right]  }  &  =\left(
\begin{array}
[c]{cccccccc}%
e^{i\alpha_{1}} & 0 & 0 & 0 & 0 & 0 & 0 & 0\\
0 & 0 & 0 & 0 & 0 & 0 & e^{i\alpha_{2}} & 0\\
0 & 0 & e^{i\alpha_{3}} & 0 & 0 & 0 & 0 & 0\\
0 & 0 & 0 & 0 & e^{i\alpha_{4}} & 0 & 0 & 0\\
0 & 0 & 0 & e^{i\alpha_{5}} & 0 & 0 & 0 & 0\\
0 & 0 & 0 & 0 & 0 & e^{i\alpha_{6}} & 0 & 0\\
0 & e^{i\alpha_{7}} & 0 & 0 & 0 & 0 & 0 & 0\\
0 & 0 & 0 & 0 & 0 & 0 & 0 & e^{i\alpha_{8}}%
\end{array}
\right)  ,\ \ I_{star2}^{\prime\left[  3\right]  }=\left(
\begin{array}
[c]{cccccccc}%
0 & 0 & 0 & 0 & 0 & 0 & 0 & e^{i\alpha_{2}}\\
0 & e^{i\alpha_{1}} & 0 & 0 & 0 & 0 & 0 & 0\\
0 & 0 & 0 & 0 & 0 & e^{i\alpha_{4}} & 0 & 0\\
0 & 0 & 0 & e^{i\alpha_{3}} & 0 & 0 & 0 & 0\\
0 & 0 & 0 & 0 & e^{i\alpha_{6}} & 0 & 0 & 0\\
0 & 0 & e^{i\alpha_{5}} & 0 & 0 & 0 & 0 & 0\\
0 & 0 & 0 & 0 & 0 & 0 & e^{i\alpha_{8}} & 0\\
e^{i\alpha_{7}} & 0 & 0 & 0 & 0 & 0 & 0 & 0
\end{array}
\right)  ,\nonumber\\
e^{2i\alpha_{1}}  &  =e^{2i\alpha_{3}}=e^{2i\alpha_{6}}=e^{2i\alpha_{8}%
}=e^{i\left(  \alpha_{2}+\alpha_{7}\right)  }=e^{i\left(  \alpha_{4}%
+\alpha_{5}\right)  }=1,\ \ \alpha_{1},\ldots,\alpha_{8}\in\mathbb{R},
\label{is}%
\end{align}
and%
\begin{align}
I_{circ1}^{\prime\left[  3\right]  }  &  =\left(
\begin{array}
[c]{cccccccc}%
e^{i\alpha_{1}} & 0 & 0 & 0 & 0 & 0 & 0 & 0\\
0 & 0 & 0 & 0 & e^{i\alpha_{2}} & 0 & 0 & 0\\
0 & 0 & 0 & 0 & 0 & 0 & 0 & e^{i\alpha_{3}}\\
0 & 0 & 0 & e^{i\alpha_{4}} & 0 & 0 & 0 & 0\\
0 & e^{i\alpha_{5}} & 0 & 0 & 0 & 0 & 0 & 0\\
0 & 0 & 0 & 0 & 0 & e^{i\alpha_{6}} & 0 & 0\\
0 & 0 & 0 & 0 & 0 & 0 & e^{i\alpha_{7}} & 0\\
0 & 0 & e^{i\alpha_{8}} & 0 & 0 & 0 & 0 & 0
\end{array}
\right)  ,\ \ I_{circ2}^{\prime\left[  3\right]  }=\left(
\begin{array}
[c]{cccccccc}%
0 & 0 & 0 & 0 & 0 & e^{i\alpha_{2}} & 0 & 0\\
0 & e^{i\alpha_{1}} & 0 & 0 & 0 & 0 & 0 & 0\\
0 & 0 & e^{i\alpha_{4}} & 0 & 0 & 0 & 0 & 0\\
0 & 0 & 0 & 0 & 0 & 0 & e^{i\alpha_{3}} & 0\\
0 & 0 & 0 & 0 & e^{i\alpha_{6}} & 0 & 0 & 0\\
e^{i\alpha_{5}} & 0 & 0 & 0 & 0 & 0 & 0 & 0\\
0 & 0 & 0 & e^{i\alpha_{8}} & 0 & 0 & 0 & 0\\
0 & 0 & 0 & 0 & 0 & 0 & 0 & e^{i\alpha_{7}}%
\end{array}
\right)  ,\nonumber\\
e^{2i\alpha_{1}}  &  =e^{2i\alpha_{4}}=e^{2i\alpha_{6}}=e^{2i\alpha_{7}%
}=e^{i\left(  \alpha_{3}+\alpha_{8}\right)  }=e^{i\left(  \alpha_{2}%
+\alpha_{5}\right)  }=1,\ \ \alpha_{1},\ldots,\alpha_{8}\in\mathbb{R},
\label{ic}%
\end{align}
such that all the identities are the $8\times8$ matrix reflections $\left(
I^{\prime\left[  3\right]  }\right)  ^{2}=I_{8}$ (see (\ref{ik})). If
$\alpha_{j}=0$, $j=1,\ldots,8$, the ternary identities (\ref{is})--(\ref{ic})
coincide with the $8\times8$ permutation matrices (\ref{cp8b})--(\ref{cp8s}),
which are solutions to the ternary braid equations (\ref{pt}).

The module structure of the $8$-vertex star-like (\ref{ns}) and circle-like
(\ref{nc}) $8\times8$ matrix sets differs from the $4\times4$ matrix sets
(\ref{sss1})--(\ref{ccc6}). Firstly, because of the absence of pentuple matrix
products (\ref{ccc1})--(\ref{ccc6}), and secondly through some differences in
the ternary closed products of sets.

We have the following triple relations between star and circle matrices
separately (the sets corresponding to modules are in brackets, and we
informally denote modules by their sets)%

\begin{align}
\mathsf{N}_{star1}^{\prime}\left(  \mathsf{N}_{star2}^{\prime}\right)
\mathsf{N}_{star1}^{\prime}  &  =\left(  \mathsf{N}_{star2}^{\prime}\right)
,\ \ \ \mathsf{N}_{circ1}^{\prime}\left(  \mathsf{N}_{circ2}^{\prime}\right)
\mathsf{N}_{circ1}^{\prime}=\left(  \mathsf{N}_{circ2}^{\prime}\right)
,\label{s8s1}\\
\mathsf{N}_{star1}^{\prime}\mathsf{N}_{star1}^{\prime}\left(  \mathsf{N}%
_{star2}^{\prime}\right)   &  =\left(  \mathsf{N}_{star2}^{\prime}\right)
,\ \ \ \mathsf{N}_{circ1}^{\prime}\mathsf{N}_{circ1}^{\prime}\left(
\mathsf{N}_{circ2}^{\prime}\right)  =\mathsf{N}_{circ2}^{\prime}%
,\label{s8s2}\\
\left(  \mathsf{N}_{star2}^{\prime}\right)  \mathsf{N}_{star1}^{\prime
}\mathsf{N}_{star1}^{\prime}  &  =\left(  \mathsf{N}_{star2}^{\prime}\right)
,\ \ \ \left(  \mathsf{N}_{circ2}^{\prime}\right)  \mathsf{N}_{circ1}^{\prime
}\mathsf{N}_{circ1}^{\prime}=\left(  \mathsf{N}_{circ2}^{\prime}\right)
,\label{s8s3}\\
\mathsf{N}_{star2}^{\prime}\mathsf{N}_{star2}^{\prime}\left(  \mathsf{N}%
_{star1}^{\prime}\right)   &  =\left(  \mathsf{N}_{star1}^{\prime}\right)
,\ \ \ \mathsf{N}_{circ2}^{\prime}\mathsf{N}_{circ2}^{\prime}\left(
\mathsf{N}_{circ1}^{\prime}\right)  =\left(  \mathsf{N}_{circ1}^{\prime
}\right)  ,\label{s8s4}\\
\mathsf{N}_{star2}^{\prime}\left(  \mathsf{N}_{star1}^{\prime}\right)
\mathsf{N}_{star2}^{\prime}  &  =\left(  \mathsf{N}_{star1}^{\prime}\right)
,\ \ \ \mathsf{N}_{circ2}^{\prime}\left(  \mathsf{N}_{circ1}^{\prime}\right)
\mathsf{N}_{circ2}^{\prime}=\left(  \mathsf{N}_{circ1}^{\prime}\right)
,\label{s8s5}\\
\left(  \mathsf{N}_{star1}^{\prime}\right)  \mathsf{N}_{star2}^{\prime
}\mathsf{N}_{star2}^{\prime}  &  =\left(  \mathsf{N}_{star1}^{\prime}\right)
,\ \ \ \left(  \mathsf{N}_{circ1}^{\prime}\right)  \mathsf{N}_{circ2}^{\prime
}\mathsf{N}_{circ2}^{\prime}=\left(  \mathsf{N}_{circ1}^{\prime}\right)  .
\label{s8s6}%
\end{align}
So we may observe the following module structures: 1) from (\ref{s8s1}%
)--(\ref{s8s3}), the sets $\mathsf{N}_{star2}^{\prime}$ ($\mathsf{N}%
_{circ2}^{\prime}$) are the middle, right and left ternary modules over
$\mathsf{N}_{star1}^{\prime}$ ($\mathsf{N}_{circ1}^{\prime}$); 2) from
(\ref{s8s4})--(\ref{s8s6}), the set $\mathsf{N}_{star1}^{\prime}$
($\mathsf{N}_{circ1}^{\prime}$) are middle, right and left ternary modules
over $\mathsf{N}_{star2}^{\prime}$ ($\mathsf{N}_{circ2}^{\prime}$);
\begin{align}
\mathsf{N}_{star1}^{\prime}\mathsf{N}_{star1}^{\prime}\left(  \mathsf{N}%
_{circ1}^{\prime}\right)   &  =\left(  \mathsf{N}_{circ1}^{\prime}\right)
,\ \ \ \left(  \mathsf{N}_{circ1}^{\prime}\right)  \mathsf{N}_{star1}^{\prime
}\mathsf{N}_{star1}^{\prime}=\left(  \mathsf{N}_{circ1}^{\prime}\right)
,\label{s8c3}\\
\mathsf{N}_{star1}^{\prime}\mathsf{N}_{star1}^{\prime}\left(  \mathsf{N}%
_{circ2}^{\prime}\right)   &  =\left(  \mathsf{N}_{circ2}^{\prime}\right)
,\ \ \ \left(  \mathsf{N}_{circ2}^{\prime}\right)  \mathsf{N}_{star1}^{\prime
}\mathsf{N}_{star1}^{\prime}=\left(  \mathsf{N}_{circ2}^{\prime}\right)  ,\\
\mathsf{N}_{star2}^{\prime}\mathsf{N}_{star2}^{\prime}\left(  \mathsf{N}%
_{circ1}^{\prime}\right)   &  =\left(  \mathsf{N}_{circ1}^{\prime}\right)
,\ \ \ \left(  \mathsf{N}_{circ1}^{\prime}\right)  \mathsf{N}_{star2}^{\prime
}\mathsf{N}_{star2}^{\prime}=\left(  \mathsf{N}_{circ1}^{\prime}\right)  ,\\
\mathsf{N}_{star2}^{\prime}\mathsf{N}_{star2}^{\prime}\left(  \mathsf{N}%
_{circ2}^{\prime}\right)   &  =\left(  \mathsf{N}_{circ2}^{\prime}\right)
,\ \ \ \left(  \mathsf{N}_{circ2}^{\prime}\right)  \mathsf{N}_{star2}^{\prime
}\mathsf{N}_{star2}^{\prime}=\left(  \mathsf{N}_{circ2}^{\prime}\right)  ,
\label{s8c6}%
\end{align}
3) from (\ref{s8c3})--(\ref{s8c6}), the sets $\mathsf{N}_{circ1,2}^{\prime}$
are right and left ternary modules over $\mathsf{N}_{star1,2}^{\prime}$;%
\begin{align}
\mathsf{N}_{circ1}^{\prime}\mathsf{N}_{circ1}^{\prime}\left(  \mathsf{N}%
_{star1}^{\prime}\right)   &  =\left(  \mathsf{N}_{star1}^{\prime}\right)
,\ \ \ \left(  \mathsf{N}_{star1}^{\prime}\right)  \mathsf{N}_{circ1}^{\prime
}\mathsf{N}_{circ1}^{\prime}=\left(  \mathsf{N}_{star1}^{\prime}\right)
,\label{c8c1}\\
\mathsf{N}_{circ1}^{\prime}\mathsf{N}_{circ1}^{\prime}\left(  \mathsf{N}%
_{star2}^{\prime}\right)   &  =\left(  \mathsf{N}_{star2}^{\prime}\right)
,\ \ \ \left(  \mathsf{N}_{star2}^{\prime}\right)  \mathsf{N}_{circ1}^{\prime
}\mathsf{N}_{circ1}^{\prime}=\left(  \mathsf{N}_{star2}^{\prime}\right)  ,\\
\mathsf{N}_{circ2}^{\prime}\mathsf{N}_{circ2}^{\prime}\left(  \mathsf{N}%
_{star1}^{\prime}\right)   &  =\left(  \mathsf{N}_{star1}^{\prime}\right)
,\ \ \ \left(  \mathsf{N}_{star1}^{\prime}\right)  \mathsf{N}_{circ2}^{\prime
}\mathsf{N}_{circ2}^{\prime}=\left(  \mathsf{N}_{star1}^{\prime}\right)  ,\\
\mathsf{N}_{circ2}^{\prime}\mathsf{N}_{circ2}^{\prime}\left(  \mathsf{N}%
_{star2}^{\prime}\right)   &  =\left(  \mathsf{N}_{star2}^{\prime}\right)
,\ \ \ \left(  \mathsf{N}_{star2}^{\prime}\right)  \mathsf{N}_{circ2}^{\prime
}\mathsf{N}_{circ2}^{\prime}=\left(  \mathsf{N}_{star2}^{\prime}\right)  ,
\label{c8c4}%
\end{align}
4) from (\ref{c8c1})--(\ref{c8c4}), the sets $\mathsf{N}_{star1,2}^{\prime}$
are right and left ternary modules over $\mathsf{N}_{circ1,2}^{\prime}$.

\subsection{Group structure of the star and circle $16$-vertex matrices}

Next we will introduce $8\times8$ matrices of a special form similar to the
star $8$-vertex matrices (\ref{star}) and the circle $8$-vertex matrices
(\ref{circ}), analyze their group structure and establish which ones could be
$16$-vertex solutions to the ternary braid equations (\ref{pt}). We will
derive the solutions in the opposite way to that for the $8$-vertex
Yang-Baxter maps, following the note after (\ref{cp2}). Indeed, the sum of the
permutation $\mathsf{bisymm}$ solutions (\ref{cp8b}) gives the shape of the
$8\times8$ star matrix $M_{star}^{\prime}$ (as in (\ref{star})), while the sum
of $\mathsf{symm}$ solutions (\ref{cp8s}) gives the $8\times8$ circle matrix
$M_{circ}^{\prime}$ (as in (\ref{circ}))%
\begin{align}
M_{star}^{\prime}  &  =\left(
\begin{array}
[c]{cccccccc}%
x & 0 & 0 & 0 & 0 & 0 & 0 & y\\
0 & z & 0 & 0 & 0 & 0 & s & 0\\
0 & 0 & t & 0 & 0 & u & 0 & 0\\
0 & 0 & 0 & v & w & 0 & 0 & 0\\
0 & 0 & 0 & a & b & 0 & 0 & 0\\
0 & 0 & c & 0 & 0 & d & 0 & 0\\
0 & f & 0 & 0 & 0 & 0 & g & 0\\
h & 0 & 0 & 0 & 0 & 0 & 0 & p
\end{array}
\right)  ,\ \ \label{m8s}\\
M_{circ}^{\prime}  &  =\left(
\begin{array}
[c]{cccccccc}%
x & 0 & 0 & 0 & 0 & y & 0 & 0\\
0 & z & 0 & 0 & s & 0 & 0 & 0\\
0 & 0 & t & 0 & 0 & 0 & 0 & u\\
0 & 0 & 0 & v & 0 & 0 & w & 0\\
0 & f & 0 & 0 & g & 0 & 0 & 0\\
h & 0 & 0 & 0 & 0 & p & 0 & 0\\
0 & 0 & 0 & a & 0 & 0 & b & 0\\
0 & 0 & c & 0 & 0 & 0 & 0 & d
\end{array}
\right)  , \label{m8c}%
\end{align}%
\begin{align}
&  \operatorname*{tr}M^{\prime}=x+z+t+v+b+d+g+p,\ \ \ \det M^{\prime
}=(bv-aw)(cu-dt)(fs-gz)(px-hy),\\
&  \text{eigenvalues}\text{: }\frac{1}{2}\left(  d+t-\sqrt{4cu+(d-t)^{2}%
}\right)  ,\frac{1}{2}\left(  d+t+\sqrt{4cu+(d-t)^{2}}\right)  ,\nonumber\\
&  \frac{1}{2}\left(  b+v-\sqrt{4aw+(b-v)^{2}}\right)  ,\frac{1}{2}\left(
b+v+\sqrt{4aw+(b-v)^{2}}\right)  ,\frac{1}{2}\left(  p+x-\sqrt{4hy+(p-x)^{2}%
}\right)  ,\nonumber\\
&  \frac{1}{2}\left(  p+x+\sqrt{4hy+(p-x)^{2}}\right)  ,\frac{1}{2}\left(
g+z-\sqrt{4fs+(g-z)^{2}}\right)  ,\frac{1}{2}\left(  g+z+\sqrt{4fs+(g-z)^{2}%
}\right)  .
\end{align}

The $16$-vertex matrices are invertible, if $\det M_{star}^{\prime}\neq0$ and
$\det M_{circ}^{\prime}\neq0$, which give the following joint conditions on
the parameters (cf. (\ref{i}))%
\begin{equation}
bv-aw\neq0,\ \ cu-dt\neq0,\ \ fs-gz\neq0,\ \ px-hy\neq0. \label{i8}%
\end{equation}

Only in this concrete parametrization (\ref{m8s}) and (\ref{m8c}) do the
matrices $M_{star}^{\prime}$ and $M_{circ}^{\prime}$ have the same trace,
determinant and eigenvalues, and they are diagonalizable and conjugate via%
\begin{equation}
U^{\prime}=\left(
\begin{array}
[c]{cccccccc}%
1 & 0 & 0 & 0 & 0 & 0 & 0 & 0\\
0 & 1 & 0 & 0 & 0 & 0 & 0 & 0\\
0 & 0 & 1 & 0 & 0 & 0 & 0 & 0\\
0 & 0 & 0 & 1 & 0 & 0 & 0 & 0\\
0 & 0 & 0 & 0 & 1 & 0 & 1 & 0\\
0 & 0 & 0 & 0 & 0 & 0 & 0 & 1\\
0 & 0 & 0 & 0 & 1 & 0 & 0 & 0\\
0 & 0 & 0 & 0 & 0 & 1 & 0 & 0
\end{array}
\right)  . \label{u1}%
\end{equation}
The matrix $U^{\prime}$ cannot be presented in the form of a triple Kronecker
product (\ref{q8}), and so two matrices $M_{star}^{\prime}$ and $M_{circ}%
^{\prime}$ are not $q$-conjugate in the parametrization (\ref{m8s}) and
(\ref{m8c}), and can lead to different solutions to the ternary braid
equations (\ref{pt}). It follows from (\ref{i8}) that $16$-vertex matrices
with all nonzero entries equal to $1$ are non-invertible, having vanishing
determinant and rank $4$ (despite each one being a sum of two permutation
matrices). In the case all the conditions (\ref{i8}) holding, the inverse
matrices become%
\begin{equation}
\left(  M_{star}^{\prime}\right)  ^{-1}=\left(
\begin{array}
[c]{cccccccc}%
\frac{p}{px-hy} & 0 & 0 & 0 & 0 & 0 & 0 & -\frac{y}{px-hy}\\
0 & -\frac{g}{fs-gz} & 0 & 0 & 0 & 0 & \frac{s}{fs-gz} & 0\\
0 & 0 & -\frac{d}{cu-dt} & 0 & 0 & \frac{u}{cu-dt} & 0 & 0\\
0 & 0 & 0 & \frac{b}{bv-aw} & -\frac{w}{bv-aw} & 0 & 0 & 0\\
0 & 0 & 0 & -\frac{a}{bv-aw} & \frac{v}{bv-aw} & 0 & 0 & 0\\
0 & 0 & \frac{c}{cu-dt} & 0 & 0 & -\frac{t}{cu-dt} & 0 & 0\\
0 & \frac{f}{fs-gz} & 0 & 0 & 0 & 0 & -\frac{z}{fs-gz} & 0\\
-\frac{h}{px-hy} & 0 & 0 & 0 & 0 & 0 & 0 & \frac{x}{px-hy}%
\end{array}
\right)  , \label{mi1}%
\end{equation}%
\begin{equation}
\left(  M_{circ}^{\prime}\right)  ^{-1}=\left(
\begin{array}
[c]{cccccccc}%
\frac{p}{px-hy} & 0 & 0 & 0 & 0 & -\frac{y}{px-hy} & 0 & 0\\
0 & -\frac{g}{fs-gz} & 0 & 0 & \frac{s}{fs-gz} & 0 & 0 & 0\\
0 & 0 & -\frac{d}{cu-dt} & 0 & 0 & 0 & 0 & \frac{u}{cu-dt}\\
0 & 0 & 0 & \frac{b}{bv-aw} & 0 & 0 & -\frac{w}{bv-aw} & 0\\
0 & \frac{f}{fs-gz} & 0 & 0 & -\frac{z}{fs-gz} & 0 & 0 & 0\\
-\frac{h}{px-hy} & 0 & 0 & 0 & 0 & \frac{x}{px-hy} & 0 & 0\\
0 & 0 & 0 & -\frac{a}{bv-aw} & 0 & 0 & \frac{v}{bv-aw} & 0\\
0 & 0 & \frac{c}{cu-dt} & 0 & 0 & 0 & 0 & -\frac{t}{cu-dt}%
\end{array}
\right)  . \label{mi2}%
\end{equation}

Denoting the sets of matrices corresponding to (\ref{m8s}) and (\ref{m8c}) by
$\mathsf{M}_{star}^{\prime}$ and $\mathsf{M}_{circ}^{\prime}$, their
multiplications are%
\begin{equation}
\mathsf{M}_{star}^{\prime}\mathsf{M}_{star}^{\prime}=\mathsf{M}_{star}%
^{\prime},\ \ \ \ \ \ \ \mathsf{M}_{circ}^{\prime}\mathsf{M}_{circ}^{\prime
}=\mathsf{M}_{circ}^{\prime}, \label{mm8}%
\end{equation}
and in term of sets $\mathsf{M}_{star}^{\prime}=\mathsf{N}_{star1}^{\prime
}\cup\mathsf{N}_{star2}^{\prime}$ and $\mathsf{M}_{circ}^{\prime}%
=\mathsf{N}_{circ1}^{\prime}\cup\mathsf{N}_{circ2}^{\prime}$, and
$\mathsf{N}_{star1}^{\prime}\cap\mathsf{N}_{star2}^{\prime}=\mathsf{D}$ and
$\mathsf{N}_{circ1}^{\prime}\cap\mathsf{N}_{circ2}^{\prime}=\mathsf{D}$ (see
(\ref{mf8})). Note that the structure (\ref{mm8}) is considerably different
from the binary case (\ref{mmm})--(\ref{mmm2}), and therefore it may not
necessarily be related to the Cartan decomposition.

The products (\ref{mm8}) mean that both $\mathsf{M}_{star}^{\prime}$ and
$\mathsf{M}_{circ}^{\prime}$ are separately closed with respect to binary
matrix multiplication $\left(  \cdot\right)  $, and therefore $\mathcal{S}%
_{16vert}^{star}=\left\langle \mathsf{M}_{star}^{\prime}\mid\cdot\right\rangle
$ and $\mathcal{S}_{16vert}^{circ}=\left\langle \mathsf{M}_{circ}^{\prime}%
\mid\cdot\right\rangle $ are semigroups. We denote their intersection by
$\mathcal{S}_{8vert}^{\operatorname*{diag}}=\mathcal{S}_{16vert}^{star}%
\cap\mathcal{S}_{16vert}^{circ}$ which is a semigroup of diagonal $8$-vertex
matrices. In case, the invertibility conditions (\ref{i8}) are fulfilled, the
sets $\mathsf{M}_{star}^{\prime}$ and $\mathsf{M}_{circ}^{\prime}$ form
subgroups $\mathcal{G}_{16vert}^{star}=\left\langle \mathsf{M}_{star}^{\prime
}\mid\cdot,\left(  \_\right)  ^{-1},I_{8}\right\rangle $ and $\mathcal{G}%
_{16vert}^{circ}=\left\langle \mathsf{M}_{circ}^{\prime}\mid\cdot,\left(
\_\right)  ^{-1},I_{8}\right\rangle $ (where $I_{8}$ is the $8\times8$
identity matrix) of $\mathrm{GL}\left(  8,\mathbb{C}\right)  $ with the
inverse elements given explicitly by (\ref{mi1})--(\ref{mi2}). Because the
elements $M_{star}^{\prime}$ and $M_{circ}^{\prime}$ in (\ref{m8s}) and
(\ref{m8c}) are conjugates by the invertible matrix $U^{\prime}$ (\ref{u1}),
the subgroups $\mathcal{G}_{16vert}^{star}$ and $\mathcal{G}_{16vert}^{circ}$
(as well as the semigroups $\mathcal{S}_{16vert}^{star}$ and $\mathcal{S}%
_{16vert}^{circ}$) are isomorphic by the obvious isomorphism%
\begin{equation}
M_{star}^{\prime}\mapsto U^{\prime}M_{circ}^{\prime}U^{\prime-1},
\end{equation}
where $U^{\prime}$ is in (\ref{u1}).

The \textquotedblleft interaction\textquotedblright\ between $\mathsf{M}%
_{star}^{\prime}$ and $\mathsf{M}_{circ}^{\prime}$ also differs from the
binary case (\ref{mmm1}), because%
\begin{align}
\mathsf{M}_{star}^{\prime}\mathsf{M}_{circ}^{\prime}  &  =\mathsf{M}%
_{quad}^{\prime},\ \ \ \mathsf{M}_{circ}^{\prime}\mathsf{M}_{star}^{\prime
}=\mathsf{M}_{quad}^{\prime},\label{mcq}\\
\mathsf{M}_{quad}^{\prime}\mathsf{M}_{quad}^{\prime}  &  =\mathsf{M}%
_{quad}^{\prime}, \label{mq}%
\end{align}
where $\mathsf{M}_{quad}^{\prime}$ is a set of $32$-vertex so called
\textit{quad-matrices} of the form%
\begin{equation}
M_{quad}^{\prime}=\left(
\begin{array}
[c]{cccccccc}%
x_{1} & 0 & y_{1} & 0 & 0 & z_{1} & 0 & s_{1}\\
0 & t_{1} & 0 & u_{1} & v_{1} & 0 & w_{1} & 0\\
a_{1} & 0 & b_{1} & 0 & 0 & c_{1} & 0 & d_{1}\\
0 & f_{1} & 0 & g_{1} & h_{1} & 0 & p_{1} & 0\\
0 & x_{2} & 0 & y_{2} & z_{2} & 0 & s_{2} & 0\\
t_{2} & 0 & u_{2} & 0 & 0 & v_{2} & 0 & w_{2}\\
0 & a_{2} & 0 & b_{2} & c_{2} & 0 & d_{2} & 0\\
f_{2} & 0 & g_{2} & 0 & 0 & h_{2} & 0 & p_{2}%
\end{array}
\right)  . \label{qu}%
\end{equation}

Because of (\ref{mq}), the set $\mathsf{M}_{quad}^{\prime}$ is closed with
respect to matrix multiplication, and therefore (for invertible matrices
$M_{quad}^{\prime}$) the group $\mathcal{G}_{32vert}^{quad}=\left\langle
\mathsf{M}_{quad}^{\prime}\mid\cdot,\left(  \_\right)  ^{-1},I_{8}%
\right\rangle $ is a subgroup of $\mathrm{GL}\left(  8,\mathbb{C}\right)  $.
So, in trying to find higher $32$-vertex solutions (having at most half as
many unknown variables as a general $8\times8$ matrix) to the ternary braid
equations (\ref{pt}) it is worthwhile to search within the class of
quad-matrices (\ref{qu}).

Thus, the group structure of the above $16$-vertex $8\times8$ matrices
(\ref{mm8})--(\ref{mq}) is considerably different to that of $8$-vertex
$4\times4$ matrices (\ref{star})--(\ref{circ})as the former contains two
isomorphic binary subgroups $\mathcal{G}_{16vert}^{star}$ and $\mathcal{G}%
_{16vert}^{circ}$ of $\mathrm{GL}\left(  8,\mathbb{C}\right)  $ (cf.
(\ref{mmm})--(\ref{mmm2}) and (\ref{mm8})).

The sets $\mathsf{M}_{star}^{\prime}$, $\mathsf{M}_{circ}^{\prime}$ and
$\mathsf{M}_{quad}^{\prime}$ are closed with respect to matrix addition as
well, and therefore (because of the distributivity of $\mathbb{C}$) they are
the matrix rings $\mathcal{R}_{16vert}^{star}$, $\mathcal{R}_{16vert}^{circ}$
and $\mathcal{R}_{32vert}^{quad}$, respectively. In the invertible case
(\ref{i8}) and $\det M_{quad}^{\prime}\neq0$, these become matrix fields.

\subsection{Pauli matrix presentation of the star and circle 16-vertex
constant matrices}

The main peculiarity of the $16$-vertex $8\times8$ matrices (\ref{mm8}%
)--(\ref{mq}) is the fact that they can be expressed as special tensor
(Kronecker) products of the Pauli matrices (see, also, \cite{kha/gla,bal/wu}).
Indeed, let%
\begin{equation}
\Sigma_{ijk}=\rho_{i}\otimes_{K}\rho_{j}\otimes_{K}\rho_{k}%
,\ \ \ \ \ i,j,k=1,2,3,4, \label{si}%
\end{equation}
where $\rho_{i}$ are Pauli matrices (we have already used the letter
\textquotedblleft$\sigma$\textquotedblright\ for the braid group generators
(\ref{bm}))%
\begin{equation}
\rho_{1}=\left(
\begin{array}
[c]{cc}%
0 & 1\\
1 & 0
\end{array}
\right)  ,\ \ \rho_{2}=\left(
\begin{array}
[c]{cc}%
0 & -i\\
i & 0
\end{array}
\right)  ,\ \ \rho_{3}=\left(
\begin{array}
[c]{cc}%
1 & 0\\
0 & -1
\end{array}
\right)  ,\ \ \rho_{4}=I_{2}=\left(
\begin{array}
[c]{cc}%
1 & 0\\
0 & 1
\end{array}
\right)  . \label{pr}%
\end{equation}

Among the total of $64$ $8\times8$ matrices $\Sigma_{ijk}$ (\ref{si}) there
are $24$ which generate $M_{star}^{\prime}$ (\ref{m8s}) and $M_{circ}^{\prime
}$ (\ref{m8c}):

$\bullet$ $8$ diagonal matrices: $\mathsf{\Sigma}_{\text{\textit{diag}}%
}=\left\{  \Sigma_{333},\Sigma_{334},\Sigma_{343},\Sigma_{344},\Sigma
_{433},\Sigma_{434},\Sigma_{443},\Sigma_{444}\right\}  ;$

$\bullet$ $8$ anti-diagonal matrices: $\mathsf{\Sigma}_{\text{\textit{adiag}}%
}=\left\{  \Sigma_{111},\Sigma_{112},\Sigma_{121},\Sigma_{122},\Sigma
_{211},\Sigma_{212},\Sigma_{221},\Sigma_{222}\right\}  ;$

$\bullet$ $8$ circle-like matrices ($M_{circ}^{\prime}$ with $0$'s on
diagonal): $\mathsf{\Sigma}_{\text{\textit{circ}}}=\left\{  \Sigma
_{131},\Sigma_{132},\Sigma_{141},\Sigma_{142},\Sigma_{231},\Sigma_{232}%
,\Sigma_{241},\Sigma_{242}\right\}  .$

Thus, in general we have the following set structure for the star and circle
16-vertex matrices (\ref{m8s}) and (\ref{m8c})
\begin{align}
&  \mathsf{M}_{star}^{\prime} =\mathsf{\Sigma}_{\text{\textit{diag}}}%
\cup\mathsf{\Sigma}_{\text{\textit{adiag}}},\\
&  \mathsf{M}_{circ}^{\prime} =\mathsf{\Sigma}_{\text{\textit{diag}}}%
\cup\mathsf{\Sigma}_{\text{\textit{circ}}},\\
&  \mathsf{M}_{star}^{\prime}\cap\mathsf{M}_{circ}^{\prime} =\mathsf{\Sigma
}_{\text{\textit{diag}}}.
\end{align}

In particular, for the $8$-vertex permutation solutions (\ref{cp8b}%
)--(\ref{cp8s}) of the ternary braid equations (\ref{pt}) we have%
\begin{align}
\tilde{c}_{rank=8}^{bisymm1,2}  &  =\frac{1}{2}\left(  \Sigma_{111}%
+\Sigma_{444}\pm\Sigma_{212}\pm\Sigma_{343}\right)  ,\\
\tilde{c}_{rank=8}^{symm1,2}  &  =\frac{1}{2}\left(  \Sigma_{141}+\Sigma
_{444}\pm\Sigma_{232}\pm\Sigma_{333}\right)  .
\end{align}

The non-invertible $16$-vertex solutions$M_{star}^{\prime}$ (\ref{m8s}) and
$M_{circ}^{\prime}$ (\ref{m8c}) having $1$'s on nonzero places are of $rank=4$
and can be presented by (\ref{si}) as follows%
\begin{equation}
M_{star}^{\prime}\left(  1\right)  =\left(
\begin{array}
[c]{cccccccc}%
1 & 0 & 0 & 0 & 0 & 0 & 0 & 1\\
0 & 1 & 0 & 0 & 0 & 0 & 1 & 0\\
0 & 0 & 1 & 0 & 0 & 1 & 0 & 0\\
0 & 0 & 0 & 1 & 1 & 0 & 0 & 0\\
0 & 0 & 0 & 1 & 1 & 0 & 0 & 0\\
0 & 0 & 1 & 0 & 0 & 1 & 0 & 0\\
0 & 1 & 0 & 0 & 0 & 0 & 1 & 0\\
1 & 0 & 0 & 0 & 0 & 0 & 0 & 1
\end{array}
\right)  =\Sigma_{111}+\Sigma_{444},
\end{equation}%
\begin{equation}
M_{circ}^{\prime}\left(  1\right)  =\left(
\begin{array}
[c]{cccccccc}%
1 & 0 & 0 & 0 & 0 & 1 & 0 & 0\\
0 & 1 & 0 & 0 & 1 & 0 & 0 & 0\\
0 & 0 & 1 & 0 & 0 & 0 & 0 & 1\\
0 & 0 & 0 & 1 & 0 & 0 & 1 & 0\\
0 & 1 & 0 & 0 & 1 & 0 & 0 & 0\\
1 & 0 & 0 & 0 & 0 & 1 & 0 & 0\\
0 & 0 & 0 & 1 & 0 & 0 & 1 & 0\\
0 & 0 & 1 & 0 & 0 & 0 & 0 & 1
\end{array}
\right)  =\Sigma_{141}+\Sigma_{444}.
\end{equation}

Similarly, one can obtain the Pauli matrix presentation for the general star
and circle 16-vertex matrices (\ref{m8s}) and (\ref{m8c}) which will contain
linear combinations of the $16$ parameters as coefficients before the $\Sigma$'s.

\subsection{Invertible and non-invertible $16$-vertex solutions to the ternary
braid equations}

First, consider the $16$-vertex solutions to (\ref{pt}) having the star matrix
shape (\ref{m8s}). We found the following $2$ one-parameter invertible
solutions%
\begin{equation}
\tilde{c}_{rank=8}^{16-vert,star}\left(  x\right)  =\left(
\begin{array}
[c]{cccccccc}%
x^{3} & 0 & 0 & 0 & 0 & 0 & 0 & -1\\
0 & x^{3} & 0 & 0 & 0 & 0 & \mp x^{2} & 0\\
0 & 0 & x^{3} & 0 & 0 & -x^{2} & 0 & 0\\
0 & 0 & 0 & x^{3} & \mp x^{4} & 0 & 0 & 0\\
0 & 0 & 0 & \pm x^{2} & x^{3} & 0 & 0 & 0\\
0 & 0 & x^{4} & 0 & 0 & x^{3} & 0 & 0\\
0 & \pm x^{4} & 0 & 0 & 0 & 0 & x^{3} & 0\\
x^{6} & 0 & 0 & 0 & 0 & 0 & 0 & x^{3}%
\end{array}
\right)  ,%
\begin{array}
[c]{c}%
\operatorname*{tr}\tilde{c}=8x^{3},\\
\det\tilde{c}=16x^{24},\ \ x\neq0,\\
\text{eigenvalues:\ }\left\{  (1+i)x^{3}\right\}  ^{\left[  4\right]
},\left\{  (1-i)x^{3}\right\}  ^{\left[  4\right]  }.
\end{array}
\label{cv16}%
\end{equation}
Both matrices in (\ref{cv16}) are diagonalizable and are conjugates via%
\begin{equation}
U_{star}=\left(
\begin{array}
[c]{cccccccc}%
1 & 0 & 0 & 0 & 0 & 0 & 0 & 0\\
0 & -1 & 0 & 0 & 0 & 0 & 0 & 0\\
0 & 0 & 1 & 0 & 0 & 0 & 0 & 0\\
0 & 0 & 0 & -1 & 0 & 0 & 0 & 0\\
0 & 0 & 0 & 0 & 1 & 0 & 0 & 0\\
0 & 0 & 0 & 0 & 0 & 1 & 0 & 0\\
0 & 0 & 0 & 0 & 0 & 0 & 1 & 0\\
0 & 0 & 0 & 0 & 0 & 0 & 0 & 1
\end{array}
\right)  ,
\end{equation}
which cannot be presented in the form of a triple Kronecker product
(\ref{q8}). Therefore, the two solutions in (\ref{cv16}) are not $q$-conjugate
and become different $16$-vertex one-parameter invertible solutions of the
braid equations (\ref{pt}).

In search of $16$-vertex solutions to the total braid equations (\ref{pt}) of
the circle matrix shape (\ref{m8c}) we found that only non-invertible ones
exist. They are the following two $2$-parameter solutions of rank $4$%
\begin{equation}
\tilde{c}_{rank=4}^{16-vert,circ}\left(  x,y\right)  =\left(
\begin{array}
[c]{cccccccc}%
\pm xy & 0 & 0 & 0 & 0 & y^{2} & 0 & 0\\
0 & \pm xy & 0 & 0 & xy & 0 & 0 & 0\\
0 & 0 & \pm xy & 0 & 0 & 0 & 0 & y^{2}\\
0 & 0 & 0 & \pm xy & 0 & 0 & xy & 0\\
0 & xy & 0 & 0 & \pm xy & 0 & 0 & 0\\
x^{2} & 0 & 0 & 0 & 0 & \pm xy & 0 & 0\\
0 & 0 & 0 & xy & 0 & 0 & \pm xy & 0\\
0 & 0 & x^{2} & 0 & 0 & 0 & 0 & \pm xy
\end{array}
\right)  ,%
\begin{array}
[c]{c}%
\operatorname*{tr}\tilde{c}=\pm8xy,\\
\text{eigenvalues: }\left\{  2xy\right\}  ^{\left[  4\right]  },\left\{
0\right\}  ^{\left[  4\right]  }.
\end{array}
\label{cv16c}%
\end{equation}
Two matrices in (\ref{cv16c}) are not even conjugates in the standard way, and
so they are different $16$-vertex two-parameter non-invertible solutions to
the braid equations (\ref{pt}).

For the only partial $13$-braid equation (\ref{p2}), there are $4$ polynomial
$16$-vertex two-parameter invertible solutions%
\begin{align}
&  \tilde{c}_{rank=8}^{16-vert,13circ}\left(  x,y\right)  =\left(
\begin{array}
[c]{cccccccc}%
x & 0 & 0 & 0 & 0 & y^{2} & 0 & 0\\
0 & xy & 0 & 0 & x & 0 & 0 & 0\\
0 & 0 & x & 0 & 0 & 0 & 0 & \pm y^{2}\\
0 & 0 & 0 & xy & 0 & 0 & \pm x & 0\\
0 & x & 0 & 0 & xy & 0 & 0 & 0\\
x^{2} & 0 & 0 & 0 & 0 & x & 0 & 0\\
0 & 0 & 0 & \pm x & 0 & 0 & xy & 0\\
0 & 0 & \pm x^{2} & 0 & 0 & 0 & 0 & x
\end{array}
\right)  ,\ \ \left(
\begin{array}
[c]{cccccccc}%
x & 0 & 0 & 0 & 0 & y^{2} & 0 & 0\\
0 & xy & 0 & 0 & -x & 0 & 0 & 0\\
0 & 0 & x & 0 & 0 & 0 & 0 & \pm y^{2}\\
0 & 0 & 0 & xy & 0 & 0 & \mp x & 0\\
0 & -x & 0 & 0 & xy & 0 & 0 & 0\\
x^{2} & 0 & 0 & 0 & 0 & x & 0 & 0\\
0 & 0 & 0 & \mp x & 0 & 0 & xy & 0\\
0 & 0 & \pm x^{2} & 0 & 0 & 0 & 0 & x
\end{array}
\right)  ,\label{c16c}\\
&  \operatorname*{tr}\tilde{c}=4x\left(  y+1\right)  ,\det\tilde{c}%
=x^{8}\left(  y^{2}-1\right)  ^{4},\ x\neq0,y\neq1,\ \text{eigenvalues:}%
\left\{  x\left(  y+1\right)  \right\}  ^{\left[  4\right]  },\left\{
x\left(  y-1\right)  \right\}  ^{\left[  2\right]  },\left\{  -x\left(
y-1\right)  \right\}  ^{\left[  2\right]  }. \label{d16}%
\end{align}

Also, for the partial $13$-braid equation (\ref{p2}), we found $4$ exotic
irrational (an analog of (\ref{c34y}) for the Yang-Baxter equation
(\ref{a123})) $16$-vertex, two-parameter invertible solutions of rank $8$ of
the form%
\begin{align}
&  \tilde{c}_{rank=8}^{16-vert,13circ,1}\left(  x,y\right) \nonumber\\
&  =\left(
\begin{array}
[c]{cccccccc}%
x(2y-1) & 0 & 0 & 0 & 0 & y^{2} & 0 & 0\\
0 & xy & 0 & 0 & x\sqrt{2(y-1)y+1} & 0 & 0 & 0\\
0 & 0 & x(2y-1) & 0 & 0 & 0 & 0 & \pm y^{2}\\
0 & 0 & 0 & xy & 0 & 0 & \pm x\sqrt{2(y-1)y+1} & 0\\
0 & x\sqrt{2(y-1)y+1} & 0 & 0 & xy & 0 & 0 & 0\\
x^{2} & 0 & 0 & 0 & 0 & x & 0 & 0\\
0 & 0 & 0 & \pm x\sqrt{2(y-1)y+1} & 0 & 0 & xy & 0\\
0 & 0 & \pm x^{2} & 0 & 0 & 0 & 0 & x
\end{array}
\right)  , \label{c16c1}%
\end{align}%
\begin{align}
&  \tilde{c}_{rank=8}^{16-vert,13circ,2}\left(  x,y\right) \nonumber\\
&  =\left(
\begin{array}
[c]{cccccccc}%
x(2y-1) & 0 & 0 & 0 & 0 & y^{2} & 0 & 0\\
0 & xy & 0 & 0 & -x\sqrt{2(y-1)y+1} & 0 & 0 & 0\\
0 & 0 & x(2y-1) & 0 & 0 & 0 & 0 & \pm y^{2}\\
0 & 0 & 0 & xy & 0 & 0 & \mp x\sqrt{2(y-1)y+1} & 0\\
0 & -x\sqrt{2(y-1)y+1} & 0 & 0 & xy & 0 & 0 & 0\\
x^{2} & 0 & 0 & 0 & 0 & x & 0 & 0\\
0 & 0 & 0 & \mp x\sqrt{2(y-1)y+1} & 0 & 0 & xy & 0\\
0 & 0 & \pm x^{2} & 0 & 0 & 0 & 0 & x
\end{array}
\right)  , \label{c16c2}%
\end{align}%
\begin{align}
\operatorname*{tr}\tilde{c}  &  =8xy,\ \ \text{\ }\det\tilde{c}=\allowbreak
x^{8}\left(  y-1\right)  ^{8},\ \ \ \ \ \ \ \ x\neq0,\ \ \ y\neq
1,\label{d16c}\\
\text{eigenvalues}  &  \text{: }\left\{  x\left(  y+\sqrt{2(y-1)y+1}\right)
\right\}  ^{\left[  4\right]  },\left\{  x\left(  y-\sqrt{2(y-1)y+1}\right)
\right\}  ^{\left[  4\right]  }. \label{e16c}%
\end{align}
The matrices in (\ref{c16c})--(\ref{c16c2}) are diagonalizable, have the same
eigenvalues (\ref{e16c}) and are pairwise conjugate by%
\begin{equation}
U_{circ}=\left(
\begin{array}
[c]{cccccccc}%
1 & 0 & 0 & 0 & 0 & 0 & 0 & 0\\
0 & 1 & 0 & 0 & 0 & 0 & 0 & 0\\
0 & 0 & -1 & 0 & 0 & 0 & 0 & 0\\
0 & 0 & 0 & -1 & 0 & 0 & 0 & 0\\
0 & 0 & 0 & 0 & 1 & 0 & 0 & 0\\
0 & 0 & 0 & 0 & 0 & 1 & 0 & 0\\
0 & 0 & 0 & 0 & 0 & 0 & 1 & 0\\
0 & 0 & 0 & 0 & 0 & 0 & 0 & 1
\end{array}
\right)  .
\end{equation}
Because $U_{circ}$ cannot be presented in the form (\ref{q8}), all solutions
in (\ref{c16c})--(\ref{c16c2}) are not mutually $q$-conjugate and become $8$
different $16$-vertex two-parameter invertible solutions to the partial
$13$-braid equation (\ref{p2}). If $y=1$, then the matrices (\ref{c16c}%
)--(\ref{c16c2}) become of rank $4$ with vanishing determinants (\ref{d16}),
(\ref{d16c}), and therefore in this case they are a $16$-vertex one-parameter
circle of non-invertible solutions to the total braid equations (\ref{pt}).

Further families of solutions could be constructed using additional
parameters: the scaling parameter $t$ in (\ref{cqt}) and the complex elements
of the matrix $q$ (\ref{qg}).

\subsection{Higher $2^{n}$-vertex constant solutions to $n$-ary braid
equations}

Next we considered the $4$-ary constant braid equations (\ref{aa1}%
)--(\ref{aa2}) and found the following $32$-vertex star solution%
\begin{equation}
\tilde{c}_{16}=\left(
\begin{array}
[c]{cccccccccccccccc}%
1 & 0 & 0 & 0 & 0 & 0 & 0 & 0 & 0 & 0 & 0 & 0 & 0 & 0 & 0 & -1\\
0 & 1 & 0 & 0 & 0 & 0 & 0 & 0 & 0 & 0 & 0 & 0 & 0 & 0 & -1 & 0\\
0 & 0 & 1 & 0 & 0 & 0 & 0 & 0 & 0 & 0 & 0 & 0 & 0 & -1 & 0 & 0\\
0 & 0 & 0 & 1 & 0 & 0 & 0 & 0 & 0 & 0 & 0 & 0 & -1 & 0 & 0 & 0\\
0 & 0 & 0 & 0 & 1 & 0 & 0 & 0 & 0 & 0 & 0 & -1 & 0 & 0 & 0 & 0\\
0 & 0 & 0 & 0 & 0 & 1 & 0 & 0 & 0 & 0 & -1 & 0 & 0 & 0 & 0 & 0\\
0 & 0 & 0 & 0 & 0 & 0 & 1 & 0 & 0 & -1 & 0 & 0 & 0 & 0 & 0 & 0\\
0 & 0 & 0 & 0 & 0 & 0 & 0 & 1 & -1 & 0 & 0 & 0 & 0 & 0 & 0 & 0\\
0 & 0 & 0 & 0 & 0 & 0 & 0 & 1 & 1 & 0 & 0 & 0 & 0 & 0 & 0 & 0\\
0 & 0 & 0 & 0 & 0 & 0 & 1 & 0 & 0 & 1 & 0 & 0 & 0 & 0 & 0 & 0\\
0 & 0 & 0 & 0 & 0 & 1 & 0 & 0 & 0 & 0 & 1 & 0 & 0 & 0 & 0 & 0\\
0 & 0 & 0 & 0 & 1 & 0 & 0 & 0 & 0 & 0 & 0 & 1 & 0 & 0 & 0 & 0\\
0 & 0 & 0 & 1 & 0 & 0 & 0 & 0 & 0 & 0 & 0 & 0 & 1 & 0 & 0 & 0\\
0 & 0 & 1 & 0 & 0 & 0 & 0 & 0 & 0 & 0 & 0 & 0 & 0 & 1 & 0 & 0\\
0 & 1 & 0 & 0 & 0 & 0 & 0 & 0 & 0 & 0 & 0 & 0 & 0 & 0 & 1 & 0\\
1 & 0 & 0 & 0 & 0 & 0 & 0 & 0 & 0 & 0 & 0 & 0 & 0 & 0 & 0 & 1
\end{array}
\right)  . \label{c16}%
\end{equation}

We may compare (\ref{c16}) with particular cases of the star solutions to the
Yang-Baxter equation (\ref{c24}) and the ternary braid equation (\ref{cv16})%
\begin{equation}
\tilde{c}_{4}=\left(
\begin{array}
[c]{cccc}%
1 & 0 & 0 & -1\\
0 & 1 & -1 & 0\\
0 & 1 & 1 & 0\\
1 & 0 & 0 & 1
\end{array}
\right)  ,\ \ \tilde{c}_{8}=\left(
\begin{array}
[c]{cccccccc}%
1 & 0 & 0 & 0 & 0 & 0 & 0 & -1\\
0 & 1 & 0 & 0 & 0 & 0 & -1 & 0\\
0 & 0 & 1 & 0 & 0 & -1 & 0 & 0\\
0 & 0 & 0 & 1 & -1 & 0 & 0 & 0\\
0 & 0 & 0 & 1 & 1 & 0 & 0 & 0\\
0 & 0 & 1 & 0 & 0 & 1 & 0 & 0\\
0 & 1 & 0 & 0 & 0 & 0 & 1 & 0\\
1 & 0 & 0 & 0 & 0 & 0 & 0 & 1
\end{array}
\right)  .
\end{equation}

Informally we call such solutions the \textquotedblleft
Minkowski\textquotedblright\ star solutions, since their legs have the
\textquotedblleft Minkowski signature\textquotedblright. Thus, we assume that
in the general case for the $n$-ary braid equation there exist $2^{n+1}%
$-vertex $2^{n}\times2^{n}$ matrix \textquotedblleft
Minkowski\textquotedblright\ star invertible solutions of the above form%
\begin{equation}
\ \tilde{c}_{2^{n}}=\left(
\begin{array}
[c]{cccccc}%
1 & 0 & 0 & 0 & 0 & -1\\
0 & \ddots & 0 & 0 & \iddots & 0\\
0 & 0 & 1 & -1 & 0 & 0\\
0 & 0 & 1 & 1 & 0 & 0\\
0 & \iddots & 0 & 0 & \ddots & 0\\
1 & 0 & 0 & 0 & 0 & 1
\end{array}
\right)  . \label{cn}%
\end{equation}

This allows us to use the general solution (\ref{cn}) as $n$-ary braiding
quantum gates with an arbitrary number of qubits.

\section{Invertible and noninvertible quantum gates}

Informally, quantum computing consists of preparation (setting up an initial
quantum state), evolution (by a quantum circuit) and measurement (projection
onto the final state). Mathematically (in the computational basis) the initial
state is a vector in a Hilbert space (multi-qubit state), the evolution is
governed by successive (quantum circuit) invertible linear transformations
(unitary matrices called quantum gates) and the measurement is made by
non-invertible projection matrices to leave only one final quantum
(multi-qubit) state. So, quantum computing is non-invertible overall, and we
may consider non-invertible transformations at each step. It was then realized
that one can \textquotedblleft invite\textquotedblright\ the Yang-Baxter
operators (solutions of the constant Yang-Baxter equation) into quantum gates
, providing a means of entangling otherwise non-entangled states. This insight
uncovered a deep connection between quantum and topological computation (see
for details, e.g. \cite{kau/lom2002,kau/lom2004}).

Here we propose extending the above picture in two directions. First, we can
treat higher braided operators as higher braiding gates. Second, we will
analyze the possible role of non-invertible linear transformations (described
by the partial unitary matrices introduced in (\ref{mm})--(\ref{mm1})), which
can be interpreted as a property of some hypothetical quantum circuit (for
instance, with specific \textquotedblleft loss\textquotedblright\ of
information, some kind of \textquotedblleft dissipativity\textquotedblright%
\ or \textquotedblleft vagueness\textquotedblright). This can be considered as
an intermediate case between standard unitary computing and the measurement
only computing of \cite{bon/fre/nay}.

To establish notation recall \cite{nie/chu}, that in the computational basis
(vector representation) and Dirac notation, a (pure) one-qubit state is
described by a vector in two-dimensional Hilbert space $V=\mathbb{C}^{2}$%
\begin{equation}
\left\vert \psi\right\rangle \equiv\left\vert \psi^{\left(  1\right)
}\right\rangle =a_{0}\left\vert 0\right\rangle +a_{1}\left\vert 1\right\rangle
,\ \ \left\vert 0\right\rangle =\left(
\begin{array}
[c]{c}%
1\\
0
\end{array}
\right)  ,\ \ \left\vert 1\right\rangle =\left(
\begin{array}
[c]{c}%
0\\
1
\end{array}
\right)  ,\ \ \left\vert a_{0}\right\vert ^{2}+\left\vert a_{1}\right\vert
^{2}=1,\ \ \ \ a_{i}\in\mathbb{C},\ \ ,i=1,2, \label{p}%
\end{equation}
where $a_{i}$ is a probability amplitude of $\left\vert i\right\rangle $.
Sometimes, for a one-qubit state it is convenient to use the Bloch unit sphere
representation (normalized up to a general unimportant and unmeasurable phase)%
\begin{equation}
\left\vert \psi\left(  \theta,\gamma\right)  \right\rangle =\cos\frac{\theta
}{2}\left\vert 0\right\rangle +e^{i\gamma}\sin\frac{\theta}{2}\left\vert
1\right\rangle ,\ \ \ 0\leq\theta\leq\pi,\ \ \ 0\leq\gamma\leq2\pi. \label{pb}%
\end{equation}
A (pure) state of $L$-qubits $\left\vert \psi^{\left(  L\right)
}\right\rangle $ is described by $2^{L}$ amplitudes, and so is a vector in
$2^{L}$-dimensional Hilbert space. If $\left\vert \psi^{\left(  L\right)
}\right\rangle $ cannot be presented as a tensor product of $L$ one-qubit
states (\ref{p}), it is called \textit{entangled}. For instance, a two-qubit
pure state%
\begin{equation}
\left\vert \psi^{\left(  2\right)  }\right\rangle =a_{00}\left\vert
00\right\rangle +a_{01}\left\vert 01\right\rangle +a_{10}\left\vert
10\right\rangle +a_{11}\left\vert 11\right\rangle ,\ \ \left\vert
a_{00}\right\vert ^{2}+\left\vert a_{01}\right\vert ^{2}+\left\vert
a_{10}\right\vert ^{2}+\left\vert a_{11}\right\vert ^{2}=1,\ \ \ \ a_{ij}%
\in\mathbb{C},\ \ ,i,j=1,2, \label{ps2}%
\end{equation}
is entangled, if $\det\left(  a_{ij}\right)  \neq0$, and the
\textit{concurrence}%
\begin{equation}
C^{\left(  2\right)  }\equiv C^{\left(  2\right)  }\left(  \left\vert
\psi^{\left(  2\right)  }\right\rangle \right)  =2\left\vert \det\left(
a_{ij}\right)  \right\vert \label{c2}%
\end{equation}
is the measure of entanglement $0\leq C^{\left(  2\right)  }\leq1$. It follows
from (\ref{p}), that the tensor product of states has vanishing concurrence
$C^{\left(  2\right)  }\left(  \left\vert \psi_{1}\right\rangle \otimes
\left\vert \psi_{2}\right\rangle \right)  =0$. An example of the maximally
entangled ($C^{\left(  2\right)  }=1$) two-qubit states is the (first) Bell
state $\left\vert \psi^{\left(  2\right)  }\right\rangle _{Bell}=(\left\vert
00\right\rangle +\left\vert 11\right\rangle )/\sqrt{2}$.

The concurrence of the three-qubit state%
\begin{equation}
\left\vert \psi^{\left(  3\right)  }\right\rangle =\sum_{i,j,k=0}^{1}%
a_{ijk}\left\vert ijk\right\rangle ,\ \ \ \ \sum_{i,j,k=0}^{1}\left\vert
a_{ijk}\right\vert ^{2}=1,\ \ a_{ijk}\in\mathbb{C}, \label{pp}%
\end{equation}
is determined by the Cayley's $2\times2\times2$ hyperdeterminant%
\begin{equation}
C^{\left(  3\right)  }=4\left\vert \mathrm{h}\det\left(  a_{ijk}\right)
\right\vert ,\ \ \ \ \ 0\leq C^{\left(  3\right)  }\leq1, \label{c3}%
\end{equation}%
\begin{align}
&  \mathrm{h}\det\left(  a_{ijk}\right)  =a_{000}^{2}a_{111}^{2}+a_{001}%
^{2}a_{110}^{2}+a_{010}^{2}a_{101}^{2}+a_{100}^{2}a_{011}^{2}-2a_{000}%
a_{001}a_{110}a_{111}\nonumber\\
&  -2a_{000}a_{010}a_{101}a_{111}-2a_{000}a_{011}a_{100}a_{111}-2a_{001}%
a_{010}a_{101}a_{111}-2a_{001}a_{011}a_{100}a_{110}\nonumber\\
&  -2a_{010}a_{011}a_{100}a_{101}+4a_{000}a_{011}a_{101}a_{110}+4a_{001}%
a_{010}a_{100}a_{111}. \label{hd3}%
\end{align}

Thus, if the three-qubit state (\ref{pp}) is not entangled, then $C^{\left(
3\right)  }=0$ (for the tensor product of one-qubit states). One of the
maximally entangled ($C^{\left(  3\right)  }=1$) three-qubit states is the GHZ
state $\left\vert \psi^{\left(  3\right)  }\right\rangle _{GHZ}=(\left\vert
000\right\rangle +\left\vert 111\right\rangle )/\sqrt{2}$.

A quantum $L$-qubit gate is a linear transformation of $2^{L}$-dimensional
Hilbert space $\left(  \mathbb{C}^{2}\right)  ^{\otimes L}\rightarrow\left(
\mathbb{C}^{2}\right)  ^{\otimes L}$ which in the computational basis
(\ref{p}) is described of the $2^{L}\times2^{L}$ matrix $U^{\left(  L\right)
}$ such that the $L$-qubit state transforms as $\left\vert \psi^{\prime\left(
L\right)  }\right\rangle =U^{\left(  L\right)  }\left\vert \psi^{\left(
L\right)  }\right\rangle $. In this way, a \textit{quantum circuit} is
described as the successive application of elementary gates to an initial
quantum state, that is the product of the corresponding matrices (for details,
see, e.g., \cite{nie/chu}). It is a standard assumption that each elementary
$L$-qubit transformation is \textit{unitary}, which implies the following
strong restriction on the corresponding matrix $U\equiv U^{\left(  L\right)
}$ as%
\begin{equation}
U^{\mathbf{\star}}U=UU^{\mathbf{\star}}=I\equiv I_{2^{L}\times2^{L}},
\label{uu}%
\end{equation}
where $I$ is the $2^{L}\times2^{L}$ identity matrix for $L$-qubit state and
the operation $\left(  \star\right)  $ is the conjugate-transposition. The
first equality in (\ref{uu}) means that the matrix $U^{\left(  L\right)  }$ is
\textit{normal} (cf. (\ref{mm})--(\ref{mm1})). The equations (\ref{uu}) follow
from the definition of the \textit{adjoint operator}%
\begin{equation}
\left\langle U\psi^{\left(  L\right)  }\mid I\varphi^{\left(  L\right)
}\right\rangle =\left\langle I\psi^{\left(  L\right)  }\mid U^{\mathbf{\star}%
}\varphi^{\left(  L\right)  }\right\rangle \label{ui}%
\end{equation}
applied to this simplest example of $L$-qubits in the $2^{L}$-dimensional
Hilbert space $\left(  \mathbb{C}^{2}\right)  ^{\otimes L}$ (for the general
case the derivation almost literally coincides), which we write in the
following special form (in Dirac notation with bra- and ket- vectors) with
explicitly added identities. Then (\ref{uu}) follows from (\ref{ui}) as%
\begin{equation}
\left\langle U^{\mathbf{\star}}U\psi^{\left(  L\right)  }\mid I\varphi
^{\left(  L\right)  }\right\rangle =\left\langle I\psi^{\left(  L\right)
}\mid UU^{\mathbf{\star}}\varphi^{\left(  L\right)  }\right\rangle
=\left\langle I\psi^{\left(  L\right)  }\mid I\varphi^{\left(  L\right)
}\right\rangle , \label{uii}%
\end{equation}
and any unitary matrix preserves the inner product%
\begin{equation}
\left\langle U\psi^{\left(  L\right)  }\mid U\varphi^{\left(  L\right)
}\right\rangle =\left\langle I\psi^{\left(  L\right)  }\mid I\varphi^{\left(
L\right)  }\right\rangle , \label{uui}%
\end{equation}
which means that unitary operators satisfying (\ref{uu}) are bounded operators
(bounded matrices in our case) and invertible with the inverse $U^{-1}%
=U^{\mathbf{\star}}$.

Let us consider a possibility of non-invertible intermediate transformations
of $L$-qubit states, i.e. \textit{non-invertible gates} which are described by
the $2^{L}\times2^{L}$ matrices $U\left(  r\right)  $ of (possibly) less than
full rank $1\leq r\leq2^{L}$. This can be related to the production of
\textquotedblleft degenerate\textquotedblright\ states (see, e.g.
\cite{jaf/oed}), \textquotedblleft particle loss\textquotedblright%
\ \cite{nev/mar/bas,fra/bra,zan/qia}, and the role of ranks in multiparticle
entanglement \cite{cho/kei/sto,bru/fri/zyc}.

In the limited cases $U\left(  r=2^{L}\right)  \equiv U=U^{\left(  L\right)
}$, and $U\left(  1\right)  $ corresponds to the measurement matrix being the
projection to one final vector $\left\vert i_{final}\right\rangle $. In this
case, for non-invertible transformations with $r<2^{L}$ instead of unitarity
(\ref{uu}) we consider partial unitarity (\ref{mm})--(\ref{mm1}) as%
\begin{align}
U\left(  r\right)  ^{\mathbf{\star}}U\left(  r\right)   &  =I_{1}\left(
r\right)  ,\label{uu1}\\
U\left(  r\right)  U\left(  r\right)  ^{\mathbf{\star}}  &  =I_{2}\left(
r\right)  , \label{uu2}%
\end{align}
where $I_{1}\left(  r\right)  $ and $I_{2}\left(  r\right)  $ are (or may be)
different partial shuffle identities having $r$ units on the diagonal. There
is an exotic limiting case, which is impossible for the identity $I$: we call
two partial identities \textit{orthogonal}, if%
\begin{equation}
I_{1}\left(  r\right)  I_{2}\left(  r\right)  =Z, \label{ii}%
\end{equation}
where $Z=Z_{2^{L}\times2^{L}}$ is the zero $2^{L}\times2^{L}$ matrix.

We propose corresponding non-invertible analogs of (\ref{ui})--(\ref{uui}) as
follows. The \textit{partial adjoint operator} $U\left(  r\right)
^{\mathbf{\star}}$ in the $2^{L}$-dimensional Hilbert space $\left(
\mathbb{C}^{2}\right)  ^{\otimes L}$ is defined by%
\begin{equation}
\left\langle U\left(  r\right)  \psi^{\left(  L\right)  }\mid I_{2}\left(
r\right)  \varphi^{\left(  L\right)  }\right\rangle =\left\langle I_{1}\left(
r\right)  \psi^{\left(  L\right)  }\mid U\left(  r\right)  ^{\mathbf{\star}%
}\varphi^{\left(  L\right)  }\right\rangle , \label{up}%
\end{equation}
such that (see (\ref{uu1})--(\ref{uu2}))%
\begin{equation}
\left\langle U\left(  r\right)  ^{\mathbf{\star}}U\left(  r\right)
\psi^{\left(  L\right)  }\mid I_{2}\left(  r\right)  \varphi^{\left(
L\right)  }\right\rangle =\left\langle I_{1}\left(  r\right)  \psi^{\left(
L\right)  }\mid U\left(  r\right)  U\left(  r\right)  ^{\mathbf{\star}}%
\varphi^{\left(  L\right)  }\right\rangle =\left\langle I_{1}\left(  r\right)
\psi^{\left(  L\right)  }\mid I_{2}\left(  r\right)  \varphi^{\left(
L\right)  }\right\rangle . \label{i12}%
\end{equation}

We call the r.h.s. of (\ref{i12}) the \textit{partial inner product}. So
instead of (\ref{uui}) we define $U\left(  r\right)  $ as the
\textit{partially bounded operator}%
\begin{equation}
\left\langle U\left(  r\right)  \psi^{\left(  L\right)  }\mid U\left(
r\right)  \varphi^{\left(  L\right)  }\right\rangle =\left\langle I_{1}\left(
r\right)  \psi^{\left(  L\right)  }\mid I_{2}\left(  r\right)  \varphi
^{\left(  L\right)  }\right\rangle . \label{uuii}%
\end{equation}

Thus, if the partial identities are orthogonal (\ref{ii}), then the partial
inner product vanishes identically, and the operator $U\left(  r\right)  $
becomes a zero norm operator in the sense of (\ref{uuii}), although
(\ref{uu1})--(\ref{uu2}) are not zero.

In case the rank $r$ is fixed, there can be $\left(  2^{L}!/r!\left(
2^{L}-r\right)  !\right)  ^{2}$ partial unitary matrices $U\left(  r\right)  $
satisfying (\ref{uu1})--(\ref{uu2}).

We define a \textit{general unitary semigroup} as a semigroup of matrices
$U\left(  r\right)  $ of rank $r$ satisfying partial regularity (\ref{uu1}%
)--(\ref{uu2}) (in the \textquotedblleft symmetric\textquotedblright\ case
$I_{1}\left(  r\right)  =I_{2}\left(  r\right)  \equiv I\left(  r\right)  $).

As an example, we consider two $2$-qubit states (\ref{ps2}) $\left\vert
\psi^{\left(  2\right)  }\right\rangle $ and $\left\vert \varphi^{\left(
2\right)  }\right\rangle $ (with $a_{ij}^{\prime}$ and $\left\vert i^{\prime
}j^{\prime}\right\rangle $) and the non-invertible transformation described by
three-parameter $4\times4$ matrices of rank $3$ (but which are not nilpotent)%
\begin{equation}
U\left(  3\right)  =U^{\left(  L=2\right)  }\left(  r=3\right)  =\left(
\begin{array}
[c]{cccc}%
0 & 0 & 0 & 0\\
0 & e^{i\beta} & 0 & 0\\
0 & 0 & 0 & e^{i\gamma}\\
e^{i\alpha} & 0 & 0 & 0
\end{array}
\right)  ,\ \ \ \alpha,\beta,\gamma\in\mathbb{R}. \label{u3}%
\end{equation}

The partial unitarity (\ref{uu1})--(\ref{uu2}) and partial identities now
become%
\begin{align}
U\left(  3\right)  ^{\mathbf{\star}}U\left(  3\right)   &  =I_{1}\left(
3\right)  =\left(
\begin{array}
[c]{cccc}%
1 & 0 & 0 & 0\\
0 & 1 & 0 & 0\\
0 & 0 & 0 & 0\\
0 & 0 & 0 & 1
\end{array}
\right)  ,\label{i1}\\
U\left(  3\right)  U\left(  3\right)  ^{\mathbf{\star}}  &  =I_{2}\left(
3\right)  =\left(
\begin{array}
[c]{cccc}%
0 & 0 & 0 & 0\\
0 & 1 & 0 & 0\\
0 & 0 & 1 & 0\\
0 & 0 & 0 & 1
\end{array}
\right)  . \label{i2}%
\end{align}
The partial identities (\ref{i1})--(\ref{i2}) are not orthogonal (\ref{ii}),
because%
\begin{equation}
I_{1}\left(  3\right)  I_{2}\left(  3\right)  =\left(
\begin{array}
[c]{cccc}%
0 & 0 & 0 & 0\\
0 & 1 & 0 & 0\\
0 & 0 & 0 & 0\\
0 & 0 & 0 & 1
\end{array}
\right)  \neq Z,
\end{equation}
which directly gives the signature of the partial inner product (\ref{i12}),
in our case of the Hilbert space $\left(  \mathbb{C}^{2}\right)  ^{\otimes2}$.

The definition of a partial adjoint operator (\ref{up}) is satisfied with both
sides being equal to $a_{00}a_{11}^{\prime}e^{i\alpha}\left\langle
00\mid1^{\prime}1^{\prime}\right\rangle +a_{01}a_{01}^{\prime}e^{i\beta
}\left\langle 01\mid0^{\prime}1^{\prime}\right\rangle +a_{11}a_{10}^{\prime
}e^{i\gamma}\left\langle 11\mid1^{\prime}0^{\prime}\right\rangle $. The
partial boundedness condition (\ref{uuii}) holds with the partial inner
product (\ref{i12}) becoming $a_{01}a_{01}^{\prime}\left\langle 01\mid
0^{\prime}1^{\prime}\right\rangle +a_{11}a_{11}^{\prime}\left\langle
11\mid1^{\prime}1^{\prime}\right\rangle $, thus $U\left(  3\right)  $
(\ref{u3}) which is a bounded partial unitary operator.

An example of a zero norm (in our sense (\ref{uuii})) operator is the
two-parameter partial unitary rank $2$ matrix%
\begin{equation}
U_{nil}\left(  2\right)  =U^{\left(  L=2\right)  }\left(  r=2\right)  =\left(
\begin{array}
[c]{cccc}%
0 & 0 & 0 & 0\\
0 & 0 & 0 & e^{i\beta}\\
e^{i\alpha} & 0 & 0 & 0\\
0 & 0 & 0 & 0
\end{array}
\right)  ,\ \ \ \alpha,\beta\in\mathbb{R}. \label{un}%
\end{equation}

The partial unitarity relations for $U_{nil}\left(  2\right)  $ have the form%
\begin{align}
U_{nil}\left(  2\right)  ^{\mathbf{\star}}U_{nil}\left(  2\right)   &
=I_{nil,1}\left(  2\right)  =\left(
\begin{array}
[c]{cccc}%
1 & 0 & 0 & 0\\
0 & 0 & 0 & 0\\
0 & 0 & 0 & 0\\
0 & 0 & 0 & 1
\end{array}
\right)  ,\\
U_{nil}\left(  2\right)  U_{nil}\left(  2\right)  ^{\mathbf{\star}}  &
=I_{nil,2}\left(  2\right)  =\left(
\begin{array}
[c]{cccc}%
0 & 0 & 0 & 0\\
0 & 1 & 0 & 0\\
0 & 0 & 1 & 0\\
0 & 0 & 0 & 0
\end{array}
\right)  .
\end{align}

It may be seen that the partial identities $I_{nil,1}\left(  2\right)  $ and
$I_{nil,2}\left(  2\right)  $ are now orthogonal (\ref{ii}), and the partial
inner product (\ref{i12}) vanishes identically, and also the boundedness
condition (\ref{uuii}) holds with the r.h.s. vanishing, despite $U_{nil}%
\left(  2\right)  $ being a nonzero nilpotent matrix (\ref{un}).

\section{Binary braiding quantum gates}

Let us consider those Yang-Baxter maps which could be linear transformations
of two-qubit spaces. We will pay attention to the most general $8$-vertex
solutions to the Yang-Baxter equations (\ref{c24})--(\ref{c22}) and
(\ref{c4c})--(\ref{c2c}) which are unitary (and invertible) or partial unitary
(\ref{mm})--(\ref{mm1}) (and non-invertible).

Consider the unitary version of the invertible star $8$-vertex solutions
(\ref{c24})--(\ref{c34y}) to the matrix Yang-Baxter equation (\ref{a123}). We
use the exponential form of the parameters%
\begin{equation}
x=r_{x}e^{i\alpha},\ \ y=r_{y}e^{i\beta},\ \ z=r_{z}e^{i\gamma}%
,\ \ \ \ r_{x,y,z},\alpha,\beta,\gamma\in\mathbb{R},\ \ \ r_{x,y,z}%
\geq0,\ \ \ \left\vert \alpha\right\vert ,\left\vert \beta\right\vert
,\left\vert \gamma\right\vert \leq2\pi. \label{x}%
\end{equation}

For (\ref{c24}), exploiting unitarity (\ref{uu}) we obtain%
\begin{align}
U_{rank=4}^{8-vert,star}\left(  \alpha,\beta\right)   &  =\frac{1}{\sqrt{2}%
}\left(
\begin{array}
[c]{cccc}%
e^{i\left(  \alpha+\beta\right)  } & 0 & 0 & e^{2i\beta}\\
0 & e^{i\left(  \alpha+\beta\right)  } & \pm e^{i\left(  \alpha+\beta\right)
} & 0\\
0 & \mp e^{i\left(  \alpha+\beta\right)  } & e^{i\left(  \alpha+\beta\right)
} & 0\\
-e^{2i\alpha} & 0 & 0 & e^{i\left(  \alpha+\beta\right)  }%
\end{array}
\right)  ,\ \ \ \
\begin{array}
[c]{c}%
\operatorname*{tr}U=2\sqrt{2}e^{i(\alpha+\beta)},\\
\det U=e^{4i(\alpha+\beta)},
\end{array}
,\label{ua1}\\
\text{eigenvalues}  &  \text{: }\left\{  -\left(  -1\right)  ^{3/4}e^{i\left(
\alpha+\beta\right)  }\right\}  ^{\left[  2\right]  },\left\{  \left(
-1\right)  ^{1/4}e^{i\left(  \alpha+\beta\right)  }\right\}  ^{\left[
2\right]  }. \label{e1}%
\end{align}

With the particular choice of parameters $\alpha=\beta=0$ and lower signs, the
solution (\ref{ua1}) coincides with the $8$-vertex braiding gate of
\cite{kau/lom2004}.

Next we search for unitary solutions among the invertible circle of $8$-vertex
traceless solutions (\ref{c4c}) to the matrix Yang-Baxter equation
(\ref{a123}) with parameters in the exponential form (\ref{x}). The unitarity
conditions (\ref{uu}) give the following equations on the parameters (\ref{x})%
\begin{align}
r  &  =r_{y}=r_{z},\ \ \ r^{2}\left(  r_{x}^{2}+r^{2}\right)  =1,\ \ \ r^{8}%
+r^{6}-2r^{4}+1=r^{2}\label{r}\\
\alpha-\beta &  =\frac{\pi}{2}. \label{a}%
\end{align}

The system of equations (\ref{r}) has two real positive (or zero) solutions%
\begin{align}
1)\ \ \ r_{x}  &  =1,\ \ \ r=\sqrt{\frac{\sqrt{5}-1}{2}},\label{r1}\\
2)\ \ \ r_{x}  &  =0,\ \ \ r=1. \label{r2}%
\end{align}

Thus, only the first solution leads to an $8$-vertex two-parameter unitary
braiding quantum gate of the form (we put $\gamma\mapsto\beta$ in (\ref{x}))%

\begin{align}
U_{rank=4}^{8-vert,circ}\left(  \alpha,\beta\right)   &  =\sqrt{\frac{\sqrt
{5}-1}{2}}\left(
\begin{array}
[c]{cccc}%
0 & e^{i\left(  \alpha+\beta\right)  } & ie^{i\left(  \alpha+\beta\right)
}\sqrt{\frac{\sqrt{5}-1}{2}} & 0\\
-e^{2i\alpha}\sqrt{\frac{\sqrt{5}-1}{2}} & 0 & 0 & e^{i\left(  \alpha
+\beta\right)  }\\
ie^{2i\alpha} & 0 & 0 & ie^{i\left(  \alpha+\beta\right)  }\sqrt{\frac
{\sqrt{5}-1}{2}}\\
0 & -e^{2i\alpha}\sqrt{\frac{\sqrt{5}-1}{2}} & ie^{2i\alpha} & 0
\end{array}
\right)  ,\label{ua2}\\
\det U  &  =e^{2i(3\alpha+\beta)}.
\end{align}

The second solution (\ref{r2}) gives $4$-vertex two-parameter unitary braiding
quantum gate (we also put $\gamma\mapsto\beta$ in (\ref{x}))%
\begin{equation}
U_{rank=4}^{4-vert,circ}\left(  \alpha,\beta\right)  =\left(
\begin{array}
[c]{cccc}%
0 & 0 & e^{i\left(  \alpha+\beta\right)  } & 0\\
e^{2i\alpha} & 0 & 0 & 0\\
0 & 0 & 0 & e^{i\left(  \alpha+\beta\right)  }\\
0 & e^{2i\alpha} & 0 & 0
\end{array}
\right)  ,\ \ \ \ \det U=-e^{2i(3\alpha+\beta)}. \label{ua3}%
\end{equation}

The non-invertible $8$-vertex circle solution (\ref{c2c}) to the Yang-Baxter
equation (\ref{a123}) cannot be partial unitary (\ref{uu1})--(\ref{uu2}) with
any values of its parameters.

\section{Higher braiding quantum gates}

In general, only special linear transformations of $2^{L}$-dimensional Hilbert
space can be treated as elementary quantum gates for an $L$-qubit state
\cite{nie/chu}. First, in the invertible case, the transformations should be
unitary (\ref{uu}), and in the hypothetical non-invertible case they can
satisfy partial unitarity (\ref{uu1})--(\ref{uu2}). Second, the braiding gates
have to be $2^{L}\times2^{L}$ matrix solutions to the constant Yang-Baxter
equation \cite{kau/lom2004} or higher braid equations (\ref{aa1}%
)--(\ref{aa2}). Here we consider (as a lowest case higher example) the ternary
braiding gates acting on $3$-qubit quantum states, i.e. $8\times8$ matrix
solutions to the ternary braid equations (\ref{pt}) which satisfy unitarity
(\ref{uu}) or partial unitarity (\ref{uu1})--(\ref{uu2}).

Note that all the permutation solutions (\ref{cp8b})--(\ref{cp8s}) are by
definition unitary, and are therefore ternary braiding gates \textquotedblleft
automatically\textquotedblright, and we call them \textit{permutation }%
$8$-\textit{vertex ternary braiding quantum gates} $U_{perm}^{8-vertex}$. By
the same reasoning the unitary version of the invertible star $8$-vertex
parameter-permutation solutions (\ref{b11})--(\ref{s22}) to the ternary braid
equations (\ref{pt}) will contain the complex numbers of unit magnitude as parameters.

Indeed, for the bisymmetric series (\ref{b11})--(\ref{b12}) of star-like
solutions we have $4$ two real parameter unitary ternary braiding quantum
gates ($\varkappa=\pm1$)%
\begin{equation}
U_{bisymm1}^{8-vertex}\left(  \alpha,\beta\right)  =\left(
\begin{array}
[c]{cccccccc}%
e^{i(\alpha+\beta)} & 0 & 0 & 0 & 0 & 0 & 0 & 0\\
0 & 0 & 0 & 0 & 0 & 0 & \varkappa e^{2i\beta} & 0\\
0 & 0 & e^{i(\alpha+\beta)} & 0 & 0 & 0 & 0 & 0\\
0 & 0 & 0 & 0 & \varkappa e^{2i\alpha} & 0 & 0 & 0\\
0 & 0 & 0 & \pm\varkappa e^{2i\beta} & 0 & 0 & 0 & 0\\
0 & 0 & 0 & 0 & 0 & e^{i(\alpha+\beta)} & 0 & 0\\
0 & \pm\varkappa e^{2i\alpha} & 0 & 0 & 0 & 0 & 0 & 0\\
0 & 0 & 0 & 0 & 0 & 0 & 0 & e^{i(\alpha+\beta)}%
\end{array}
\right)  ,\ \ \ \ \alpha,\beta\in\mathbb{R},\ \ \ \left\vert \alpha\right\vert
,\left\vert \beta\right\vert \leq2\pi, \label{u8b1}%
\end{equation}
which is a ternary analog of the first parameter-permutation solution to the
Yang-Baxter equation from (\ref{cp1}). The ternary analog of the second star
solution is the following unitary version of the bisymmetric series
(\ref{b21})--(\ref{b22})%
\begin{equation}
U_{bisymm2}^{8-vertex}\left(  \alpha,\beta\right)  =\left(
\begin{array}
[c]{cccccccc}%
0 & 0 & 0 & 0 & 0 & 0 & 0 & e^{6i\alpha}\\
0 & \varkappa e^{3i(\alpha+\beta)} & 0 & 0 & 0 & 0 & 0 & 0\\
0 & 0 & 0 & 0 & 0 & e^{2i(2\alpha+\beta)} & 0 & 0\\
0 & 0 & 0 & \varkappa e^{3i(\alpha+\beta)} & 0 & 0 & 0 & 0\\
0 & 0 & 0 & 0 & \varkappa e^{3i(\alpha+\beta)} & 0 & 0 & 0\\
0 & 0 & \pm e^{2i(\alpha+2\beta)} & 0 & 0 & 0 & 0 & 0\\
0 & 0 & 0 & 0 & 0 & 0 & \varkappa e^{3i(\alpha+\beta)} & 0\\
\pm e^{6i\beta} & 0 & 0 & 0 & 0 & 0 & 0 & 0
\end{array}
\right)  . \label{u8b2}%
\end{equation}

The same unitary ternary analogs of the symmetric series (\ref{s11}%
)--(\ref{s22}) for the first and the second circle-like solutions from
(\ref{cp2}) are%

\begin{equation}
U_{symm1}^{8-vertex}\left(  \alpha,\beta\right)  =\left(
\begin{array}
[c]{cccccccc}%
e^{i(\alpha+\beta)} & 0 & 0 & 0 & 0 & 0 & 0 & 0\\
0 & 0 & 0 & 0 & \varkappa e^{i(\alpha+\beta)} & 0 & 0 & 0\\
0 & 0 & 0 & 0 & 0 & 0 & 0 & e^{2i\beta}\\
0 & 0 & 0 & e^{i(\alpha+\beta)} & 0 & 0 & 0 & 0\\
0 & \pm\varkappa e^{i(\alpha+\beta)} & 0 & 0 & 0 & 0 & 0 & 0\\
0 & 0 & 0 & 0 & 0 & e^{i(\alpha+\beta)} & 0 & 0\\
0 & 0 & 0 & 0 & 0 & 0 & e^{i(\alpha+\beta)} & 0\\
0 & 0 & \pm e^{2i\alpha} & 0 & 0 & 0 & 0 & 0
\end{array}
\right)  , \label{u8s1}%
\end{equation}
and%
\begin{equation}
U_{symm2}^{8-vertex}\left(  \alpha,\beta\right)  =\left(
\begin{array}
[c]{cccccccc}%
0 & 0 & 0 & 0 & 0 & e^{2i\beta} & 0 & 0\\
0 & e^{i(\alpha+\beta)} & 0 & 0 & 0 & 0 & 0 & 0\\
0 & 0 & e^{i(\alpha+\beta)} & 0 & 0 & 0 & 0 & 0\\
0 & 0 & 0 & 0 & 0 & 0 & \varkappa e^{i(\alpha+\beta)} & 0\\
0 & 0 & 0 & 0 & e^{i(\alpha+\beta)} & 0 & 0 & 0\\
\pm e^{2i\alpha} & 0 & 0 & 0 & 0 & 0 & 0 & 0\\
0 & 0 & 0 & \pm\varkappa e^{i(\alpha+\beta)} & 0 & 0 & 0 & 0\\
0 & 0 & 0 & 0 & 0 & 0 & 0 & e^{i(\alpha+\beta)}%
\end{array}
\right)  , \label{u8s2}%
\end{equation}
respectively.

The invertible $16$-vertex star-like solutions (\ref{cv16}) to the ternary
braid equations (\ref{pt}) lead to the following two unitary one-parameter
ternary braiding quantum gates (cf. the binary case (\ref{ua1}))%
\begin{equation}
U_{3-qubits\pm}^{16-vertex}\left(  \alpha\right)  =\frac{1}{\sqrt{2}}\left(
\begin{array}
[c]{cccccccc}%
e^{3i\alpha} & 0 & 0 & 0 & 0 & 0 & 0 & -1\\
0 & e^{3i\alpha} & 0 & 0 & 0 & 0 & \mp e^{2i\alpha} & 0\\
0 & 0 & e^{3i\alpha} & 0 & 0 & -e^{2i\alpha} & 0 & 0\\
0 & 0 & 0 & e^{3i\alpha} & \mp e^{4i\alpha} & 0 & 0 & 0\\
0 & 0 & 0 & \pm e^{2i\alpha} & e^{3i\alpha} & 0 & 0 & 0\\
0 & 0 & e^{4i\alpha} & 0 & 0 & e^{3i\alpha} & 0 & 0\\
0 & \pm e^{4i\alpha} & 0 & 0 & 0 & 0 & e^{3i\alpha} & 0\\
e^{6i\alpha} & 0 & 0 & 0 & 0 & 0 & 0 & e^{3i\alpha}%
\end{array}
\right)  . \label{u16}%
\end{equation}

The braiding gate (\ref{u16}) is a ternary analog of (\ref{ua1}), and
therefore with $\alpha=0$ it can be treated as a ternary analog of the
$8$-vertex braiding gate considered in \cite{kau/lom2004}. Note that the
solution $U_{3-qubits+}^{16-vertex}\left(  0\right)  $ is transpose to the
so-called generalized Bell matrix \cite{row/zha/wu/ge}. Comparing (\ref{m8s})
and (\ref{u16}), we observe that the ternary braiding quantum gates (acting on
$3$ qubits) are those elements of the $16$-vertex star semigroup
$\mathcal{G}_{16vert}^{star}$ (\ref{mm8}), which satisfy unitarity (\ref{uu}).

In the same way, the$\ 32$-vertex analog the $8$-vertex binary braiding gate
of \cite{kau/lom2004} (now acting on $4$ qubits) is the following constant
$4$-ary braiding unitary quantum gate%
\begin{equation}
U_{4-qubits}^{32-vertex}=\frac{1}{\sqrt{2}}\left(
\begin{array}
[c]{cccccccccccccccc}%
1 & 0 & 0 & 0 & 0 & 0 & 0 & 0 & 0 & 0 & 0 & 0 & 0 & 0 & 0 & -1\\
0 & 1 & 0 & 0 & 0 & 0 & 0 & 0 & 0 & 0 & 0 & 0 & 0 & 0 & -1 & 0\\
0 & 0 & 1 & 0 & 0 & 0 & 0 & 0 & 0 & 0 & 0 & 0 & 0 & -1 & 0 & 0\\
0 & 0 & 0 & 1 & 0 & 0 & 0 & 0 & 0 & 0 & 0 & 0 & -1 & 0 & 0 & 0\\
0 & 0 & 0 & 0 & 1 & 0 & 0 & 0 & 0 & 0 & 0 & -1 & 0 & 0 & 0 & 0\\
0 & 0 & 0 & 0 & 0 & 1 & 0 & 0 & 0 & 0 & -1 & 0 & 0 & 0 & 0 & 0\\
0 & 0 & 0 & 0 & 0 & 0 & 1 & 0 & 0 & -1 & 0 & 0 & 0 & 0 & 0 & 0\\
0 & 0 & 0 & 0 & 0 & 0 & 0 & 1 & -1 & 0 & 0 & 0 & 0 & 0 & 0 & 0\\
0 & 0 & 0 & 0 & 0 & 0 & 0 & 1 & 1 & 0 & 0 & 0 & 0 & 0 & 0 & 0\\
0 & 0 & 0 & 0 & 0 & 0 & 1 & 0 & 0 & 1 & 0 & 0 & 0 & 0 & 0 & 0\\
0 & 0 & 0 & 0 & 0 & 1 & 0 & 0 & 0 & 0 & 1 & 0 & 0 & 0 & 0 & 0\\
0 & 0 & 0 & 0 & 1 & 0 & 0 & 0 & 0 & 0 & 0 & 1 & 0 & 0 & 0 & 0\\
0 & 0 & 0 & 1 & 0 & 0 & 0 & 0 & 0 & 0 & 0 & 0 & 1 & 0 & 0 & 0\\
0 & 0 & 1 & 0 & 0 & 0 & 0 & 0 & 0 & 0 & 0 & 0 & 0 & 1 & 0 & 0\\
0 & 1 & 0 & 0 & 0 & 0 & 0 & 0 & 0 & 0 & 0 & 0 & 0 & 0 & 1 & 0\\
1 & 0 & 0 & 0 & 0 & 0 & 0 & 0 & 0 & 0 & 0 & 0 & 0 & 0 & 0 & 1
\end{array}
\right)  .
\end{equation}

Thus, in general, the \textquotedblleft Minkowski\textquotedblright\ star
solutions for $n$-ary braid equations correspond to $2^{n}$-vertex braiding
unitary quantum gates as $2^{L}\times2^{L}$ matrices acting on $L=n$ qubits%
\begin{equation}
U_{L-qubits}^{2^{L}-vertex}=\frac{1}{\sqrt{2}}\left(
\begin{array}
[c]{cccccc}%
1 & 0 & 0 & 0 & 0 & -1\\
0 & \ddots & 0 & 0 & \iddots & 0\\
0 & 0 & 1 & -1 & 0 & 0\\
0 & 0 & 1 & 1 & 0 & 0\\
0 & \iddots & 0 & 0 & \ddots & 0\\
1 & 0 & 0 & 0 & 0 & 1
\end{array}
\right)  . \label{ul}%
\end{equation}

The braiding gate (\ref{ul}) can be treated as a polyadic ($n$-ary)
generalization of the GHZ generator (see, e.g., \cite{row/zha/wu/ge,bal/wu})
acting on $L=n$ qubits.

\section{Entangling braiding gates}

Entangled quantum states are obtained from separable states by acting with
special quantum gates on two-qubit states and multi-qubit states
\cite{jaf/oed,wal/gro/eis}. Here we consider the concrete form of braiding
gates which can be entangling or not entangling. There are general
considerations on these subjects for the Yang-Baxter maps
\cite{kau/lom2004,bal/san,pad/sug/tra2021} and generalized Yang-Baxter maps
\cite{che2012,vas/wan/won,row/zha/wu/ge,pad/sug/tra2020}. We present the
solutions for the binary and ternary braid maps introduced above, which
connect the parameters of the gate and the state.

\subsection{Entangling binary braiding gates}

Let us first examine, how the $8$-vertex star binary braiding gate
$U_{s}\left(  \alpha,\beta\right)  \equiv U_{rank=4}^{8-vert,star}\left(
\alpha,\beta\right)  $ (\ref{ua1}) acts on the product of one-qubit states
concretely. We use the Bloch representation (\ref{pb}) to obtain the
expression for the transformed concurrence (\ref{c2})%
\begin{equation}
C_{s\pm}^{\left(  2\right)  }\left(  U_{s}\left(  \alpha,\beta\right)
\left\vert \psi\left(  \theta_{1},\gamma_{1}\right)  \right\rangle
\otimes\left\vert \psi\left(  \theta_{2},\gamma_{2}\right)  \right\rangle
\right)  =\left\vert \left(  e^{i\left(  \beta+2\gamma_{1}\right)  }\sin
^{2}\frac{\theta_{1}}{2}\pm e^{i\alpha}\cos^{2}\frac{\theta_{1}}{2}\right)
\left(  e^{i\left(  \beta+2\gamma_{2}\right)  }\sin^{2}\frac{\theta_{2}}{2}\mp
e^{i\alpha}\cos^{2}\frac{\theta_{2}}{2}\right)  \right\vert . \label{cu}%
\end{equation}

In general, a braiding gate is \textit{entangling} if the transformed
concurrence (\ref{cu}) does not vanish, and its roots give the values of the
gate parameters $U\left(  \alpha,\beta\right)  $ for which the gate is
\textit{not entangling} for a given two-qubit state. In search of the
solutions for the transformed concurrence $C_{s\pm}^{\left(  2\right)  }=0$,
we observe that one of the qubits has to be on the Bloch sphere equator
$\theta_{1}=\frac{\pi}{2}$ (or $\theta_{2}=\frac{\pi}{2}$). Only in this case
can the first (or second) bracket in (\ref{cu}) vanish, and we obtain%
\begin{align}
\text{1) }C_{s+}^{\left(  2\right)  }  &  =0\text{, if }\theta_{1}=\frac{\pi
}{2}\text{ and }\alpha-\beta=2\gamma_{1}-\pi\text{, or }\theta_{2}=\frac{\pi
}{2}\text{ and }\alpha-\beta=2\gamma_{2}\text{;}\label{cc1}\\
\text{2) }C_{s-}^{\left(  2\right)  }  &  =0\text{, if }\theta_{1}=\frac{\pi
}{2}\text{ and }\alpha-\beta=2\gamma_{1}\text{, or }\theta_{2}=\frac{\pi}%
{2}\text{ and }\alpha-\beta=2\gamma_{2}-\pi. \label{cc2}%
\end{align}

Therefore the $8$-vertex star binary braiding gates (\ref{ua1}) with the
parameters fixed by (\ref{cc1})--(\ref{cc2}) are not entangling.

For the $8$-vertex circle binary braiding gate $U_{c}\left(  \alpha
,\beta\right)  \equiv U_{rank=4}^{8-vert,circ}\left(  \alpha,\beta\right)  $
(\ref{ua2}) we obtain%
\begin{align}
C_{c}^{\left(  2\right)  }\left(  U_{c}\left(  \alpha,\beta\right)  \left\vert
\psi\left(  \theta_{1},\gamma_{1}\right)  \right\rangle \otimes\left\vert
\psi\left(  \theta_{1},\gamma_{1}\right)  \right\rangle \right)   &
=W\left\vert \left(  e^{i\left(  \beta+2\gamma_{1}\right)  }\sin^{2}%
\frac{\theta_{1}}{2}-ie^{i\alpha}\cos^{2}\frac{\theta_{1}}{2}\right)  \left(
e^{i\left(  \beta+2\gamma_{2}\right)  }\sin^{2}\frac{\theta_{2}}%
{2}-ie^{i\alpha}\cos^{2}\frac{\theta_{2}}{2}\right)  \right\vert ,\nonumber\\
W  &  =\frac{\left(  \sqrt{5}-1\right)  ^{\frac{3}{2}}}{\sqrt{2}}=0.971\,74.
\end{align}

Analogously to (\ref{cc1})--(\ref{cc2}), the concurrence of the states
transformed by the $8$-vertex circle binary braiding gate (\ref{ua2}) can
vanish if%
\begin{equation}
C_{c}^{\left(  2\right)  }=0\text{, if }\theta_{1}=\frac{\pi}{2}\text{ and
}\alpha-\beta=2\gamma_{1}-\frac{\pi}{2}\text{, or }\theta_{2}=\frac{\pi}%
{2}\text{ and }\alpha-\beta=2\gamma_{2}-\frac{\pi}{2}. \label{cc3}%
\end{equation}

Thus, the $8$-vertex circle binary braiding gates (\ref{ua2}) are not
entangling if the parameters satisfy (\ref{cc3}).

In the case of the $4$-vertex circle binary braiding gate (\ref{ua3}) the
transformed concurrence vanishes identically, and therefore this gate is not
entangling for any values of its parameters.

\subsection{Entangling ternary braiding gates}

Let us consider the tensor product of three qubit pure states $\left\vert
\psi\left(  \theta_{1},\gamma_{1}\right)  \right\rangle \otimes\left\vert
\psi\left(  \theta_{2},\gamma_{2}\right)  \right\rangle \otimes\left\vert
\psi\left(  \theta_{3},\gamma_{3}\right)  \right\rangle $ (in the Bloch
representation (\ref{pb})), which obviously has zero concurrence $C^{\left(
3\right)  }$ (\ref{c3}), because of the vanishing of the hyperdeterminant
(\ref{hd3}). After transforming by the $16$-vertex star ternary braiding gates
$U_{16}\left(  \alpha\right)  \equiv U_{3-qubits}^{16-vertex}\left(
\alpha\right)  $ (\ref{u16}) the concurrence becomes%
\begin{align}
&  C_{16\pm}^{\left(  3\right)  }\left(  U_{16}\left(  \alpha\right)
\left\vert \psi\left(  \theta_{1},\gamma_{1}\right)  \right\rangle
\otimes\left\vert \psi\left(  \theta_{2},\gamma_{2}\right)  \right\rangle
\otimes\left\vert \psi\left(  \theta_{3},\gamma_{3}\right)  \right\rangle
\right)  =\label{c316}\\
&  \frac{1}{64}\left\vert \left(  \left(  e^{2i\alpha}\pm e^{2i\gamma_{1}%
}+(e^{2i\alpha}\mp e^{2i\gamma_{1}})\cos\theta_{1}\right)  \left(
e^{2i\alpha}-e^{2i\gamma_{2}}+(e^{2i\alpha}+e^{2i\gamma_{2}})\cos\theta
_{2}\right)  \left(  e^{2i\alpha}\mp e^{2i\gamma_{3}}+(e^{2i\alpha}\pm
e^{2i\gamma_{3}})\cos\theta_{3}\right)  \right)  ^{2}\right\vert .\nonumber
\end{align}

We observe that the ternary concurrence (\ref{c316}) vanishes if any of the
brackets is equal to zero. Because the domain of all angles is $\mathbb{R}$,
we have solutions only for fixed discrete $\theta_{k}=\pi,-\pi,\pi/2$,
$k=1,2,3$, which means that on the Bloch sphere the quantum states should be
on the equator (as in the binary case), or additionally at the poles. In this
case, $e^{i\alpha}=\pm e^{i\gamma_{k}}$, and%
\begin{equation}
\alpha=\left\{
\begin{array}
[c]{c}%
\gamma_{k}\\
\gamma_{k}+\pi
\end{array}
\right.  ,\ \ k=1,2,3. \label{ag}%
\end{equation}

Thus, for a fixed three-qubit product state one (or more) of which is at a
pole or the equator of the Bloch sphere, those ternary braiding gates
$U_{16}\left(  \alpha\right)  $ satisfying the conditions (\ref{ag}) are not
entangling $C_{16\pm}^{\left(  3\right)  }=0$, whereas in other cases they are
entangling $C_{16\pm}^{\left(  3\right)  }\neq0$.

By analogy, a similar action of the $8$-vertex bisymmetric (star-like) ternary
braiding gates $U_{8b1,2}\left(  \alpha,\beta\right)  \equiv U_{bisymm1,2}%
^{8-vertex}\left(  \alpha,\beta\right)  $ (\ref{u8b1})--(\ref{u8b2}) gives%
\begin{align}
C_{8b1}^{\left(  3\right)  }\left(  U_{8b1}\left(  \alpha,\beta\right)
\left\vert \psi\left(  \theta_{1},\gamma_{1}\right)  \right\rangle
\otimes\left\vert \psi\left(  \theta_{2},\gamma_{2}\right)  \right\rangle
\otimes\left\vert \psi\left(  \theta_{3},\gamma_{3}\right)  \right\rangle
\right)   &  =\left\vert \sin^{2}\theta_{1}\sin^{2}\theta_{3}\left(
e^{2i\left(  \beta+\gamma_{2}\right)  }\sin^{2}\frac{\theta_{2}}%
{2}-e^{2i\alpha}\cos^{2}\frac{\theta_{2}}{2}\right)  ^{2}\right\vert ,\\
C_{8b2}^{\left(  3\right)  }\left(  U_{8b2}\left(  \alpha,\beta\right)
\left\vert \psi\left(  \theta_{1},\gamma_{1}\right)  \right\rangle
\otimes\left\vert \psi\left(  \theta_{2},\gamma_{2}\right)  \right\rangle
\otimes\left\vert \psi\left(  \theta_{3},\gamma_{3}\right)  \right\rangle
\right)   &  =\left\vert \sin^{2}\theta_{1}\sin^{2}\theta_{3}\left(
e^{2i\left(  \alpha+\gamma_{2}\right)  }\sin^{2}\frac{\theta_{2}}%
{2}-e^{2i\beta}\cos^{2}\frac{\theta_{2}}{2}\right)  ^{2}\right\vert .
\end{align}
Their solutions coincide with the binary case (\ref{cc1})--(\ref{cc2}) applied
to the middle qubit $\left\vert \psi\left(  \theta_{2},\gamma_{2}\right)
\right\rangle $ and $\gamma_{2}\rightarrow2\gamma_{2}$.

The action of the $8$-vertex symmetric (circle-like) ternary braiding gates
$U_{8s}\left(  \alpha,\beta\right)  \equiv U_{symm1,2}^{8-vertex}\left(
\alpha,\beta\right)  $ (\ref{u8s1})--(\ref{u8s2}) leads to the transformed
concurrence%
\begin{align}
&  C_{8s}^{\left(  3\right)  }\left(  U_{8s}\left(  \alpha,\beta\right)
\left\vert \psi\left(  \theta_{1},\gamma_{1}\right)  \right\rangle
\otimes\left\vert \psi\left(  \theta_{2},\gamma_{2}\right)  \right\rangle
\otimes\left\vert \psi\left(  \theta_{3},\gamma_{3}\right)  \right\rangle
\right) \nonumber\\
&  =\left\vert \sin^{2}\theta_{2}\left(  e^{i\left(  \beta+2\gamma_{1}\right)
}\sin^{2}\frac{\theta_{1}}{2}-e^{i\alpha}\cos^{2}\frac{\theta_{1}}{2}\right)
\left(  e^{i\left(  \beta+2\gamma_{3}\right)  }\sin^{2}\frac{\theta_{3}}%
{2}-e^{i\alpha}\cos^{2}\frac{\theta_{3}}{2}\right)  \right\vert .
\end{align}
The conditions for this to vanish (i.e. when the gate $U_{8s}\left(
\alpha,\beta\right)  $ becomes not entangling) coincide with those for the
binary case (\ref{cc1})--(\ref{cc2}), applied here to the first or the third qubit.

Thus we have shown that the braiding binary and ternary quantum gates can be
either entangling or not entangling, depending on how their parameters are
related to the concrete quantum state on which they act. The constructions
presented here could be used, e.g. in the entanglement-free protocols
\cite{deb/bar,reh/shi} and some experiments \cite{alm/fed/bro,hig/ber/bar}.
This can also allow us to build quantum networks without any entangling at all
(\textit{non-entangling networks}), when the next gate depends upon the
previous state in such a way that at each step there is no entangling, as the
separable, but different, final state is received from a separable initial state.

\bigskip

\textbf{Acknowledgement}. The first author (S.D.) is grateful to Vladimir
Akulov, Mike Hewitt, Mikhail Krivoruchenko, Grigorij Kurinnoy, Thomas Nordahl,
Sergey Prokushkin, Vladimir Tkach, Alexander Voronov and Wend Werner for
fruitful discussions.

\end{document}